\documentclass[reprint,amsmath,amssymb,onecolumn,aps,11pt]{revtex4-2}

\usepackage[utf8]{inputenc} %
\usepackage{graphicx}%
\usepackage{dcolumn}%
\usepackage{bm}%
\usepackage{booktabs}%
\usepackage{xcolor}

\graphicspath{{figures/}}
\newcommand{\Mmat}{\mathbf{M}}

\begin{document}

\title{A linear stability method based on generalized hydrodynamics: application to high-speed plane Couette flow beyond continuum regime}

\author{Zheng Pang}
\author{Jihui Ou}
\email{Corresponding author: oujihui@tju.edu.cn}
\affiliation{Department of Mechanics, School of Mechanical Engineering, Tianjin University, Tianjin 300072, China}

\begin{abstract}

Linear stability analysis of compressible shear flows beyond continuum regime attracts increasing attention in recent years accounting for high-altitude applications. In this paper, the temporal stability of high-speed plane Couette flow in near-continuum regime is addressed. A novel linear stability method based on the generalized hydrodynamic equations (GHE) with slip boundary conditions is developed within the 13-moment framework, and the corresponding numerical method is formulated. Meanwhile, the linear stability equations of the nonlinear coupled constitutive relations (NCCR) and the Navier--Stokes (NS) equations are also given within the moment framework. Based on the developed methods, the linear stability of plane Couette flows at $\mathit{Ma}=2.5$--10 and $\mathit{Kn}=2.7\times 10^{-5}$--0.07 are analyzed, and the rarefaction effects on the base flow, eigenspectrum, mode synchronization and least stable mode are investigated. The results show that the GHE with slip boundary conditions agrees well with direct simulation Monte Carlo (DSMC) results in predicting base flow profiles of velocity, temperature, all stress and heat flux components, while the NS equations show considerable differences, especially for capturing the nonlinear normal stress and tangential heat flux. The GHE-based linear stability theory (LST) recovers to the classical NS-based LST in the continuum limit, while they show considerable differences when $\mathit{Kn}$ reaches $O(0.01)$. As rarefaction degree increases, the synchronization between acoustic modes and viscous modes becomes weak, and all modes become stable in rarefied plane Couette flow. The surface slip effect and nonlinear transport increase the disturbance phase speeds, while they have different effects on the growth rate of different modes. Generally, rarefaction effects increase the growth rate of the least stable Mode II at low wavenumbers, and show more pronounced effects on higher modes and high-wavenumber disturbances. %

\end{abstract}


\maketitle

\section{Introduction}\label{sec:1}

Linear stability analysis of compressible shear flows is a fundamental problem in fluid mechanics, which is not only of academic interest, but also very important for understanding and predicting transition in associated engineering applications~\cite{ref1,ref2,ref3}. Generally, the linear stability problem focuses on the evolution of infinitesimal disturbances superimposed on a steady base flow. The stability characteristics of different types of shear flows at high Reynolds numbers have been revealed in the past decades, such as plane Couette and Poiseuille flows (wall-bounded shear flows)~\cite{ref4,ref5,ref6}, boundary layers (external flows)~\cite{ref2,ref7,ref8}, mixing layers and wakes (free shear flows)~\cite{ref9,ref10,ref11}. However, in high-altitude applications or microflow systems, the local flow may involve rarefaction effects, which are usually induced by combined effects of low density, small system size and large flow gradients. The linear stability of such rarefied shear flows has received attention in recent years~\cite{refWang2025}. However, in such flows, the conventional linear stability theory (LST) based on the Navier--Stokes (NS) equations breaks down, and the rarefaction effects on stability are not well understood. The plane Couette flow is a classical problem in fluid mechanics, and also serves as a typical simplified prototype for studying stability of shear flows. Herein, the linear stability of plane Couette flow at high speeds and beyond continuum regime is addressed.

The linear stability of plane Couette flow at high Reynolds numbers has been largely studied in the past decades. Morawetz~\cite{ref12} analyzed the eigenvalue spectrum of incompressible plane Couette flow in detail, and Romanov~\cite{ref13} theoretically showed that the flow is linearly stable to both inviscid and viscous disturbances in all parameter ranges. Duck et al.~\cite{ref4} investigated the inviscid stability of compressible plane Couette flow, and showed that supersonic plane Couette flow admits multiple inviscid instability modes, which are acoustic modes in nature. These modes are classified into two families, i.e. the odd modes I, III, V, \ldots{} and the even modes II, IV, VI, \ldots{}, according to their phase speeds as the wavenumber tends to zero. Hu and Zhong~\cite{ref5} analyzed the viscous stability of compressible plane Couette flow and pointed out that Mode II is the dominant instability mode. Meanwhile, viscosity has both stabilizing and destabilizing effects on Mode II. Malik et al.~\cite{ref14} studied the effect of viscosity stratification on the stability of plane Couette flow, and Ramachandran et al.~\cite{ref15} investigated the effect of the Prandtl number on stability. Deka et al.~\cite{ref16,ref17} theoretically analyzed the viscous instability of compressible channel and pipe flows using asymptotic theory, and showed that the viscous instability of Modes II, IV, etc. arises from the existence of a critical point in the flow profile, across which the shear production term exceeds viscous dissipation. More recently, Ou~\cite{ref6} performed a detailed analysis of the instability of compressible plane Couette and Poiseuille flows, and showed that the discrete acoustic modes in plane Couette flow can be synchronized with the viscous branch and the acoustic modes, leading to complex branching topology (involving multiple modes) and viscous or inviscid instability modes. Meanwhile, the dominant instability mode from low speed to high speed in plane Couette and Poiseuille flows are revealed, and the results show that the viscous or inviscid Mode II is usually most unstable in plane Couette flow. 

Considering high-altitude and microflow applications, the linear stability of near-continuum flows has attracted increasing attention in recent years. The stability characteristics of some simple shear flows were analyzed based on the gas kinetic theory.
Yoshida and Aoki~\cite{ref18} analyzed the stability of rarefied cylindrical Couette flow by solving the initial--boundary-value problem of the linearized Bhatnagar-Gross-Krook (BGK) model equation under small disturbance assumptions, and obtained the instability modes at different parameters. Zou et al.~\cite{ref19,ref20,ref21} derived linear stability equations from the BGK equation and investigated the stability characteristics of rarefied plane Couette and Poiseuille flows by solving the eigenvalue problem in temporal mode. The results show that the BGK-based stability analyses recover to the conventional LST based on the NS equations in the continuum regime. As flow becomes rarefied, the NS-based stability analysis underestimated the growth rate of the least stable mode, though the slip boundary conditions are adopted.
Karpuzcu et al.~\cite{refKarpuzcu} also developed a BGK-based linear stability method, validated it against the results of Zou et al.~\cite{ref21} for plane Couette flow, and further analyzed the temporal stability of the one-dimensional shock structure.
Wu~\cite{ref22} developed an iterative method for solving the eigenvalue problem of kinetic disturbance equations by combining the shifted inverse power method with the general synthetic iterative scheme (GSIS), and applied it to analyze planar sound waves and low-speed plane Couette flow. On the whole, the above studies illustrate the performance of Boltzmann-type equations for simple shear-flow stability analysis, however, the computational cost and storage requirement of such methods are enormous, especially for hypersonic flows, since the Boltzmann equation involves both physical-space and velocity-space discretizations. 

From a macroscopic view, the rarefaction effects on flow instability of typical shear flows have also been investigated, by accounting for surface slip effects and nonlinear constitutive relations. Many studies have analyzed the stability of low-speed slip Couette and Poiseuille flows, and the results show that slip increases the critical Reynolds number and suppresses the instability~\cite{ref23,ref24,ref25}. Liu and Zhang~\cite{ref26} studied the slip effect on the stability and transition of the incompressible flat-plate boundary layer, and found that slip effect shifts the neutral curve of the TS wave and the transition location downstream. For compressible boundary layers, Ou and Chen~\cite{ref27} analyzed the effects of velocity slip and temperature jump on the stability of supersonic boundary layers, and showed that surface slip stabilizes the second mode while destabilizes the first mode. For more rarefied flows, Ou et al.~\cite{ref28} incorporated an effective constitutive model into the stability equations, and found that the rarefaction effects suppress the second mode instability by changing the base flow while enhance it through modifying the stability equation. In these studies, the conventional LST is extended to slightly rarefied flow by incorporating surface slip effects and nonlinear transport effects within the conventional stability framework.

In rarefied gas dynamics, various extended hydrodynamic equations have been derived from the Boltzmann equation through Chapman-Enskog expansion or moment method, such as the Burnett equation~\cite{ref29,ref30}, Grad's 13 moment equation (G13)~\cite{ref31}, regularized 13 moment equation (R13)~\cite{ref32}, generalized hydrodynamic equations (GHE)~\cite{ref33} and nonlinear coupled constitutive relations (NCCR)~\cite{ref34}. These extended hydrodynamic equations are generally applicable to early transition regime, with Knudsen number $\mathit{Kn}\sim 1$. Some of these equations were also used to analyze the linear stability of simple flows.
Jiang et al.~\cite{ref35} analyzed the linear stability of the GHE and NCCR equations in rest and uniform flow state, and the results show that unlike the Burnett equations, the GHE and NCCR equations are unconditionally linearly stable to small disturbances of all wavenumbers and frequencies in both states, exhibiting the same stability behavior as the NS equations.
Khayat and Eu~\cite{ref36} studied the stability of cylindrical Couette flow based on the GHE, and pointed out that as rarefaction degree increases, the stability results obtained from the GHE differ from those by NS stability theory. At present, there are very few studies that apply the extended hydrodynamic equations to the stability analysis of high-speed shear flows.

The present study mainly focuses on the linear stability of high-speed plane Couette flow in near-continuum regime. Generally speaking, the shear-flow instability is a near-continuum flow phenomenon, with the critical Knudsen number not exceeding $O(0.1)$. In such parameter range, extended hydrodynamic equations are generally applicable. Developing a linear stability method based on extended hydrodynamic equations can not only capture rarefaction effects, but also achieve high computational efficiency compared with solving Boltzmann-type equations, which might be a suitable option for near-continuum flow stability analysis, especially at high Mach numbers. Hence, the present study aims to develop a linear stability method based on the GHE along with surface slip models, and to investigate the rarefaction effects on the stability characteristics of near-continuum plane Couette flow, including eigenvalue spectrum, mode synchronization and the least stable Mode II.
The remainder of this paper is organized as follows. Section~\ref{sec:2} introduces the governing equations. Section~\ref{sec:3} gives the problem description and obtains the steady base flow. Section~\ref{sec:4} formulates the linear stability theory based on the GHE. Section~\ref{sec:5} presents the base-flow and linear stability results for plane Couette flow at different Mach and Knudsen numbers.
Finally, conclusions are drawn in Section~\ref{sec:6}.

\section{Governing equations}\label{sec:2}

The Boltzmann equation provides a unified description of gas flows from continuous to rarefied. For a single-component monatomic gas, the evolution equation of the molecular velocity distribution function $f^{*}\left( \boldsymbol{c}^{*},\;\boldsymbol{x}^{*},\;t^{*} \right)$ reads
\begin{equation}
\left[ \frac{\partial}{\partial t^{*}}+\boldsymbol{c}^{*}\cdot\nabla \right]f^{*}\left( \boldsymbol{c}^{*},\;\boldsymbol{x}^{*},\;t^{*} \right)=J\left[ f^{*},f_{1}^{*} \right]
\label{eq:1}
\end{equation}
where the superscript `*' denotes dimensional quantities, $t^{*}$ is the time, $\boldsymbol{c}^{*}$ is the molecular velocity, $\boldsymbol{x}^{*}$ is the position vector, and $J\left[ f^{*},f_{1}^{*} \right]$ is the collision integral of the Boltzmann equation.
Multiplying the Boltzmann equation by $m^{*}$, $m^{*}\boldsymbol{c}^{*}$ and $1/2\,m^{*}c^{*2}$, where $m^{*}$ is the molecular mass, and integrating over the velocity space yields the conservation equations of mass, momentum and energy
\begin{equation}
\left\{ \begin{aligned}&\frac{D\rho^{*}}{Dt^{*}}+\rho^{*}\nabla\cdot\boldsymbol{u}^{*}=0 \\ &\rho^{*}\frac{D\boldsymbol{u}^{*}}{Dt^{*}}+\nabla p^{*}+\nabla\cdot\boldsymbol{\Pi}^{*}=0 \\ &\rho^{*}c_{p}^{*}\frac{DT^{*}}{Dt^{*}}-\frac{Dp^{*}}{Dt^{*}}+\boldsymbol{\Pi}^{*}:\nabla\boldsymbol{u}^{*}+\nabla\cdot\boldsymbol{Q}^{*}=0\end{aligned} \right.
\label{eq:2}
\end{equation}
where $\rho^{*}$ is the gas density, $\boldsymbol{u}^{*}$ is the velocity vector, $T^{*}$ is the temperature, $p^{*}$ is the pressure, $c_{p}^{*}$ is the specific heat at constant pressure, $\boldsymbol{\Pi}^{*}$ is the viscous stress tensor, $\boldsymbol{Q}^{*}$ is the heat flux vector, and $D/Dt^{*}$ is the material derivative.
These conservation equations are obtained by directly integrating the Boltzmann equation without any simplification, and therefore hold for both continuous and rarefied flows.
However, the stress tensor $\boldsymbol{\Pi}^{*}$ and the heat flux vector $\boldsymbol{Q}^{*}$ in Eq.~\eqref{eq:2} are not closed.
In the continuum regime, the conservation equations can be closed by the generalized Newton's law of viscosity and Fourier's law of heat conduction, written as
\begin{equation}
\boldsymbol{\Pi}_{NS}^{*}=-2\mu^{*}\left[ \nabla\boldsymbol{u}^{*} \right]^{(2)},\quad \boldsymbol{Q}_{NS}^{*}=-k^{*}\nabla T^{*}
\label{eq:3}
\end{equation}
where $\mu^{*}$ and $k^{*}$ are the viscosity and thermal conductivity, respectively, and $\left[ \nabla\boldsymbol{u}^{*} \right]^{(2)}$ is the traceless symmetric part of $\nabla\boldsymbol{u}^{*}$, defined as
\begin{equation}
\left[ \nabla\boldsymbol{u}^{*} \right]^{(2)}=\frac{1}{2}\left( \frac{\partial u_{i}^{*}}{\partial x_{j}^{*}}+\frac{\partial u_{j}^{*}}{\partial x_{i}^{*}} \right)-\frac{1}{3}\delta_{ij}\frac{\partial u_{k}^{*}}{\partial x_{k}^{*}}
\label{eq:4}
\end{equation}
where $\delta_{ij}$ is the Kronecker delta and $(i,j,k)=1,2,3$.
The conservation equations \eqref{eq:2} combined with the linear constitutive relations \eqref{eq:3} constitute the conventional NS equations, which can also be rigorously derived from the Boltzmann equation through the Chapman--Enskog expansion.
However, the NS equations are only valid in the continuum regime, and gradually break down as flow becomes rarefied. For rarefied transitional flows, various moment equations, describing the evolution of the stress tensor and heat flux vector, have been proposed. Eu~\cite{ref33} has proposed a modified moment method by combining the concepts of irreversible thermodynamics with the nonequilibrium ensemble method. The generalized hydrodynamic equations are developed based on a nonequilibrium canonical distribution function and a cumulant expansion of the collisional integral, with kinematic and collision terms both truncated to second order.
For a single-component gas without body forces and bulk viscosity, the generalized hydrodynamic equations are written as~\cite{ref33}
\begin{equation}
\frac{D\boldsymbol{\Pi}^{*}}{Dt^{*}}+\boldsymbol{\Pi}^{*}\nabla\cdot\boldsymbol{u}^{*}+2[\boldsymbol{\Pi}^{*}\cdot\nabla\boldsymbol{u}^{*}]^{(2)}+2p^{*}[\nabla\boldsymbol{u}^{*}]^{(2)}=-\frac{p^{*}}{\mu^{*}}\boldsymbol{\Pi}^{*}q\left( \kappa \right),
\label{eq:5}
\end{equation}
\begin{equation}
\frac{D\boldsymbol{Q}^{*}}{Dt^{*}}+\boldsymbol{Q}^{*}\nabla\cdot\boldsymbol{u}^{*}+\frac{D\boldsymbol{u}^{*}}{Dt^{*}}\cdot\boldsymbol{\Pi}^{*}+\left( \boldsymbol{\Pi}^{*}+p^{*}\mathbf{I} \right)\cdot c_{p}^{*}\nabla T^{*}+\boldsymbol{Q}^{*}\cdot\nabla\boldsymbol{u}^{*}=-\frac{p^{*}c_{p}^{*}}{k^{*}}\boldsymbol{Q}^{*}q\left( \kappa \right)
\label{eq:6}
\end{equation}
where $\mathbf{I}$ is the unit tensor, and $[\boldsymbol{\Pi}^{*}\cdot\nabla\boldsymbol{u}^{*}]^{(2)}$ is given by
\begin{equation}
[\boldsymbol{\Pi}^{*}\cdot\nabla\boldsymbol{u}^{*}]^{(2)}=\frac{1}{2}\left( \Pi_{ki}^{*}\frac{\partial u_{j}^{*}}{\partial x_{k}^{*}}+\Pi_{lj}^{*}\frac{\partial u_{i}^{*}}{\partial x_{l}^{*}} \right)-\frac{1}{3}\Pi_{kl}^{*}\frac{\partial u_{k}^{*}}{\partial x_{l}^{*}}\delta_{ij}
\label{eq:7}
\end{equation}
Compared with Grad's moment equations (such as G13), the unique feature of GHE is the nonlinear factor $q\left( \kappa \right)$, which is related to entropy production, where $\kappa$ is the modified Rayleigh--Onsager dissipation function, written as
\begin{equation}
q\left( \kappa \right)=\frac{\sinh (\kappa)}{\kappa},\quad \kappa=\frac{\left( m^{*}k_{B}^{*}T^{*} \right)^{1/4}}{\sqrt{2}d^{*}}\frac{1}{p^{*}}\left[ \frac{\boldsymbol{\Pi}^{*}:\boldsymbol{\Pi}^{*}}{2\mu^{*}}+\frac{\boldsymbol{Q}^{*}\cdot\boldsymbol{Q}^{*}}{k^{*}T^{*}} \right]^{1/2}
\label{eq:8}
\end{equation}
where $k_{B}^{*}$ is the Boltzmann constant and $d^{*}$ is the molecular diameter.
Note that $\kappa$ and $q\left( \kappa \right)$ are both dimensionless.
For a continuous flow, $\kappa \to 0$ and $q\left( \kappa \right) \to 1$. As flow becomes rarefied, $\kappa > 0$ and $q\left( \kappa \right)$ is always greater than 1.
The state equation of ideal gas is written as
\begin{equation}
p^{*}=\rho^{*}R^{*}T^{*}
\label{eq:9}
\end{equation}
where $R^{*}$ is the gas constant.

The GHE are nondimensionalized as follows:
\begin{equation}
\begin{aligned}&t=\frac{t^{*}}{L^{*}/u_{r}^{*}},\;\;\boldsymbol{x}=\frac{\boldsymbol{x}^{*}}{L^{*}},\;\;\rho=\frac{\rho^{*}}{\rho_{r}^{*}},\;\;\boldsymbol{u}=\frac{\boldsymbol{u}^{*}}{u_{r}^{*}},\;\;T=\frac{T^{*}}{T_{r}^{*}},\;\;p=\frac{p^{*}}{\rho_{r}^{*}u_{r}^{*2}}, \\ &\mu=\frac{\mu^{*}}{\mu_{r}^{*}},\;\;k=\frac{k^{*}}{k_{r}^{*}},\;\;\boldsymbol{\Pi}=\frac{\boldsymbol{\Pi}^{*}}{\mu_{r}^{*}u_{r}^{*}/L^{*}},\;\;\boldsymbol{Q}=\frac{\boldsymbol{Q}^{*}}{k_{r}^{*}T_{r}^{*}/L^{*}}\end{aligned}
\label{eq:10}
\end{equation}
where the subscript `$r$' denotes the corresponding reference values and $L^{*}$ is the reference length.
The nondimensional generalized hydrodynamic equations are
\begin{equation}
\left\{ \begin{aligned}&\frac{\partial\rho}{\partial t}+\boldsymbol{u}\cdot\nabla\rho+\rho\nabla\cdot\boldsymbol{u}=0 \\ &\rho\left( \frac{\partial\boldsymbol{u}}{\partial t}+\boldsymbol{u}\cdot\nabla\boldsymbol{u} \right)+\nabla p+\frac{1}{\mathit{Re}}\nabla\cdot\boldsymbol{\Pi}=0 \\ &\rho\left( \frac{\partial T}{\partial t}+\boldsymbol{u}\cdot\nabla T \right)-\left( \gamma-1 \right)\mathit{Ma}^{2}\left( \frac{\partial p}{\partial t}+\boldsymbol{u}\cdot\nabla p \right)+\frac{\left( \gamma-1 \right)\mathit{Ma}^{2}}{\mathit{Re}}\boldsymbol{\Pi}:\nabla\boldsymbol{u}+\frac{1}{\mathit{Pr}\mathit{Re}}\nabla\cdot\boldsymbol{Q}=0\end{aligned} \right.
\label{eq:11}
\end{equation}
and
\begin{equation}
\frac{\partial\boldsymbol{\Pi}}{\partial t}+\boldsymbol{u}\cdot\nabla\boldsymbol{\Pi}+\boldsymbol{\Pi}\nabla\cdot\boldsymbol{u}+2[\boldsymbol{\Pi}\cdot\nabla\boldsymbol{u}]^{(2)}+\underline{\mathit{Re}2p[\nabla\boldsymbol{u}]^{(2)}}=\underline{-\mathit{Re}\frac{p}{\mu}\boldsymbol{\Pi}q\left( \kappa \right)},
\label{eq:12}
\end{equation}
\begin{equation}
\begin{aligned}&\frac{\partial\boldsymbol{Q}}{\partial t}+\boldsymbol{u}\cdot\nabla\boldsymbol{Q}+\boldsymbol{Q}\nabla\cdot\boldsymbol{u}+\left( \gamma-1 \right)\mathit{Pr}\,\mathit{Ma}^{2}\left( \frac{\partial\boldsymbol{u}}{\partial t}+\boldsymbol{u}\cdot\nabla\boldsymbol{u} \right)\cdot\boldsymbol{\Pi}+\mathit{Pr}\boldsymbol{\Pi}\cdot\nabla T+\boldsymbol{Q}\cdot\nabla\boldsymbol{u} \\ &+\underline{\mathit{Re}\mathit{Pr}p\nabla T}=\underline{-\mathit{Re}\mathit{Pr}\frac{p}{k}\boldsymbol{Q}q\left( \kappa \right)}\end{aligned}
\label{eq:13}
\end{equation}
The gas state equation is nondimensionalized as
\begin{equation}
p=\frac{1}{\gamma \mathit{Ma}^{2}}\rho T
\label{eq:14}
\end{equation}
In the above equations, $\gamma$ is the specific heat ratio, and the Mach number, Reynolds number and Prandtl number are defined as
\begin{equation}
\mathit{Ma}=\frac{u_{r}^{*}}{\sqrt{\gamma R^{*}T_{r}^{*}}},\quad \mathit{Re}=\frac{\rho_{r}^{*}u_{r}^{*}H^{*}}{\mu_{r}^{*}},\quad \mathit{Pr}=\frac{\mu_{r}^{*}c_{p}^{*}}{k_{r}^{*}}
\label{eq:15}
\end{equation}
The nonlinear factor and the Rayleigh--Onsager dissipation function are
\begin{equation}
q\left( \kappa \right)=\frac{\sinh \left( \kappa \right)}{\kappa},\quad \kappa=\frac{c}{\mathit{Re}\,p}\sqrt{\boldsymbol{\Pi}:\boldsymbol{\Pi}+\;\frac{2\varepsilon}{T}\boldsymbol{Q}\cdot\boldsymbol{Q}}
\label{eq:16}
\end{equation}
where $c=\left[ \left( 2\sqrt{\pi}/5 \right)A_{2}\left( \nu \right)\Gamma\left[ 4-2/\left( \nu-1 \right) \right] \right]^{1/2}$ is a constant, depending only on the inverse-power-law exponent $\nu$, and having a value between 1.0138 and 1.2232. $\varepsilon=\frac{1}{\left( \gamma-1 \right)\mathit{Pr}\,\mathit{Ma}^{2}}$ is a dimensionless parameter.
The gas is assumed to be calorically perfect, with $\gamma$, $\mathit{Pr}$, $c_{p}^{*}$ and $R^{*}$ being constants.
The viscosity is computed by the power law or Sutherland's formula, and the thermal conductivity is computed from the viscosity through the Prandtl number.
The nondimensional viscosity and thermal conductivity are given by
\begin{equation}
\left\{ \begin{aligned}&\mu=k=T^{s}, &&\text{Power law} \\ &\mu=k=T^{3/2}\frac{1+C}{T+C}, &&\text{Sutherland law}\end{aligned} \right.
\label{eq:17}
\end{equation}
where $s$ is the viscosity--temperature exponent and $C$ is the Sutherland constant.
In addition, the molecular mean free path is computed as
\begin{equation}
\lambda^{*}=\frac{\mu^{*}}{p^{*}}\sqrt{\frac{\pi R^{*}T^{*}}{2}}
\label{eq:18}
\end{equation}
The Knudsen number is defined as follows and related to the Mach number and the Reynolds number as
\begin{equation}
\mathit{Kn}=\frac{\lambda_{r}^{*}}{L^{*}}=\sqrt{\frac{\pi\gamma}{2}}\frac{\mathit{Ma}}{\mathit{Re}}
\label{eq:19}
\end{equation}

The GHE, i.e., Eqs.~\eqref{eq:12} and \eqref{eq:13}, are hyperbolic partial differential equations of ($\boldsymbol{\Pi},\;\boldsymbol{Q}$) in a function of position and time, which are difficult to be dealt with.
Eu~\cite{ref33} has introduced an adiabatic approximation, i.e., assuming that the nonconserved variables ($\boldsymbol{\Pi},\;\boldsymbol{Q}$) vary much faster than the conserved variables ($\rho,\;\boldsymbol{u},\;E$), the time-derivative term $\partial/\partial t$ and the convective term $\left( \boldsymbol{u}\cdot\nabla \right)\left( \boldsymbol{\Pi},\;\boldsymbol{Q} \right)$ are neglected.
Myong~\cite{ref34} further neglected the terms $\frac{D\boldsymbol{u}}{Dt}\cdot\boldsymbol{\Pi}$ and $\left( \boldsymbol{\Pi},\boldsymbol{Q} \right)\left( \nabla\cdot\boldsymbol{u} \right)$, and finally obtained a set of algebraic equations, known as the nonlinear coupled constitutive relations (NCCR), written as
\begin{equation}
2[\boldsymbol{\Pi}\cdot\nabla\boldsymbol{u}]^{(2)}+\underline{\mathit{Re}2p[\nabla\boldsymbol{u}]^{(2)}}=\underline{-\mathit{Re}\frac{p}{\mu}\boldsymbol{\Pi}q\left( \kappa \right)},
\label{eq:20}
\end{equation}
\begin{equation}
\mathit{Pr}\boldsymbol{\Pi}\cdot\nabla T+\boldsymbol{Q}\cdot\nabla\boldsymbol{u}+\underline{\mathit{Re}\mathit{Pr}p\nabla T}=\underline{-\mathit{Re}\mathit{Pr}\frac{p}{\mu}\boldsymbol{Q}q\left( \kappa \right)}
\label{eq:21}
\end{equation}
Unlike the GHE, Eqs.~\eqref{eq:20} and \eqref{eq:21} are a set of nonlinear algebraic equations, which greatly simplifies the mathematical treatment.
Compared with the NS linear constitutive relations, the GHE and NCCR are more complex. However, the magnitudes of the different terms in the GHE and NCCR are different.
When a flow approaches the continuum limit (i.e. $\mathit{Re}\gg 1$), the last two underlined terms in the GHE (see Eqs.~\eqref{eq:12} and~\eqref{eq:13}) and NCCR (Eqs.~\eqref{eq:20} and~\eqref{eq:21}) become dominant.
Keep these two terms and neglect other higher-order terms, the GHE and NCCR recover to the NS linear constitutive relations.

In rarefied gas flows, velocity slip and temperature jump phenomena appear at the wall, which invalidate the no-slip assumption. In this paper, both the no-slip and isothermal conditions and the Maxwell's velocity slip and Smoluchowski's temperature jump conditions~\cite{ref37,ref38} are adopted at the wall. Meanwhile, an adiabatic wall condition is also imposed at the lower wall in some cases. The wall boundary conditions are written as
\begin{equation}
\begin{alignedat}{2}&\text{No-slip:}\quad &&u^{*}=0,\;\;T^{*}=T_{w}^{*} \\ &\text{Slip:}\quad &&\left\{ \begin{aligned}&u^{*}-u_{w}^{*}=\frac{2-\sigma_{u}}{\sigma_{u}}\lambda^{*}\left( -\frac{\Pi_{xy}^{*}}{\mu^{*}} \right)+\frac{3}{4}\frac{\left( \gamma-1 \right)\mathit{Pr}}{\gamma}\left( \frac{-Q_{x}^{*}}{p^{*}} \right) \\ &T^{*}-T_{w}^{*}=\frac{2-\sigma_{T}}{\sigma_{T}}\frac{2\gamma}{\left( \gamma+1 \right)\mathit{Pr}}\lambda^{*}\left( -\frac{Q_{y}^{*}}{k^{*}} \right)\end{aligned} \right. \\ &\text{Adiabatic wall:}\quad &&Q_{y}^{*}=0\end{alignedat}
\label{eq:22}
\end{equation}
The above boundary conditions are nondimensionalized as
\begin{equation}
\begin{alignedat}{2}&\text{No-slip:}\quad &&u=0,\;\;T=T_{w} \\ &\text{Slip:}\quad &&\left\{ \begin{aligned}&u-u_{w}=-C_{m}\frac{\sqrt{T}}{p}\Pi_{xy}-C_{s}\frac{Q_{x}}{p} \\ &T-T_{w}=-C_{t}\frac{\sqrt{T}}{p}Q_{y}\end{aligned} \right. \\ &\text{Adiabatic wall:}\quad &&Q_{y}=0\end{alignedat}
\label{eq:23}
\end{equation}
where
\begin{equation}
C_{m}=\frac{2-\sigma_{u}}{\sigma_{u}}\sqrt{\frac{\pi}{2\gamma}}\frac{1}{\mathit{Ma}\,\mathit{Re}},\;\;C_{t}=\frac{2-\sigma_{T}}{\sigma_{T}}\frac{\sqrt{2\pi\gamma}}{\left( \gamma+1 \right)\mathit{Pr}}\frac{1}{\mathit{Ma}\,\mathit{Re}},\;\;C_{s}=\frac{3}{4}\frac{1}{\gamma \mathit{Ma}^{2}\mathit{Re}}
\label{eq:24}
\end{equation}
where $\sigma_u$ is the tangential momentum accommodation coefficient, and $\sigma_T$ is the thermal accommodation coefficient.

\section{Steady base flow}\label{sec:3}

\subsection{Problem description}\label{sec:3.1}

\begin{figure}[htbp]
  \centering
  \includegraphics[width=0.55\textwidth]{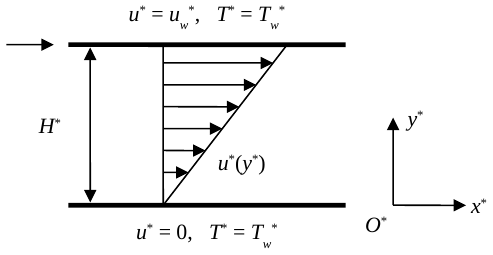}
  \caption{Schematic of plane Couette flow.}
  \label{fig:1}
\end{figure}

The schematic of the plane Couette flow is shown in Fig.~\ref{fig:1}.
The lower plate is stationary, and the upper plate moves at a velocity $u_{w}^{*}$. The two plates are separated by a distance $H^{*}$.
The temperatures of both plates are fixed at $T_{w}^{*}$. In some cases, the lower plate is also set as adiabatic.
Usually, the plane Couette flow is a strictly parallel flow, in which the velocity profile is quasi-linear and the temperature profile is approximately parabolic.
In the present study, the reference length is chosen as the channel width, i.e. $L^{*}=H^{*}$, the reference velocity as the upper-wall velocity $u_{r}^{*}=u_{w}^{*}$, the reference temperature as the upper-wall temperature $T_{r}^{*}=T_{w}^{*}$, and the reference density as the mean gas density in the channel $\rho_{r}^{*}=\frac{1}{H^{*}}\int_{0}^{H^{*}} \rho^{*}dy^{*}$.

\subsection{Governing equations and numerical procedures}\label{sec:3.2}

For steady plane Couette flow, the partial derivatives of flow variables with respect to time $t$ and streamwise coordinate $x$ are zero, and the velocity normal to the wall also vanishes, such that all flow quantities are only functions of the wall-normal coordinate $y$, i.e.,
\begin{equation}
\frac{\partial}{\partial t}=0,\quad \frac{\partial}{\partial x}=0,\quad v=0
\label{eq:25}
\end{equation}
With above assumptions, the continuity equation is automatically satisfied, and the conservation equations~\eqref{eq:11} can be simplified as
\begin{equation}
\left\{ \begin{aligned}&\frac{d\Pi_{xy}}{dy}=0 \\ &\frac{d\left( \mathit{Re}p+\Pi_{yy} \right)}{dy}=0 \\ &\frac{dQ_{y}}{dy}+\left( \gamma-1 \right)\mathit{Pr}\,\mathit{Ma}^{2}\Pi_{xy}\frac{du}{dy}=0\end{aligned} \right.
\label{eq:26}
\end{equation}
The constitutive equations for the stress and heat flux of the  GHE model (Eqs.~\eqref{eq:12} and \eqref{eq:13}) reduce to
\begin{equation}
\begin{aligned}&\mathit{Re}\frac{p}{\mu}\left[ \begin{array}{@{}l@{}}\displaystyle \Pi_{xx} \\[2pt] \displaystyle \Pi_{yy} \\[2pt] \displaystyle \Pi_{xy} \\[2pt] \displaystyle Q_{x} \\[2pt] \displaystyle Q_{y}\end{array} \right]q\left( \kappa \right)=\left[ \begin{array}{@{}l@{}}\displaystyle -(4/3)\Pi_{xy}\partial u/\partial y \\[2pt] \displaystyle (2/3)\Pi_{xy}\partial u/\partial y \\[2pt] \displaystyle -\left( \mathit{Re}p+\Pi_{yy} \right)\partial u/\partial y \\[2pt] \displaystyle -\left( \Pi_{xy}\partial T/\partial y+Q_{y}(\partial u/\partial y)/\mathit{Pr} \right) \\[2pt] \displaystyle -\left( \mathit{Re}p+\Pi_{yy} \right)\partial T/\partial y\end{array} \right] \\ &\text{and}\;\;\; \Pi_{zz}=-\Pi_{xx}-\Pi_{yy},\;\;\Pi_{xz}=\Pi_{yz}=Q_{z}=0\end{aligned}
\label{eq:27}
\end{equation}
Note that for the NCCR model, the simplified equations are identical to those above.
From the above equations \eqref{eq:26} and \eqref{eq:27}, it can be deduced that $\Pi_{xy}$, $\Pi_{yy}$ and $p$ are constants, and
\begin{equation}
\begin{aligned}&\left\{ \begin{aligned}&\frac{\Pi_{xy}}{\Pi_{xy,NS}}=F\left( K,\kappa \right), \\ &\;\frac{Q_{y}}{Q_{y,NS}}=F\left( K,\kappa \right),\end{aligned} \right.\quad\; \text{and} \\ &\left\{ \begin{aligned}&\frac{\Pi_{yy}}{\Pi_{xy}}=\frac{2}{3}\frac{K}{q\left( \kappa \right)},\quad \Pi_{xx}=-2\Pi_{yy},\quad \Pi_{zz}=\Pi_{yy} \\ &\frac{Q_{x}}{Q_{y}}=-\left( 1+\frac{1}{\mathit{Pr}} \right)\frac{K}{q\left( \kappa \right)}\end{aligned} \right.\end{aligned}
\label{eq:28}
\end{equation}
where the NS shear stress and heat flux are
\begin{equation}
\Pi_{xy,NS}=-\mu\frac{du}{dy},\quad\quad\quad\;\; Q_{y,NS}=-k\frac{dT}{dy},
\label{eq:29}
\end{equation}
$F\left( K,\kappa \right)$ is a nonlinear function, given by
\begin{equation}
\begin{aligned}&F\left( K,\kappa \right)=\frac{1}{q\left( \kappa \right)+\frac{2}{3q\left( \kappa \right)}K^{2}},\quad \text{where} \\ &\kappa=\frac{c}{\mathit{Re}p}\sqrt{6\Pi_{yy}^{2}+2\Pi_{xy}^{2}+\left( Q_{x}^{2}+Q_{y}^{2} \right)2\varepsilon/T},\quad K=\frac{1}{\mathit{Re}}\frac{\mu}{p}\frac{du}{dy}\end{aligned}
\label{eq:30}
\end{equation}
where $\kappa$ is the modified Rayleigh--Onsager dissipation function, $K$ is the shear nonequilibrium parameter that characterizes the local shear nonequilibrium degree in the flow field.
In fact, in the current dimensionless form (i.e. Eq.~\eqref{eq:10}), the magnitude of different quantities in the Couette flow can be estimated as: $\Pi_{ij} \sim \mu \partial u/\partial y \sim 1, \; Q_i \sim k\partial T/\partial y \sim 1$, and $\rho \sim 1, \; T \sim 1, \; p\sim1/\gamma \mathit{Ma}^{2}$. Hence, the magnitude of $\kappa$ and $K$ is estimated as: $\kappa,\;K\sim1/\left( \mathit{Re}\cdot p \right)\sim \mathit{Ma}^{2}/\mathit{Re}\sim \mathit{Ma}\cdot\mathit{Kn}$.
According to Eq.~\eqref{eq:28}, the shear stress and normal heat flux in rarefied Couette flow can be regarded as the products of the corresponding NS stress/heat flux and the nonlinear function $F\left( K,\kappa \right)$, which is governed by the dissipation function $\kappa$ and the shear nonequilibrium parameter $K$.
Meanwhile, in the rarefied nonequilibrium state, the normal stress and tangential heat flux are nonzero, and their ratios to the shear stress and normal heat flux are on the order of $K$.
In a continuum flow, $\kappa, K \to 0$ and $F \to 1$, the shear stress $\Pi_{xy}$ and normal heat flux $Q_{y}$ described by Eq.~\eqref{eq:28} recover to the NS linear constitutive relations, and the normal stress $\Pi_{yy}$ and tangential heat flux $Q_{x}$ tend to 0.
In rarefied flows, $K,\;\kappa \sim O(1)$ and $F<1$, the shear stress $\Pi_{xy}$ and normal heat flux $Q_{y}$ are always smaller than the corresponding NS constitutive values (so-called shear-thinning phenomenon), and the normal stress $\Pi_{yy}$ and the tangential heat flux $Q_{x}$ increase to finite values, with the same order as $\Pi_{xy}$ and $Q_{y}$.

The boundary conditions at the two plates are set as no-slip or Maxwell slip conditions, written as
\begin{equation}
\begin{alignedat}{2}&\text{No-slip:}\quad &&\left\{ \begin{aligned}y=0:\quad &u=0,\;\;T=1 \\ y=1:\quad &u=1,\;\;T=1\end{aligned} \right. \\ &\text{Slip:}\quad &&\left\{ \begin{aligned}y=0:\quad &u=0-C_{m}\frac{\sqrt{T}}{p}\Pi_{xy}-C_{s}\frac{Q_{x}}{p},\;\;T=1-C_{t}\frac{\sqrt{T}}{p}Q_{y} \\ y=1:\quad &u=1+C_{m}\frac{\sqrt{T}}{p}\Pi_{xy}-C_{s}\frac{Q_{x}}{p},\;\;T=1+C_{t}\frac{\sqrt{T}}{p}Q_{y}\end{aligned} \right.\end{alignedat}
\label{eq:31}
\end{equation}
where $C_{m}$, $C_{t}$ and $C_{s}$ are given by Eq.~\eqref{eq:24}.
Meanwhile, an adiabatic boundary condition may be also imposed at the lower wall when specified, i.e., $\left. Q_{y} \right|_{y=0}=0$. From Eq. \eqref{eq:31}, the surface slip degree is on the order of $K_{s}\sim C_{m,t}\sqrt{T}/p\sim\left( 2-\sigma_{u,T} \right)/\sigma_{u,T}\cdot\mathit{Kn}$.

Equations~\eqref{eq:26}--\eqref{eq:30} with the boundary conditions~\eqref{eq:31} constitute a two-point boundary-value problem for a set of ordinary differential equations, which can be solved by the double shooting method. Let
\begin{equation}
f_{1}=u,\;\;f_{2}=\Pi_{xy},\;\;f_{3}=T,\;\;f_{4}=Q_{y},
\label{eq:32}
\end{equation}
The conservation equations \eqref{eq:26} can then be rewritten as
\begin{equation}
\left\{ \begin{aligned}&\frac{df_{1}}{dy}=-\frac{f_{2}}{\mu}/F\left( K,\;\kappa \right), \\ &\frac{df_{2}}{dy}=0, \\ &\frac{df_{3}}{dy}=-\frac{f_{4}}{\mu}/F\left( K,\;\kappa \right), \\ &\frac{df_{4}}{dy}=(\gamma-1)\mathit{Pr}\,\mathit{Ma}^{2}\frac{f_{2}^{2}}{\mu}/F\left( K,\;\kappa \right)\end{aligned} \right.
\label{eq:33}
\end{equation}
The numerical procedures to solve the above equation are as follows. First, guess the initial values $\Pi_{xy}\left( 0 \right)$ and $Q_{y}\left( 0 \right)$ at the lower wall, and then integrate Eq. \eqref{eq:33} along the $y$ direction to obtain the gas velocity and temperature at the upper wall. The fourth-order Runge--Kutta method can be used for the integration. Second, obtain the accurate $\Pi_{xy}\left( 0 \right)$ and $Q_{y}\left( 0 \right)$ by shooting the slip velocity and jump temperature at the upper wall, and the secant method can be used for the iteration.
In the integration procedure, the local $\kappa$ and $K$ are implicitly solved from Eq.~\eqref{eq:30} with Eq.~\eqref{eq:28}.
Once the base flow profiles of velocity $u$, temperature $T$, shear stress $\Pi_{xy}$ and normal heat flux $Q_{y}$ are obtained, the normal stress $\Pi_{xx},\;\Pi_{yy}$ and the tangential heat flux $Q_{x}$ are computed from Eq.~\eqref{eq:28}. 
It should be noted that the above equations~\eqref{eq:26}--\eqref{eq:30} with the boundary conditions~\eqref{eq:31} cannot be solved uniquely.
In this paper, the conservation of total mass within the channel is adopted as a supplementary condition, written as
\begin{equation}
\int_{0}^{1} \rho dy=1\quad \text{or}\quad \int_{0}^{1} \frac{\gamma \mathit{Ma}^{2}p}{T}dy=1
\label{eq:34}
\end{equation}
The above condition is a physical constraint. The NS base flow is also computed by the shooting method, with replacing the shear stress and heat flux of GHE by the NS linear constitutive relations.

\subsection{Nonlinear transport in rarefied Couette flow}\label{sec:3.3}

Herein, the nonlinear transport properties of rarefied plane Couette flow are analyzed based on the derived relations above.
Equations~\eqref{eq:28}--\eqref{eq:30} essentially describe the simplified relations among the components of the stress tensor and heat flux vector.
These relations are coupled through $F\left( K,\kappa \right)$, where the shear nonequilibrium parameter $K$ plays a crucial role and the dissipation function $\kappa$ links nonlinear transport with entropy production.
To make these transport relations more concise while still retaining the main nonlinear features, we neglect the heat flux contributions to $\kappa$ in Eq.~\eqref{eq:30}, which gives $F$ as a function of $K$ and $\kappa$ as
\begin{equation}
\begin{aligned}&F=\frac{1}{q\left( \kappa \right)+\frac{2}{3q\left( \kappa \right)}K^{2}},\quad \text{where} \\ &\kappa=\frac{c}{\mathit{Re}p}\sqrt{6\Pi_{yy}^{2}+2\Pi_{xy}^{2}}=cFK\sqrt{\frac{8K^{2}}{3q^{2}\left( \kappa \right)}+2}\end{aligned}
\label{eq:35}
\end{equation}
Eliminating $\kappa$ from the above equation gives $F$ as an implicit function of $K$.
For a given $K$, $F$ can be solved from the Eq.~\eqref{eq:35} by Newton iteration, and then the stress and heat flux components are obtained by substituting $F\left( K,\kappa \right)$ into Eq.~\eqref{eq:28}.
If the effects of $\kappa$ are all neglected, i.e., setting $q\left( \kappa \right)=1$, Eqs.~\eqref{eq:28} and \eqref{eq:30} become even simpler:
\begin{equation}
\left\{ \begin{aligned}&\frac{\Pi_{xy}}{\Pi_{xy,NS}}=\frac{Q_{y}}{Q_{y,NS}}=F\left( K \right), \\ &\frac{\Pi_{yy}}{\Pi_{xy}}=\frac{2}{3}K,\quad \frac{Q_{x}}{Q_{y}}=-\left( 1+\frac{1}{\mathit{Pr}} \right)K,\end{aligned} \right.\quad\; \text{where}\quad F=\frac{1}{1+2K^{2}/3},
\label{eq:36}
\end{equation}
The above equation explicitly gives the dimensionless stress and heat flux as functions of $K$.

\begin{figure}[htbp]
  \centering
  \includegraphics[width=0.46\textwidth]{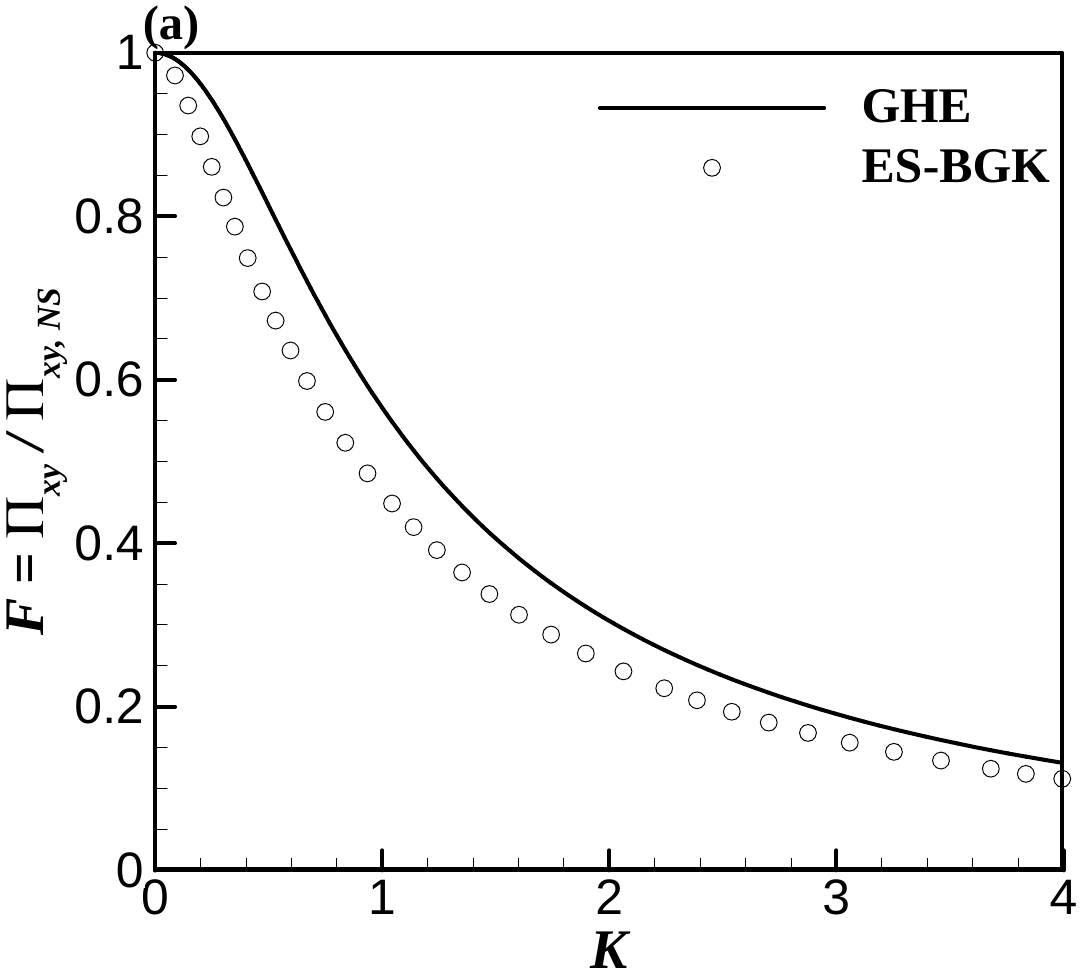}\hspace{0.03\textwidth}
  \includegraphics[width=0.46\textwidth]{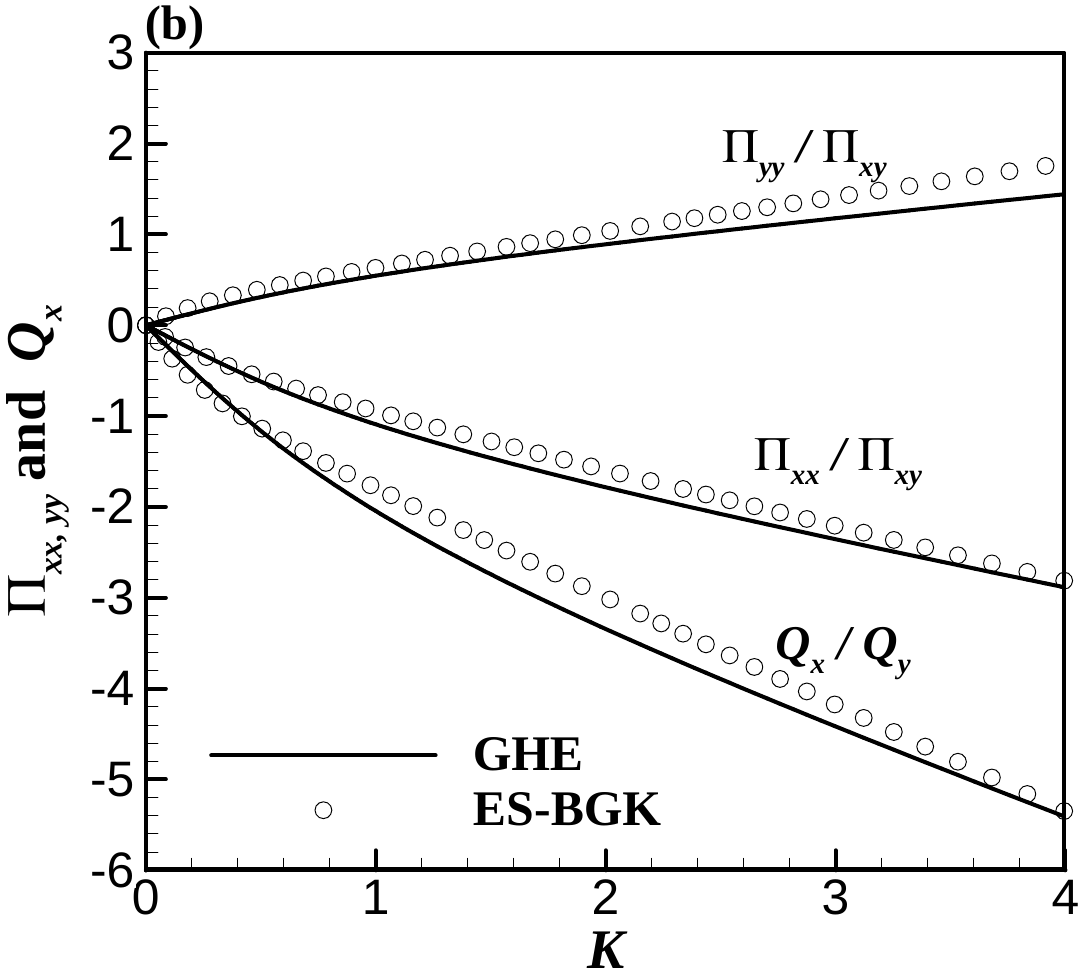}
  \caption{Various transport functions in shear-nonequilibrium flow by GHE model (Eqs.~\eqref{eq:35} and \eqref{eq:28}) compared with the ES-BGK normal solution: (a) dimensionless shear stress $F=\Pi_{xy}/\Pi_{xy,NS}$; (b) dimensionless normal stresses $\Pi_{yy}/\Pi_{xy}$ and $\Pi_{xx}/\Pi_{xy}$, and tangential heat flux $Q_{x}/Q_{y}$.}
  \label{fig:2}
\end{figure}

Garzó and Santos~\cite{ref39} analyzed the nonlinear transport properties of shear nonequilibrium flow, which can be found in the center region of high-speed rarefied plane Couette flow or by eliminating the boundary effects, based on the Ellipsoidal-Statistical-BGK (ES-BGK) model equation, and obtained a normal solution.
The results show that the nonlinear transport functions for viscous stress and heat flux are only dependent on the reduced shear rate $a\sim K$, and various implicit transport functions were given.
Herein, the ES-BGK normal solution is compared with the GHE transport relations to assess the accuracy of the GHE model and to analyze the nonlinear transport of shear nonequilibrium flows.
Figure~\ref{fig:2} compares the dimensionless shear stress ($F=\Pi_{xy}/\Pi_{xy,NS}$), normal stress ($\Pi_{yy}/\Pi_{xy}$ and $\Pi_{xx}/\Pi_{xy}$) and tangential heat flux ($Q_{x}/Q_{y}$) given by the GHE model with the ES-BGK normal solution, where the GHE results are computed from Eqs.~\eqref{eq:35} and \eqref{eq:28}.
As $K$ increases, $F$ decreases monotonically, whereas the magnitudes of $\Pi_{yy}/\Pi_{xy}$, $\Pi_{xx}/\Pi_{xy}$ and $Q_{x}/Q_{y}$ increase monotonically.
The variations of the shear stress, normal stresses and tangential heat flux with $K$ predicted by the GHE model agree well with the ES-BGK normal solution, both qualitatively and quantitatively.
At $K=1$, the GHE model gives $F\approx 0.56$, and $\Pi_{yy}/\Pi_{xy}$ and $Q_{x}/Q_{y}$ approximately 0.54 and -2.04, respectively, indicating that the actual shear stress is considerably smaller than the NS linear constitutive value, and the normal stress and tangential heat flux reach the same order as the shear stress and normal heat flux.
Although the GHE model is closed at second order, its simplified transport functions for shear nonequilibrium flow compare well with the ES-BGK normal solution.
In the following, the base flow solutions of plane Couette flow at different Mach and Knudsen numbers are further obtained by the GHE model and compared with DSMC simulation results, to further verify the accuracy of GHE.

\section{Linear stability theory based on generalized hydrodynamics}\label{sec:4}

\subsection{Linear stability equations}\label{sec:4.1}

The present study mainly investigates the two-dimension temporal stability problem of high-speed plane Couette flow.
Essentially, the linear stability theory concerns the evolution of infinitesimal disturbances superimposed on a steady base flow.
Assume that an instantaneous quantity is the sum of a mean quantity and a small disturbance, i.e.,
\begin{equation}
\begin{aligned}&\rho=\rho_{0}+\tilde{\rho},\quad u=u_{0}+\tilde{u},\quad v=v_{0}+\tilde{v},\quad T=T_{0}+\tilde{T}, \\ &p=p_{0}+\tilde{p},\quad \mu=\mu_{0}+\tilde{\mu},\quad k=k_{0}+\tilde{k}, \\ &\Pi_{ij}=\Pi_{ij0}+\tilde{\Pi}_{ij},\quad\; Q_{i}=Q_{i0}+\tilde{Q}_{i}\end{aligned}
\label{eq:37}
\end{equation}
where the subscript `0' denotes the base flow and the superscript `\textasciitilde{}' denotes disturbances.
Substituting the above equation into the conservation equations~\eqref{eq:11} and the constitutive equations of GHE (Eqs.~\eqref{eq:12} and \eqref{eq:13}), subtracting the base flow and neglecting higher-order disturbance terms, the two-dimensional disturbance equations of the GHE are obtained as

\noindent Continuity equation:
\begin{equation}
\frac{\partial\tilde{\rho}}{\partial t}+\frac{\partial}{\partial x}(\rho_{0}\tilde{u}+\tilde{\rho}u_{0})+\frac{\partial}{\partial y}(\rho_{0}\tilde{v}+\tilde{\rho}v_{0})=0
\label{eq:38}
\end{equation}

\noindent $x$-momentum equation:
\begin{equation}
\begin{aligned}&\rho_{0}\left( \frac{\partial\tilde{u}}{\partial t}+\tilde{u}\frac{\partial u_{0}}{\partial x}+\tilde{v}\frac{\partial u_{0}}{\partial y}+u_{0}\frac{\partial\tilde{u}}{\partial x}+v_{0}\frac{\partial\tilde{u}}{\partial y} \right)+\tilde{\rho}\left( u_{0}\frac{\partial u_{0}}{\partial x}+v_{0}\frac{\partial u_{0}}{\partial y} \right) \\ &+\frac{\partial\tilde{p}}{\partial x}+\frac{1}{\mathit{Re}}\left( \frac{\partial\tilde{\Pi}_{xx}}{\partial x}+\frac{\partial\tilde{\Pi}_{xy}}{\partial y} \right)=0,\end{aligned}
\label{eq:39}
\end{equation}

\noindent $y$-momentum equation:
\begin{equation}
\begin{aligned}&\rho_{0}\left( \frac{\partial\tilde{v}}{\partial t}+\tilde{u}\frac{\partial v_{0}}{\partial x}+\tilde{v}\frac{\partial v_{0}}{\partial y}+u_{0}\frac{\partial\tilde{v}}{\partial x}+v_{0}\frac{\partial\tilde{v}}{\partial y} \right)+\tilde{\rho}\left( u_{0}\frac{\partial v_{0}}{\partial x}+v_{0}\frac{\partial v_{0}}{\partial y} \right) \\ &+\frac{\partial\tilde{p}}{\partial y}+\frac{1}{\mathit{Re}}\left( \frac{\partial\tilde{\Pi}_{xy}}{\partial x}+\frac{\partial\tilde{\Pi}_{yy}}{\partial y} \right)=0\end{aligned}
\label{eq:40}
\end{equation}

\noindent Energy equation:
\begin{equation}
\begin{aligned}&\rho_{0}\left( \frac{\partial\tilde{T}}{\partial t}+\tilde{u}\frac{\partial T_{0}}{\partial x}+\tilde{v}\frac{\partial T_{0}}{\partial y}+u_{0}\frac{\partial\tilde{T}}{\partial x}+v_{0}\frac{\partial\tilde{T}}{\partial y} \right)+\tilde{\rho}\left( u_{0}\frac{\partial T_{0}}{\partial x}+v_{0}\frac{\partial T_{0}}{\partial y} \right) \\ &-(\gamma-1)\mathit{Ma}^{2}\left( \frac{\partial\tilde{p}}{\partial t}+\tilde{u}\frac{\partial p_{0}}{\partial x}+\tilde{v}\frac{\partial p_{0}}{\partial y}+u_{0}\frac{\partial\tilde{p}}{\partial x}+v_{0}\frac{\partial\tilde{p}}{\partial y} \right)+\frac{1}{\mathit{Pr}\mathit{Re}}\left( \frac{\partial\tilde{Q}_{x}}{\partial x}+\frac{\partial\tilde{Q}_{y}}{\partial y} \right) \\ &+\frac{\left( \gamma-1 \right)\mathit{Ma}^{2}}{\mathit{Re}}\left[ \begin{array}{@{}l@{}}\displaystyle \tilde{\Pi}_{xx}\frac{\partial u_{0}}{\partial x}+\tilde{\Pi}_{yy}\frac{\partial v_{0}}{\partial y}+\tilde{\Pi}_{xy}\left( \frac{\partial u_{0}}{\partial y}+\frac{\partial v_{0}}{\partial x} \right) \\[2pt] \displaystyle +\Pi_{xx0}\frac{\partial\tilde{u}}{\partial x}+\Pi_{yy0}\frac{\partial\tilde{v}}{\partial y}+\Pi_{xy0}\left( \frac{\partial\tilde{u}}{\partial y}+\frac{\partial\tilde{v}}{\partial x} \right)\end{array} \right]=0\end{aligned}
\label{eq:41}
\end{equation}

\noindent Stress equation $\tilde{\Pi}_{xx}$:
\begin{equation}
\begin{aligned}&\frac{\partial\tilde{\Pi}_{xx}}{\partial t}+\left( \tilde{u}\frac{\partial\Pi_{xx0}}{\partial x}+\tilde{v}\frac{\partial\Pi_{xx0}}{\partial y}+u_{0}\frac{\partial\tilde{\Pi}_{xx}}{\partial x}+v_{0}\frac{\partial\tilde{\Pi}_{xx}}{\partial y} \right) \\ &+\tilde{\Pi}_{xx}\left( \frac{\partial u_{0}}{\partial x}+\frac{\partial v_{0}}{\partial y} \right)+\Pi_{xx0}\left( \frac{\partial\tilde{u}}{\partial x}+\frac{\partial\tilde{v}}{\partial y} \right) \\ &+\left[ \frac{4}{3}\left( \tilde{\Pi}_{xx}\frac{\partial u_{0}}{\partial x}+\tilde{\Pi}_{xy}\frac{\partial u_{0}}{\partial y} \right)-\frac{2}{3}\left( \tilde{\Pi}_{xy}\frac{\partial v_{0}}{\partial x}+\tilde{\Pi}_{yy}\frac{\partial v_{0}}{\partial y} \right) \right] \\ &+\left[ \frac{4}{3}\left( \Pi_{xx0}\frac{\partial\tilde{u}}{\partial x}+\Pi_{xy0}\frac{\partial\tilde{u}}{\partial y} \right)-\frac{2}{3}\left( \Pi_{xy0}\frac{\partial\tilde{v}}{\partial x}+\Pi_{yy0}\frac{\partial\tilde{v}}{\partial y} \right) \right] \\ &+\mathit{Re}\,\tilde{p}\left( \frac{4}{3}\frac{\partial u_{0}}{\partial x}-\frac{2}{3}\frac{\partial v_{0}}{\partial y} \right)+\mathit{Re}\,p_{0}\left( \frac{4}{3}\frac{\partial\tilde{u}}{\partial x}-\frac{2}{3}\frac{\partial\tilde{v}}{\partial y} \right) \\ &=-\mathit{Re}\left[ \frac{\tilde{p}}{\mu_{0}}\Pi_{xx0}q\left( \kappa_{0} \right)-\frac{\tilde{\mu}}{\mu_{0}^{2}}p_{0}\Pi_{xx0}q\left( \kappa_{0} \right)+\tilde{\Pi}_{xx}\frac{p_{0}}{\mu_{0}}q\left( \kappa_{0} \right)+\frac{p_{0}}{\mu_{0}}\Pi_{xx0}\left. \frac{dq}{d\kappa} \right|_{0}\tilde{\kappa} \right],\end{aligned}
\label{eq:42}
\end{equation}

\noindent Stress equation $\tilde{\Pi}_{yy}$:
\begin{equation}
\begin{aligned}&\frac{\partial\tilde{\Pi}_{yy}}{\partial t}+\left( \tilde{u}\frac{\partial\Pi_{yy0}}{\partial x}+\tilde{v}\frac{\partial\Pi_{yy0}}{\partial y}+u_{0}\frac{\partial\tilde{\Pi}_{yy}}{\partial x}+v_{0}\frac{\partial\tilde{\Pi}_{yy}}{\partial y} \right) \\ &+\left[ \tilde{\Pi}_{yy}\left( \frac{\partial u_{0}}{\partial x}+\frac{\partial v_{0}}{\partial y} \right)+\Pi_{yy0}\left( \frac{\partial\tilde{u}}{\partial x}+\frac{\partial\tilde{v}}{\partial y} \right) \right] \\ &+\left[ \frac{4}{3}\left( \tilde{\Pi}_{xy}\frac{\partial v_{0}}{\partial x}+\tilde{\Pi}_{yy}\frac{\partial v_{0}}{\partial y} \right)-\frac{2}{3}\left( \tilde{\Pi}_{xx}\frac{\partial u_{0}}{\partial x}+\tilde{\Pi}_{xy}\frac{\partial u_{0}}{\partial y} \right) \right] \\ &+\left[ \frac{4}{3}\left( \Pi_{xy0}\frac{\partial\tilde{v}}{\partial x}+\Pi_{yy0}\frac{\partial\tilde{v}}{\partial y} \right)-\frac{2}{3}\left( \Pi_{xx0}\frac{\partial\tilde{u}}{\partial x}+\Pi_{xy0}\frac{\partial\tilde{u}}{\partial y} \right) \right] \\ &+\mathit{Re}\;\tilde{p}\left( \frac{4}{3}\frac{\partial v_{0}}{\partial y}-\frac{2}{3}\frac{\partial u_{0}}{\partial x} \right)+\mathit{Re}\;p_{0}\left( \frac{4}{3}\frac{\partial\tilde{v}}{\partial y}-\frac{2}{3}\frac{\partial\tilde{u}}{\partial x} \right) \\ &=-\mathit{Re}\left[ \frac{\tilde{p}}{\mu_{0}}\Pi_{yy0}q\left( \kappa_{0} \right)-\frac{\tilde{\mu}}{\mu_{0}^{2}}p_{0}\Pi_{yy0}q\left( \kappa_{0} \right)+\tilde{\Pi}_{yy}\frac{p_{0}}{\mu_{0}}q\left( \kappa_{0} \right)+\frac{p_{0}}{\mu_{0}}\Pi_{yy0}\left. \frac{dq}{d\kappa} \right|_{0}\tilde{\kappa} \right],\end{aligned}
\label{eq:43}
\end{equation}

\noindent Stress equation $\tilde{\Pi}_{xy}$:
\begin{equation}
\begin{aligned}&\frac{\partial\tilde{\Pi}_{xy}}{\partial t}+\left( \tilde{u}\frac{\partial\Pi_{xy0}}{\partial x}+\tilde{v}\frac{\partial\Pi_{xy0}}{\partial y}+u_{0}\frac{\partial\tilde{\Pi}_{xy}}{\partial x}+v_{0}\frac{\partial\tilde{\Pi}_{xy}}{\partial y} \right) \\ &+\tilde{\Pi}_{xy}\left( \frac{\partial u_{0}}{\partial x}+\frac{\partial v_{0}}{\partial y} \right)+\Pi_{xy0}\left( \frac{\partial\tilde{u}}{\partial x}+\frac{\partial\tilde{v}}{\partial y} \right) \\ &+\left( \tilde{\Pi}_{xx}\frac{\partial v_{0}}{\partial x}+\tilde{\Pi}_{xy}\frac{\partial v_{0}}{\partial y}+\tilde{\Pi}_{xy}\frac{\partial u_{0}}{\partial x}+\tilde{\Pi}_{yy}\frac{\partial u_{0}}{\partial y} \right) \\ &+\left( \Pi_{xx0}\frac{\partial\tilde{v}}{\partial x}+\Pi_{xy0}\frac{\partial\tilde{v}}{\partial y}+\Pi_{xy0}\frac{\partial\tilde{u}}{\partial x}+\Pi_{yy0}\frac{\partial\tilde{u}}{\partial y} \right) \\ &+\mathit{Re}\;\tilde{p}\left( \frac{\partial u_{0}}{\partial y}+\frac{\partial v_{0}}{\partial x} \right)+\mathit{Re}\;p_{0}\left( \frac{\partial\tilde{u}}{\partial y}+\frac{\partial\tilde{v}}{\partial x} \right) \\ &=-\mathit{Re}\left[ \frac{\tilde{p}}{\mu_{0}}\Pi_{xy0}q\left( \kappa_{0} \right)-\frac{\tilde{\mu}}{\mu_{0}^{2}}p_{0}\Pi_{xy0}q\left( \kappa_{0} \right)+\tilde{\Pi}_{xy}\frac{p_{0}}{\mu_{0}}q\left( \kappa_{0} \right)+\frac{p_{0}}{\mu_{0}}\Pi_{xy0}\left. \frac{dq}{d\kappa} \right|_{0}\tilde{\kappa} \right],\end{aligned}
\label{eq:44}
\end{equation}

\noindent Heat flux equation $\tilde{Q}_{x}$:
\begin{equation}
\begin{aligned}&\frac{\partial\tilde{Q}_{x}}{\partial t}+\left( \tilde{u}\frac{\partial Q_{x0}}{\partial x}+\tilde{v}\frac{\partial Q_{x0}}{\partial y}+u_{0}\frac{\partial\tilde{Q}_{x}}{\partial x}+v_{0}\frac{\partial\tilde{Q}_{x}}{\partial y} \right) \\ &+\tilde{Q}_{x}\left( \frac{\partial u_{0}}{\partial x}+\frac{\partial v_{0}}{\partial y} \right)+Q_{x0}\left( \frac{\partial\tilde{u}}{\partial x}+\frac{\partial\tilde{v}}{\partial y} \right) \\ &+\left( \gamma-1 \right)\mathit{Pr}\,\mathit{Ma}^{2}\left( \frac{D\tilde{u}}{Dt}\Pi_{xx0}+\frac{D\tilde{v}}{Dt}\Pi_{xy0}+\frac{Du_{0}}{Dt}\tilde{\Pi}_{xx}+\frac{Dv_{0}}{Dt}\tilde{\Pi}_{xy} \right) \\ &+\mathit{Pr}\left( \tilde{\Pi}_{xx}\frac{\partial T_{0}}{\partial x}+\tilde{\Pi}_{xy}\frac{\partial T_{0}}{\partial y}+\Pi_{xx0}\frac{\partial\tilde{T}}{\partial x}+\Pi_{xy0}\frac{\partial\tilde{T}}{\partial y} \right) \\ &+\left( \tilde{Q}_{x}\frac{\partial u_{0}}{\partial x}+\tilde{Q}_{y}\frac{\partial u_{0}}{\partial y}+Q_{x0}\frac{\partial\tilde{u}}{\partial x}+Q_{y0}\frac{\partial\tilde{u}}{\partial y} \right)+\mathit{Pr}\mathit{Re}\left( \tilde{p}\frac{\partial T_{0}}{\partial x}+\;p_{0}\frac{\partial\tilde{T}}{\partial x} \right) \\ &=-\mathit{Pr}\mathit{Re}\left[ \frac{\tilde{p}}{k_{0}}Q_{x0}q\left( \kappa_{0} \right)-\frac{\tilde{k}}{k_{0}^{2}}p_{0}Q_{x0}q\left( \kappa_{0} \right)+\frac{p_{0}}{k_{0}}\tilde{Q}_{x}q\left( \kappa_{0} \right)+\frac{p_{0}}{k_{0}}Q_{x0}\left. \frac{dq}{d\kappa} \right|_{0}\tilde{\kappa} \right],\end{aligned}
\label{eq:45}
\end{equation}

\noindent Heat flux equation $\tilde{Q}_{y}$:
\begin{equation}
\begin{aligned}&\frac{\partial\tilde{Q}_{y}}{\partial t}+\left( \tilde{u}\frac{\partial Q_{y0}}{\partial x}+\tilde{v}\frac{\partial Q_{y0}}{\partial y}+u_{0}\frac{\partial\tilde{Q}_{y}}{\partial x}+v_{0}\frac{\partial\tilde{Q}_{y}}{\partial y} \right) \\ &+\tilde{Q}_{y}\left( \frac{\partial u_{0}}{\partial x}+\frac{\partial v_{0}}{\partial y} \right)+Q_{y0}\left( \frac{\partial\tilde{u}}{\partial x}+\frac{\partial\tilde{v}}{\partial y} \right) \\ &+\left( \gamma-1 \right)\mathit{Pr}\,\mathit{Ma}^{2}\left( \frac{D\tilde{u}}{Dt}\Pi_{xy0}+\frac{D\tilde{v}}{Dt}\Pi_{yy0}+\frac{Du_{0}}{Dt}\tilde{\Pi}_{xy}+\frac{Dv_{0}}{Dt}\tilde{\Pi}_{yy} \right) \\ &+\mathit{Pr}\left( \tilde{\Pi}_{xy}\frac{\partial T_{0}}{\partial x}+\tilde{\Pi}_{yy}\frac{\partial T_{0}}{\partial y}+\Pi_{xy0}\frac{\partial\tilde{T}}{\partial x}+\Pi_{yy0}\frac{\partial\tilde{T}}{\partial y} \right) \\ &+\left( \tilde{Q}_{x}\frac{\partial v_{0}}{\partial x}+\tilde{Q}_{y}\frac{\partial v_{0}}{\partial y}+Q_{x0}\frac{\partial\tilde{v}}{\partial x}+Q_{y0}\frac{\partial\tilde{v}}{\partial y} \right)+\mathit{Pr}\mathit{Re}\left( \tilde{p}\frac{\partial T_{0}}{\partial y}+p_{0}\frac{\partial\tilde{T}}{\partial y} \right) \\ &=-\mathit{Pr}\mathit{Re}\left[ \frac{\tilde{p}}{k_{0}}Q_{y0}q\left( \kappa_{0} \right)-\frac{\tilde{k}}{k_{0}^{2}}p_{0}Q_{y0}q\left( \kappa_{0} \right)+\frac{p_{0}}{k_{0}}\tilde{Q}_{y}q\left( \kappa_{0} \right)+\frac{p_{0}}{k_{0}}Q_{y0}\left. \frac{dq}{d\kappa} \right|_{0}\tilde{\kappa} \right],\end{aligned}
\label{eq:46}
\end{equation}
In the above disturbance equations, the perturbation quantities $\tilde{\rho},\;\tilde{u},\;\tilde{v},\;\tilde{T},\;\tilde{p},\;\tilde{\mu},\;\tilde{k},\;\tilde{\Pi}_{ij},\;\tilde{Q}_i$ are not completely independent. For two-dimensional stability analysis, nine unknown variables can be chosen, such as $\tilde{\phi}=\left[\tilde{u},\;\tilde{v},\;\tilde{T},\;\tilde{p},\;\tilde{\Pi}_{xx},\;\tilde{\Pi}_{yy},\;\tilde{\Pi}_{xy},\;\tilde{Q}_{x},\;\tilde{Q}_{y} \right]^{\text{T}}$. If we consider three-dimensional problem, the number of unknown variables should be 13, with $\tilde{\Pi}_{xz},\;\tilde{\Pi}_{yz},\;\tilde{\Pi}_{zz},\;\tilde{Q}_{z}$ added. The selection of above unknown variables are based on 13-moment framework.

The plane Couette flow is a parallel flow, satisfying the following assumptions:
\begin{equation}
\begin{aligned}&\rho_{0}=\rho_{0}\left( y \right),\;\;u_{0}=u_{0}\left( y \right),\;\;v_{0}=0,\;\;T_{0}=T_{0}\left( y \right), \\ &p_{0}=p_{0}\left( y \right),\;\;\Pi_{ij0}=\Pi_{ij0}\left( y \right),\;\;Q_{i0}=Q_{i0}\left( y \right)\end{aligned}
\label{eq:47}
\end{equation}
With above assumptions, $D\tilde{u}/{Dt}$, $D\tilde{v}/{Dt}$ and $Du_0/{Dt}$ in Eqs.~\eqref{eq:45} and \eqref{eq:46} are
\begin{equation}
\frac{D\tilde{u}}{Dt}=\frac{\partial\tilde{u}}{\partial t}+u_{0}\frac{\partial\tilde{u}}{\partial x}+\tilde{v}\frac{\partial u_{0}}{\partial y},\quad \frac{D\tilde{v}}{Dt}=\frac{\partial\tilde{v}}{\partial t}+u_{0}\frac{\partial\tilde{v}}{\partial x},\quad \frac{Du_{0}}{Dt}=\frac{Dv_{0}}{Dt}=0
\label{eq:48}
\end{equation}
The density disturbance can be expressed in terms of the pressure and temperature disturbances through the disturbance equation of the gas state equation, as
\begin{equation}
\tilde{\rho}=\gamma\mathit{Ma}^{2}\frac{\tilde{p}}{T_{0}}-\frac{\rho_{0}}{T_{0}}\tilde{T}
\label{eq:49}
\end{equation}
In addition, noting that
\begin{equation}
\begin{aligned}&\tilde{\mu}=\left. \frac{d\mu}{dT} \right|_{0}\tilde{T},\quad \tilde{k}=\;\left. \frac{dk}{dT} \right|_{0}\tilde{T}, \\ &\tilde{q}\left( \kappa \right)=\left. \frac{dq}{d\kappa} \right|_{0}\tilde{\kappa}=\frac{\kappa_{0}\cosh \kappa_{0}-\sinh \kappa_{0}}{\kappa_{0}^{2}}\tilde{\kappa},\end{aligned}
\label{eq:50}
\end{equation}
and considering that $\kappa=\kappa\left( p,\;T,\;\Pi_{ij},\;Q_{i} \right)$, $\tilde{\kappa}$ can be written as
\begin{equation}
\begin{aligned}&\tilde{\kappa}=\left. \frac{\partial\kappa}{\partial p} \right|_{0}\tilde{p}+\left. \frac{\partial\kappa}{\partial T} \right|_{0}\tilde{T}+\left. \frac{\partial\kappa}{\partial\Pi_{ij}} \right|_{0}\tilde{\Pi}_{ij}+\left. \frac{\partial\kappa}{\partial Q_{i}} \right|_{0}\tilde{Q}_{i} \\ &\;\;\; =\kappa_{0}\left( -\frac{\tilde{p}}{p_{0}} \right)+\frac{1}{\kappa_{0}}\left( \frac{c}{\mathit{Re}\,p_{0}} \right)^{2}\left\{ \begin{array}{@{}l@{}}\displaystyle \left( 2\Pi_{xx0}+\Pi_{yy0} \right)\tilde{\Pi}_{xx}+\left( 2\Pi_{yy0}+\Pi_{xx0} \right)\tilde{\Pi}_{yy}+2\Pi_{xy0}\tilde{\Pi}_{xy} \\[2pt] \displaystyle +2\varepsilon\left[ \frac{Q_{x0}}{T_{0}}\tilde{Q}_{x}+\frac{Q_{y0}}{T_{0}}\tilde{Q}_{y}-\frac{Q_{x0}^{2}+Q_{y0}^{2}}{2T_{0}^{2}}\tilde{T} \right]\end{array} \right\}\end{aligned}
\label{eq:51}
\end{equation}
Substituting the above equations \eqref{eq:47}--\eqref{eq:51} into the disturbance equations~\eqref{eq:38}--\eqref{eq:46} gives
\begin{equation}
L\left( \partial_{t},\;\partial_{x},\;\partial_{y} \right)\tilde{\phi}=0
\label{eq:52}
\end{equation}
where $\tilde{\phi}=\left[ \tilde{u},\;\tilde{v},\;\tilde{T},\;\tilde{p},\;\tilde{\Pi}_{xx},\;\tilde{\Pi}_{yy},\;\tilde{\Pi}_{xy},\;\tilde{Q}_{x},\;\tilde{Q}_{y} \right]^{\text{T}}$, $\partial_{t}=\partial/\partial t,\;\partial_{x}=\partial/\partial x,\;\partial_{y}=\partial/\partial y$, and $L$ is a linear operator containing the base flows and their first-order derivatives.

Furthermore, the disturbances are assumed as harmonic waves
\begin{equation}
\tilde{\phi}\left( x,y,t \right)=\hat{\phi}\left( y \right)\exp\left[ i\left( \alpha x-\omega t \right) \right]+\mathrm{C}.\mathrm{C}.,
\label{eq:53}
\end{equation}
where the superscript ` $\hat{}$ ' represents eigenfunctions, $\alpha$ is the dimensionless streamwise wave number, $\omega$ is the dimensionless circular frequency, i.e. $\alpha  = {\alpha ^*}{L^*},\;\omega  = {\omega ^*}{L^*}/u_r^*$, C.C. denotes the complex conjugate. Substituting the above harmonic-wave form into the disturbance equations~\eqref{eq:52} yields the linear stability equations of GHE
\begin{equation}
\mathbf{A}\frac{d\hat{\phi}}{dy}+\mathbf{B}\hat{\phi}=\omega\,\mathbf{C}\hat{\phi}
\label{eq:54}
\end{equation}
where $\hat{\phi}=\left[ \hat{u},\;\hat{v},\;\hat{T},\;\hat{p},\;\hat{\Pi}_{xx},\;\hat{\Pi}_{yy},\;\hat{\Pi}_{xy},\;\hat{Q}_{x},\;\hat{Q}_{y} \right]^{\text{T}}$, and $\mathbf{A}$, $\mathbf{B}$ and $\mathbf{C}$ are coefficient matrices that contain the base flows, $\alpha$, $\mathit{Ma}$, $\mathit{Re}$, whose elements are given in the Appendix.
For the temporal problem, $\alpha$ is a given real number, and $\omega=\omega_{r}+i\omega_{i}$ is the complex eigenvalue to be determined, with $\omega_{i}$ the disturbance growth rate. If $\omega_{i}>0$, the disturbance grows in time. The complex phase speed of the disturbance is defined as $c=\omega/\alpha$.
Its real part $c_{r}$ is the propagation speed of the disturbance, and its imaginary part $c_{i}=\omega_{i}/\alpha$ has the same sign as $\omega_{i}$. Hence, the disturbance grows when $c_{i}>0$ and decays when $c_{i}<0$.
For the NS equations and NCCR model, the linear stability equations can be given in a similar form as Eq.~\eqref{eq:54}, and their coefficient matrices $\mathbf{A}$, $\mathbf{B}$ and $\mathbf{C}$ are also given in the Appendix.

In a continuous flow, the disturbances are fully accommodated to the wall, so the velocity and temperature disturbances are usually assumed zero at the wall.
As flow becomes rarefied, the disturbances exhibit velocity slip and temperature jump phenomena at the wall. Herein, the disturbance slip boundary conditions are obtained by linearizing the Maxwell slip conditions for the base flow.
The no-slip and slip boundary conditions for disturbances are written as
\begin{equation}
\begin{alignedat}{2}&\text{No-slip:}\quad &&\left\{ \begin{aligned}y=0:\quad &\hat{u}=0,\quad \hat{v}=0,\quad \hat{T}=0 \\ y=1:\quad &\hat{u}=0,\quad \hat{v}=0,\quad \hat{T}=0\end{aligned} \right. \\ &\text{Slip:}\quad &&\left\{ \begin{aligned}y=0:\quad &\hat{u}=-C_{m}\left( \frac{\sqrt{T_{0}}}{p_{0}}\hat{\Pi}_{xy}+\frac{\Pi_{xy0}}{2\sqrt{T_{0}}\,p_{0}}\hat{T}-\frac{\sqrt{T_{0}}\,\Pi_{xy0}}{p_{0}^{2}}\hat{p} \right)-C_{s}\left( \frac{\hat{Q}_{x}}{p_{0}}-\frac{Q_{x0}}{p_{0}^{2}}\hat{p} \right), \\ &\hat{v}=0, \\ &\hat{T}=-C_{t}\left( \frac{\sqrt{T_{0}}}{p_{0}}\hat{Q}_{y}+\frac{Q_{y0}}{2\sqrt{T_{0}}\,p_{0}}\hat{T}-\frac{\sqrt{T_{0}}\,Q_{y0}}{p_{0}^{2}}\hat{p} \right), \\ y=1:\quad &\hat{u}=C_{m}\left( \frac{\sqrt{T_{0}}}{p_{0}}\hat{\Pi}_{xy}+\frac{\Pi_{xy0}}{2\sqrt{T_{0}}\,p_{0}}\hat{T}-\frac{\sqrt{T_{0}}\,\Pi_{xy0}}{p_{0}^{2}}\hat{p} \right)-C_{s}\left( \frac{\hat{Q}_{x}}{p_{0}}-\frac{Q_{x0}}{p_{0}^{2}}\hat{p} \right), \\ &\hat{v}=0, \\ &\hat{T}=C_{t}\left( \frac{\sqrt{T_{0}}}{p_{0}}\hat{Q}_{y}+\frac{Q_{y0}}{2\sqrt{T_{0}}\,p_{0}}\hat{T}-\frac{\sqrt{T_{0}}\,Q_{y0}}{p_{0}^{2}}\hat{p} \right),\end{aligned} \right.\end{alignedat}
\label{eq:55}
\end{equation}
where $C_{m}$, $C_{t}$ and $C_{s}$ are constants given by Eq.~\eqref{eq:24}.
In some cases, an adiabatic condition is also imposed at the lower wall when specified, i.e., $\hat{Q}_y|_{y=0}=0$.
We should state that for the present GHE stability equations, all flow variables $\hat{u},\;\hat{v},\;\hat{T},\;\hat{p},\; \hat{\Pi}_{ij},\;\hat{Q}_i$ only involve first-order derivatives with respect to $y$, exhibiting the same order as the conventional NS-based stability equations. Hence, no higher-order boundary conditions are needed, and the disturbance boundary conditions can be same as solving the NS-based stability equations. 
Additionally, for two-dimensional problem, the linear stability equations~\eqref{eq:54} form a sixth-order differential equations. Equation \eqref{eq:55} has given six physical boundary conditions, i.e., $\hat{u},\ \hat{v},\ \hat{T}$ at each wall, which is enough to determine the problem. However, in numerical computation, the disturbances of pressure, all stress and heat flux components at the wall should also be specified.
To determine the surface pressure perturbation, a reasonable approach is to adopt the normal momentum equation
\begin{equation}
\frac{\partial\hat{p}}{\partial y}+\frac{1}{\mathit{Re}}i\alpha\hat{\Pi}_{xy}+\frac{1}{\mathit{Re}}\frac{\partial\hat{\Pi}_{yy}}{\partial y}=0
\label{eq:56}
\end{equation}
In fact, the pressure perturbation can also be determined by $\partial\hat{p}/\partial y|_{y=0}=0$, which also yields the same accurate results.
The perturbations of the viscous stress $\hat{\Pi}_{ij}$ and heat flux $\hat{Q}_i$ at the wall are determined from the corresponding disturbance equations of stress and heat flux. The numerical implementations are described in Sec.~\ref{sec:4.2}.

\subsection{Numerical methods}\label{sec:4.2}

The linear stability equations~\eqref{eq:54} with the homogeneous boundary conditions~\eqref{eq:55} form a generalized eigenvalue problem.
In this paper, the Chebyshev spectral collocation method is adopted to solve this problem.
The Chebyshev polynomial $T_{N}$ is defined in $-1\le \xi\le 1$, and the collocation points $\xi_{j}$ are chosen as the extrema of $T_{N}$
\begin{equation}
\xi_{j}=\cos \frac{\pi j}{N},\quad j=0,\;1,\;\ldots ,\;N.
\label{eq:57}
\end{equation}
The grid points in physical space are obtained through the following coordinate transformation
\begin{equation}
y_{j}=-\frac{\xi_{j}-1}{2}
\label{eq:58}
\end{equation}
The first derivative of a physical quantity with respect to $y$ is approximated by
\begin{equation}
\left. \frac{d\phi}{dy} \right|_{j}=\sum\limits_{k=0}^{N} F_{jk}\phi_{k}
\label{eq:59}
\end{equation}
where $F$ is the derivative matrix, whose elements are given by
\begin{equation}
F_{jk}=S\,E_{jk},\quad \quad \text{where}\quad \quad S=\frac{d\xi}{dy}=-2, \quad E_{jk}=\left\{ \begin{array}{ll}\displaystyle \frac{c_{j}}{c_{k}}\frac{{(-1)}^{j+k}}{\xi_{j}-\xi_{k}}, & \displaystyle j\ne k \\[9pt] \displaystyle -\frac{\xi_{j}}{2(1-\xi_{j}^{2})}, & \displaystyle 0<j=k<N \\[9pt] \displaystyle \frac{2N^{2}+1}{6}, & \displaystyle j=k=0 \\[9pt] \displaystyle -\frac{2N^{2}+1}{6}, & \displaystyle j=k=N\end{array} \right.
\label{eq:60}
\end{equation}
Here $c_{0}=c_{N}=2$ and $c_{j}=1$ ($0<j<N$).
$E$ is the derivative matrix of the Chebyshev interpolation polynomial evaluated at the collocation points, and $S$ is the coordinate-transformation factor.
The grid spacing between adjacent Chebyshev collocation points is on the order of $O(N^{-2})$ near the endpoints and $O(N^{-1})$ in the middle of the interval. The resulting grid is clustered near the upper and lower walls and is coarser in the channel interior, making it naturally suitable for the plane Couette flow.

Discretizing the linear stability equations~\eqref{eq:54} by the Chebyshev collocation method gives the following discrete form of the stability equation
\begin{equation}
\mathbf{A}_{j}\sum\limits_{k=0}^{N} F_{jk}\hat{\phi}_{k}+\mathbf{B}_{j}\hat{\phi}_{j}=\omega\,\mathbf{C}_{j}\hat{\phi}_{j},\quad \quad j=0,1,\ldots,N
\label{eq:61}
\end{equation}
where $\mathbf{A}_{j}$, $\mathbf{B}_{j}$ and $\mathbf{C}_{j}$ are evaluated from the base flow at $y_{j}$.
Note that in the present stability framework, all perturbation quantities involve only first-order derivatives, so only the first-derivative matrix $F$ is required.
At the wall, i.e., $j=0$ and $j=N$, the boundary conditions for $\hat{u},\;\hat{v},\;\hat{T},\;\hat{p}$ are implemented to cover the first four rows of $\mathbf{A}_{0}$, $\mathbf{B}_{0}$, $\mathbf{C}_{0}$ and $\mathbf{A}_{N}$, $\mathbf{B}_{N}$, $\mathbf{C}_{N}$. For the perturbations of viscous stress and heat flux, the discrete equations of $\hat{\Pi}_{ij}$ and heat flux $\hat{Q}_i$ are applied at the wall, same as at the interior nodes, and no additional boundary conditions for these variables are specified.
The resulting homogeneous linear system can be written as
\begin{equation}
\bar{\mathbf{A}}\mathbf{X}=\omega\bar{\mathbf{B}}\mathbf{X}\;\;\;\; \text{or}\;\;\;\; \Mmat(\omega)\mathbf{X}=0,\;\;\;\; \text{where}\quad \quad \mathbf{X}={[\hat{\phi}_{0}^{T},\hat{\phi}_{1}^{T},\ldots,\hat{\phi}_{N}^{T}]}^{T}
\label{eq:62}
\end{equation}
where $\bar{\mathbf{A}}$ and $\bar{\mathbf{B}}$ are square matrices of $9(N+1) \times 9(N+1)$. A nontrivial solution exists only if $|\Mmat(\omega)|=0$, which defines the discrete dispersion relation.

Two solution strategies are employed in this paper.
The first one is a global method.
For the generalized eigenvalue problem $\bar{\mathbf{A}}\mathbf{X}=\omega\bar{\mathbf{B}}\mathbf{X}$, the QZ algorithm is used to compute all eigenvalues. Sometimes the matrix $\bar{\mathbf{B}}$ may be singular, the QZ method can be still used, with the infinite eigenvalues omitted.
The second one is a local method.
Given an initial eigenvalue estimate $\omega^{(0)}$ (from the global method or the eigenvalue at the preceding wavenumber), $\Mmat(\omega^{(0)})$ is factorized by LU decomposition.
The shift-invert Arnoldi method is then used to compute eigenvalues near $\omega^{(0)}$, and the corresponding eigenfunction is a vector in the null space of $\Mmat(\omega)$.
This method requires only one LU factorization per step and is much faster than the global QZ method, which is used for mode tracking and eigenfunction computation in this paper.

\subsection{Code validation}\label{sec:4.3}

\begin{figure}[htbp]
  \centering
  \includegraphics[width=0.46\textwidth]{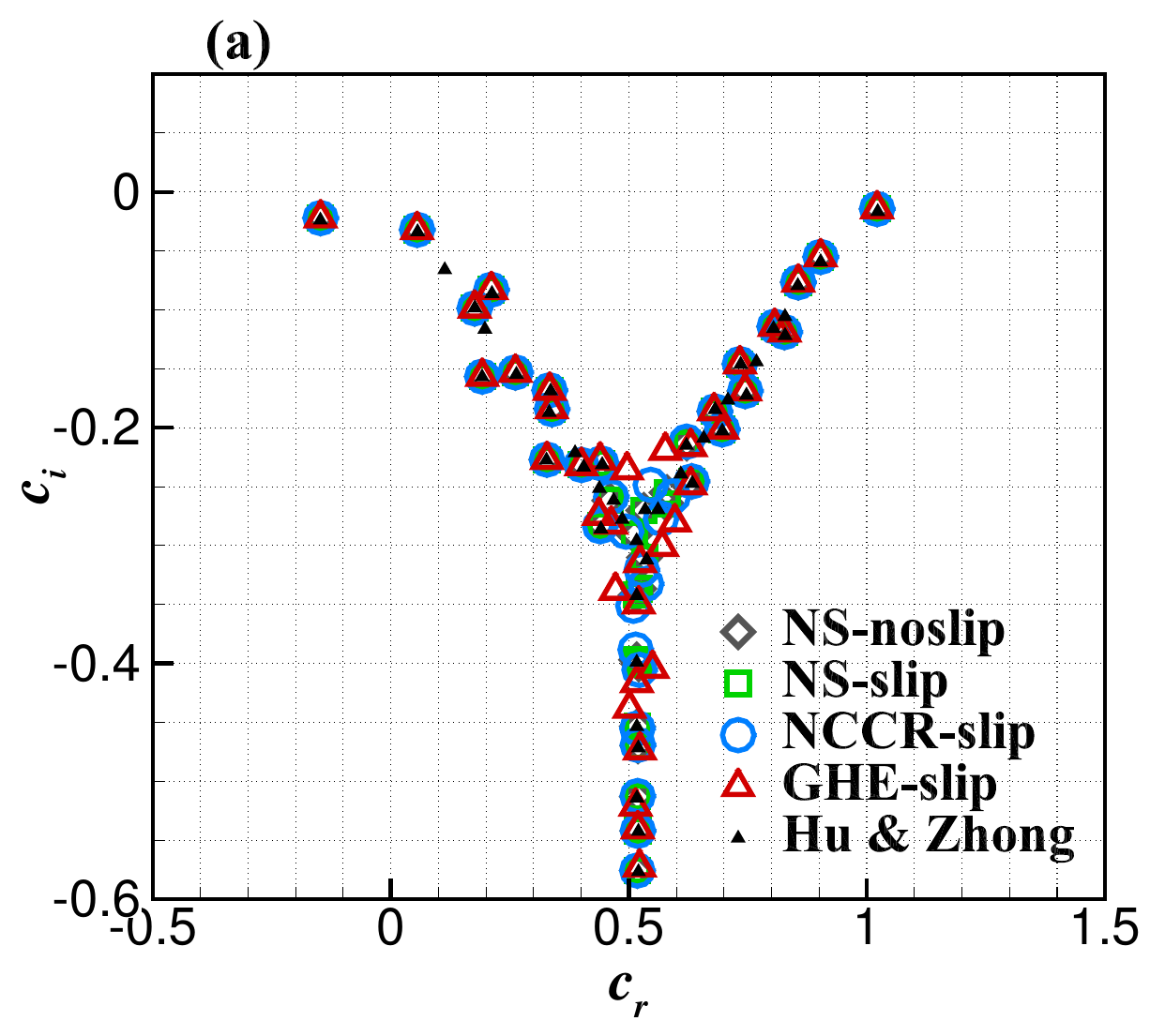}\hspace{0.03\textwidth}
  \includegraphics[width=0.46\textwidth]{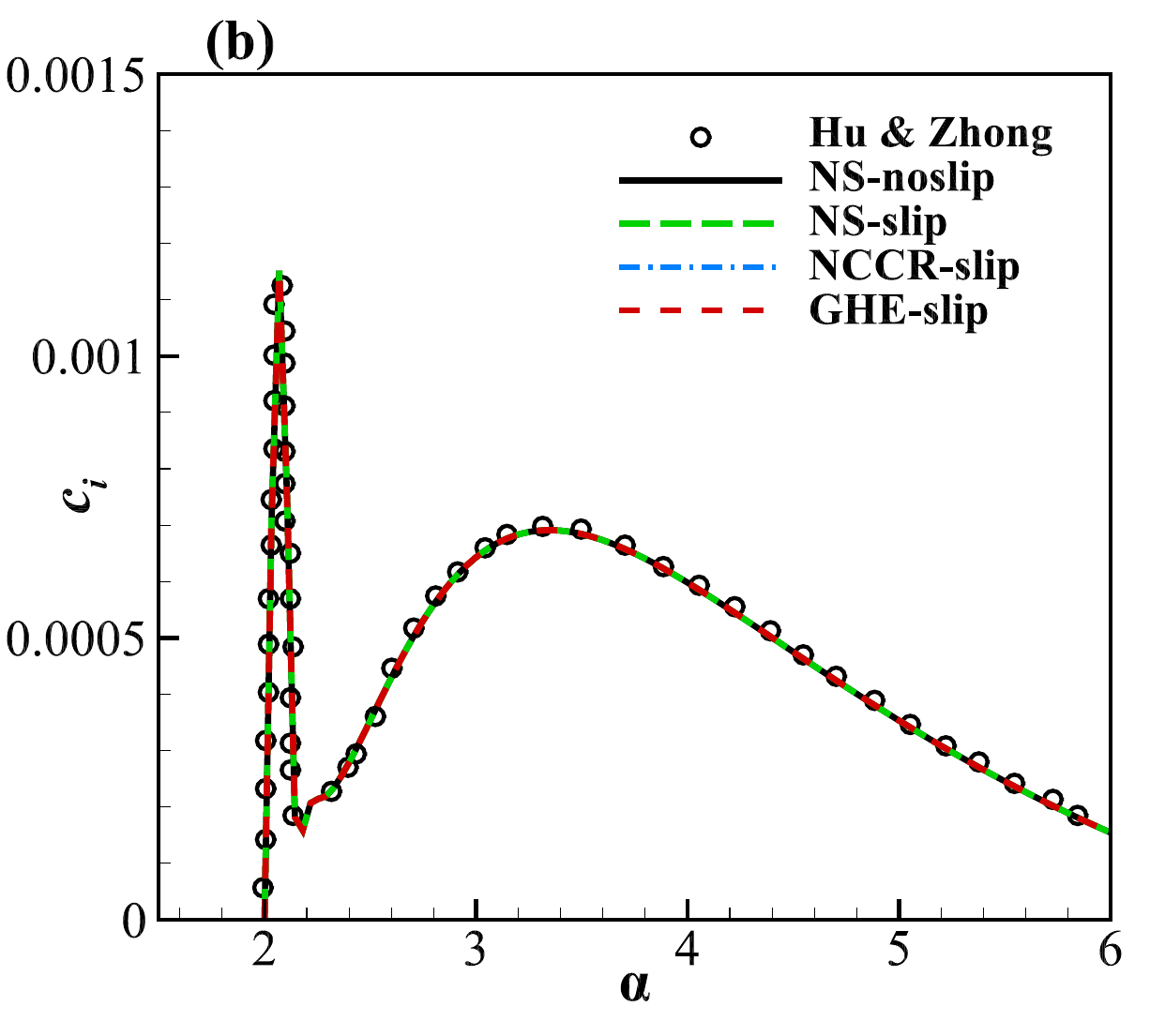}
  \caption{Comparison of the present results with those of Hu and Zhong~\cite{ref5} (adiabatic lower wall): (a) eigenvalue spectrum at $\mathit{Ma}=5$, $\mathit{Re}=5\times 10^{5}$ and $\alpha=0.1$; (b) variation of the imaginary part of the phase speed $c_{i}$ with $\alpha$ at $\mathit{Ma}=5$ and $\mathit{Re}=5\times 10^{6}$.}
  \label{fig:3}
\end{figure}

To verify the present GHE stability equations and computational code, two plane Couette flow cases are analyzed and compared with results in the literature.
First, the continuum cases of Hu and Zhong~\cite{ref5} at $\mathit{Ma}=5$, $\mathit{Re}=5\times 10^{5}$ and $\mathit{Re}=5\times 10^{6}$ with adiabatic lower wall are computed. The working gas is air, with $\gamma=1.4$, $\mathit{Pr}=0.72$ and Sutherland's law for viscosity. The number of collocation points is $N=300$.
Figure~\ref{fig:3} compares the eigenspectra at $\alpha=0.1$ and the variations of $c_{i}$ with $\alpha$ computed by the present NS-noslip, NS-slip, NCCR-slip and GHE-slip models with the results of Hu and Zhong~\cite{ref5}.
For such continuous cases, the present NS-slip, NCCR-slip and GHE-slip stability results recover to the conventional NS-noslip results and agree very well with those of Hu and Zhong~\cite{ref5}.

\begin{figure}[htbp]
  \centering
  \includegraphics[width=0.46\textwidth]{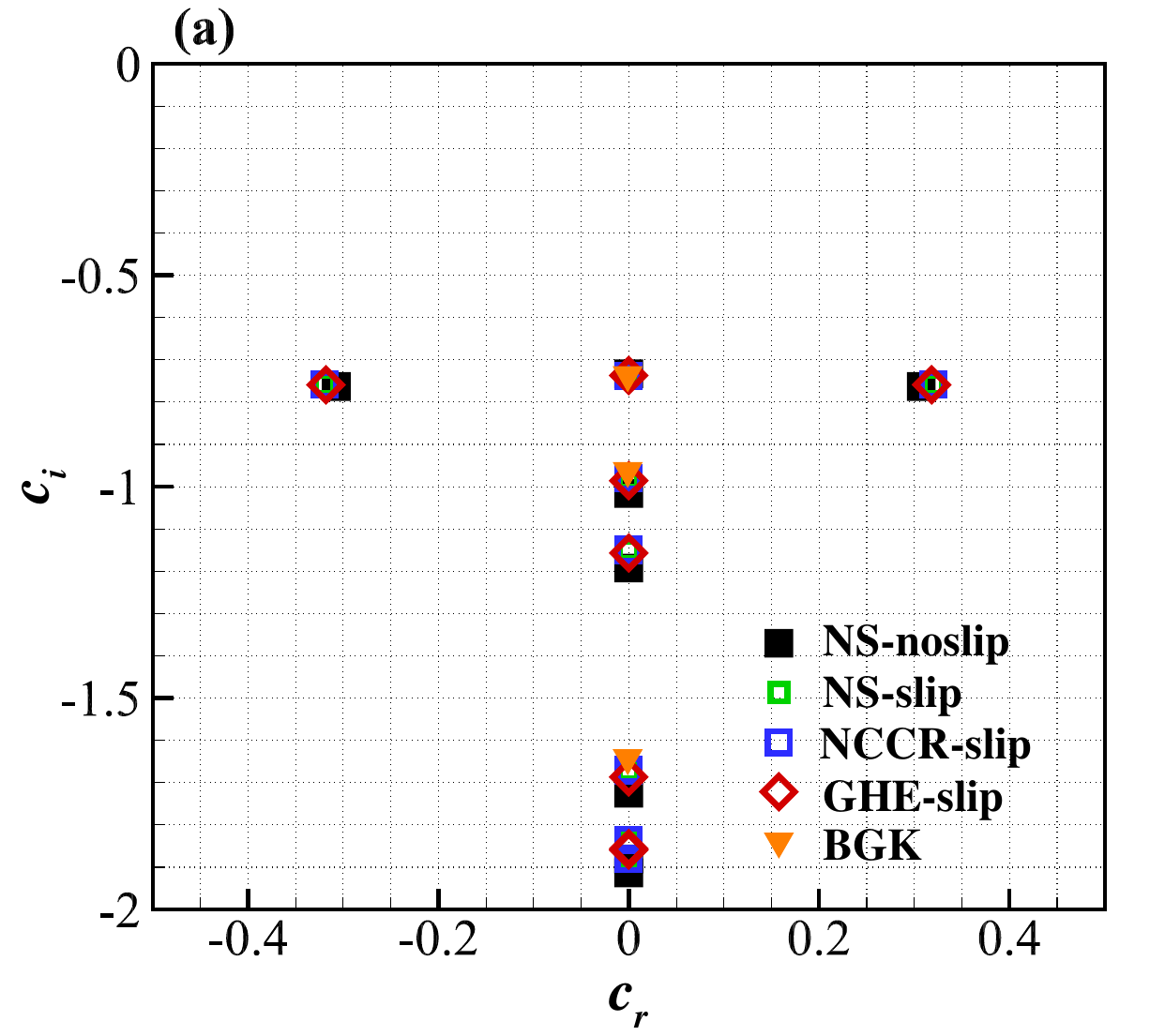}\hspace{0.03\textwidth}
  \includegraphics[width=0.46\textwidth]{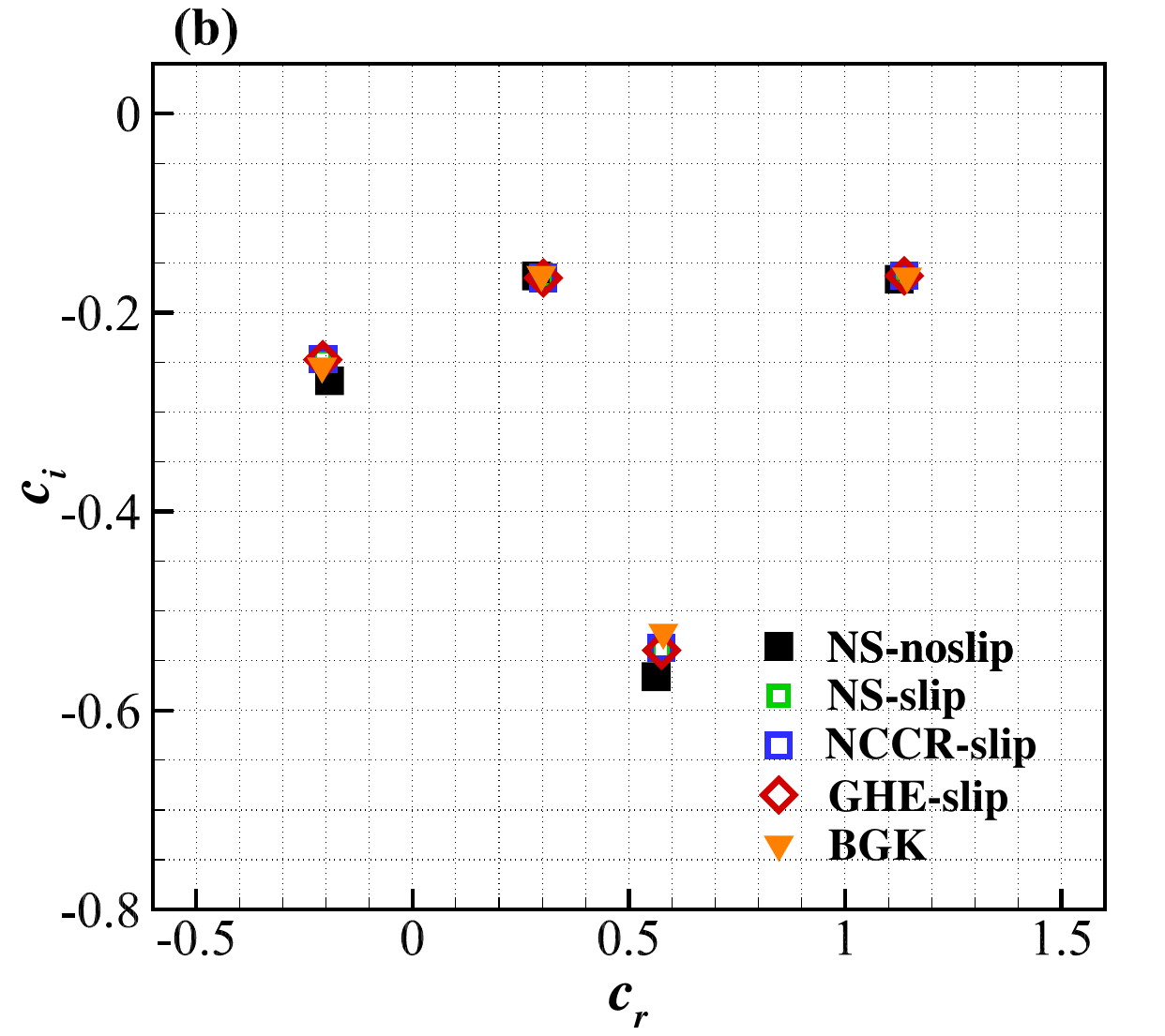}
  \caption{Comparison of the present eigenvalue spectra with the BGK-based stability results of Zou et al.~\cite{ref20,ref21}: (a) $\mathit{Ma}=0.1$, $\mathit{Kn}=0.02$ and $\alpha=3.0$; (b) $\mathit{Ma}=2.0$, $\mathit{Kn}=0.02$ and $\alpha=0.3$.}
  \label{fig:4}
\end{figure}

Second, the cases of Zou et al.~\cite{ref20,ref21} at $\mathit{Ma}=0.1$, $\mathit{Kn}=0.02$ and $\mathit{Ma}=2$, $\mathit{Kn}=0.02$ are computed. The working gas is argon, with $\gamma=5/3$ and $\mathit{Pr}=1$. For these two cases, Zou et al.~\cite{ref20,ref21} presented linear stability results of the BGK equations.
Figure~\ref{fig:4} compares the disturbance eigenspectra computed by the present NS-noslip, NS-slip, NCCR-slip and GHE-slip models with the BGK stability results of Zou et al.~\cite{ref20,ref21}. For these two cases, the flow is beyond continuum regime, and the conventional NS-noslip stability results deviate noticeably from the BGK results. By adopting slip boundary conditions and nonlinear constitutive relations, the GHE-slip model agrees well with the BGK stability analysis.
In fact, these two cases are mainly in the slip regime, so the GHE/NCCR-slip results differ only slightly from the NS-slip results.
For flows at higher Mach and Knudsen numbers, the effects of shear nonequilibrium on stability become more pronounced, as will be shown later.

\begin{figure}[htbp]
  \centering
  \includegraphics[width=0.46\textwidth]{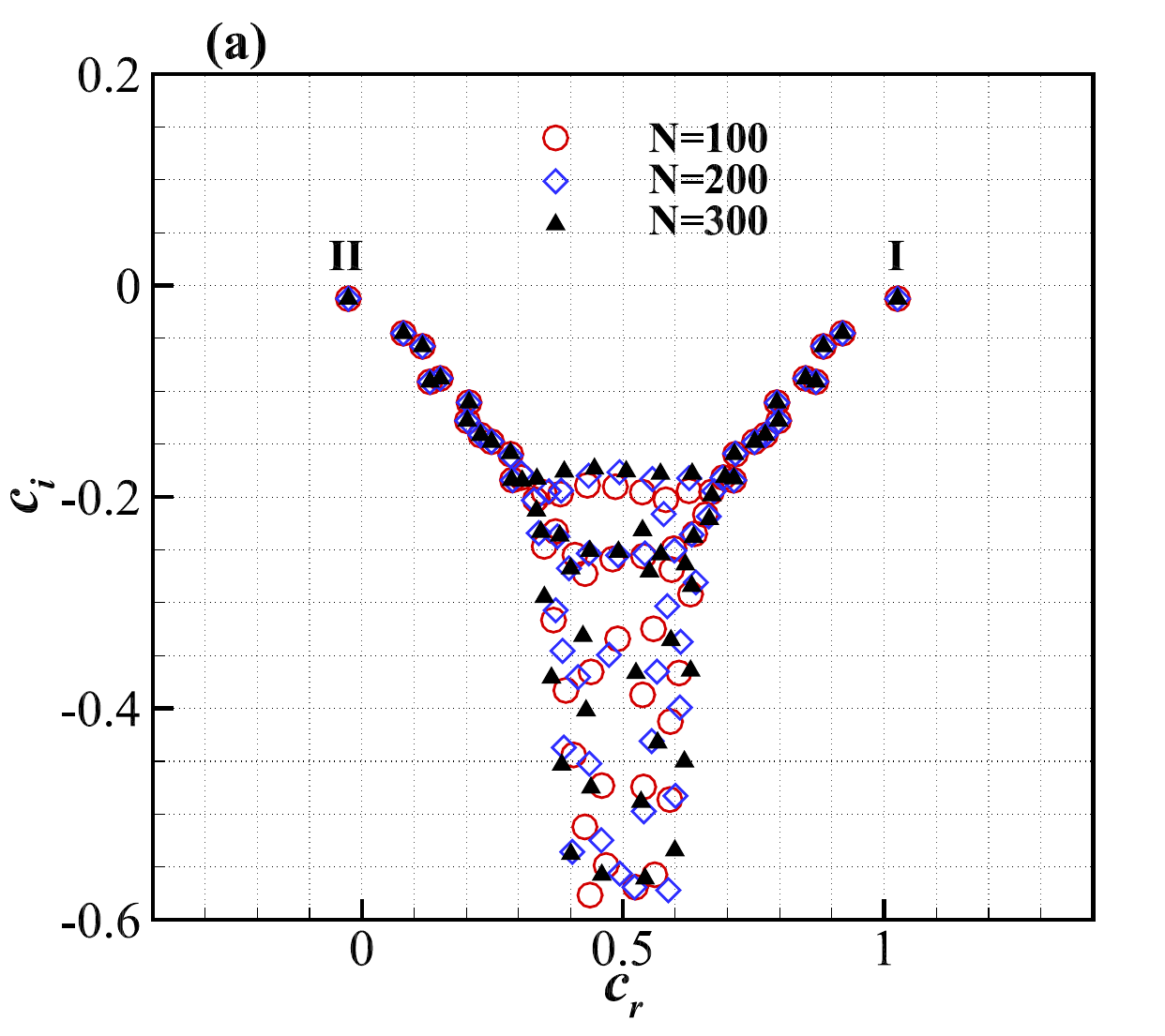}\hspace{0.03\textwidth}
  \includegraphics[width=0.46\textwidth]{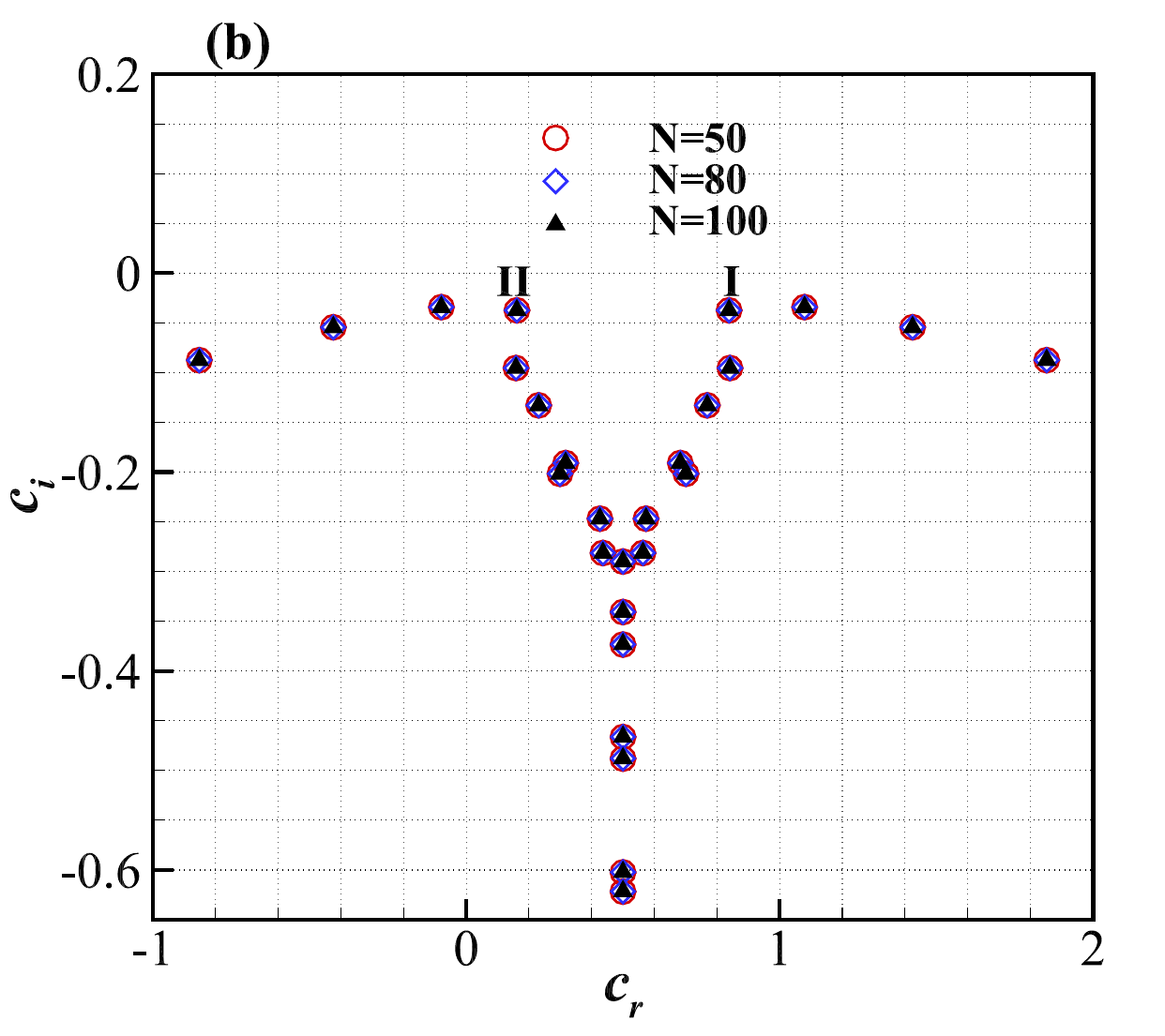}
  \caption{Eigenvalue spectra at $\mathit{Ma}=5$ for different numbers of collocation points $N$: (a) $\mathit{Re}=3\times10^5$ ($\mathit{Kn}=2.7\times 10^{-5}$) and $\alpha=0.1$; (b) $\mathit{Kn}=0.005$ and $\alpha=2$.}
  \label{fig:5}
\end{figure}

Finally, a grid-independence study is performed for the cases of $\mathit{Ma}=5$, $\mathit{Re}=3\times10^5$ ($\mathit{Kn}=2.7\times 10^{-5}$) and $\mathit{Kn}=0.005$. The eigenspectra computed by the GHE-slip stability equations with different collocation points are shown in Fig.~\ref{fig:5}.
The working gas is argon, with $\gamma=5/3$, $\mathit{Pr}=2/3$, and a power-law viscosity $\mu=T^{0.81}$. The Chebyshev collocation points adopted for the $\mathit{Kn}=2.7\times 10^{-5}$ case are 100, 200 and 300, and those for the $\mathit{Kn}=0.005$ case are 50, 80 and 100.
As can be seen, the eigenspectra computed by different collocation points agree very well.
For continuum flows, 200 collocation points are generally sufficient to resolve the acoustic modes (e.g., Modes I and II) and the upper part of the viscous spectrum (i.e., the ``V''-shaped branches), whereas more points are required to resolve the viscous spectrum below the triple point, particularly at high Reynolds numbers and high wavenumbers.
For rarefied flows, fewer collocation points are usually required (see Fig.~\ref{fig:5}(b)).
Unless otherwise stated, 200 collocation points are used in this study to compute the eigenvalues of the acoustic modes and the least stable viscous mode of plane Couette flow.

\section{Results and discussions}\label{sec:5}

\subsection{Base flow solutions}\label{sec:5.1}

First, the base-flow profiles of the GHE and NS equations at different Mach and Knudsen numbers are computed using the shooting method and compared with DSMC results.
The base-flow profiles of the NCCR model are identical to those of the GHE model, and are not shown here.
The working gas is argon, with $\gamma=5/3$, $\mathit{Pr}=2/3$, and a power-law viscosity $\mu=T^{0.81}$. In the DSMC simulation, the variable hard sphere (VHS) model is adopted for intermolecular collisions, and the fully diffuse reflection is implemented at the wall, with surface accommodation coefficients $\sigma_u=\sigma_T=1$.

\begin{figure}[htbp]
  \centering
  \includegraphics[width=0.46\textwidth]{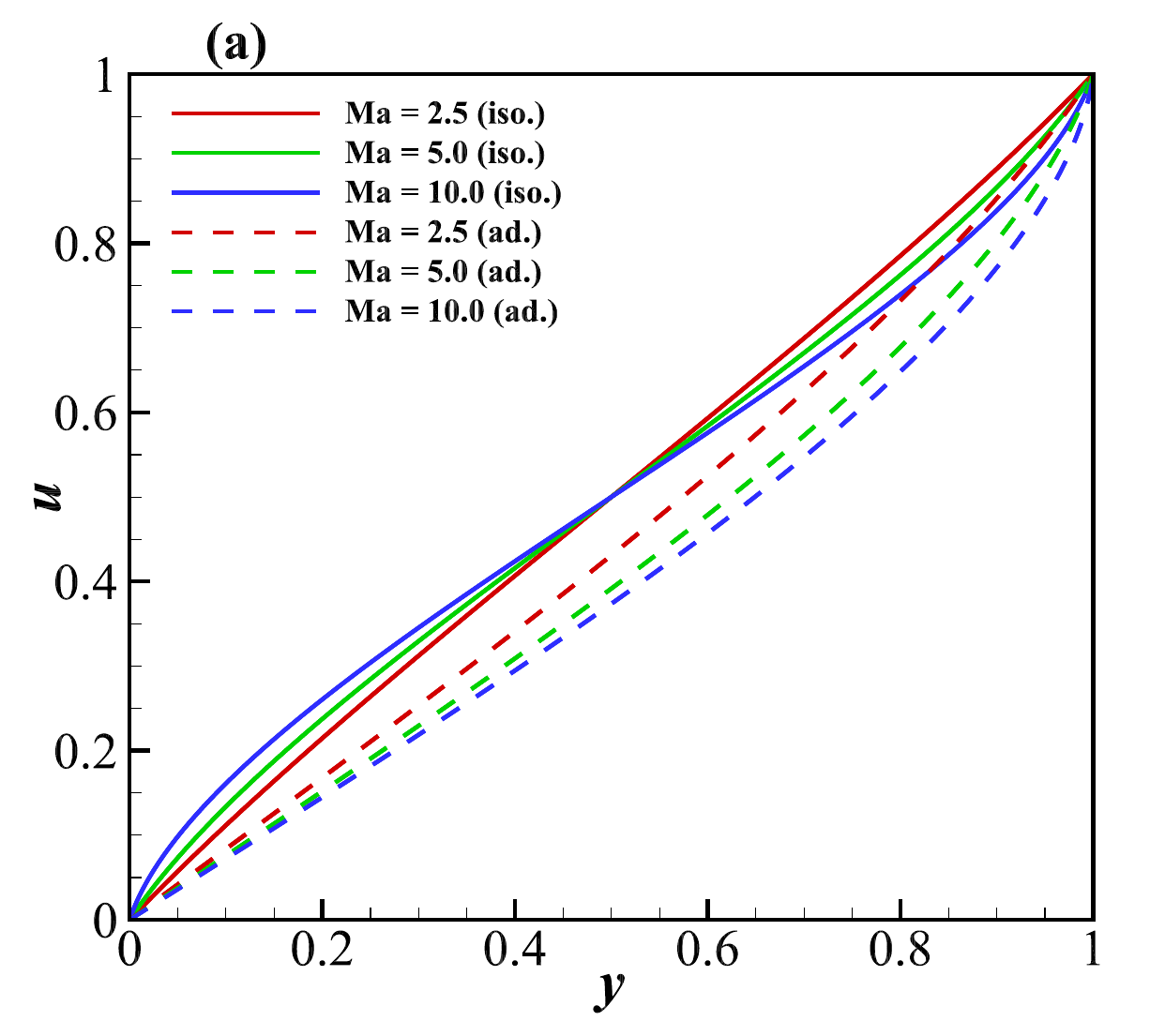}\hspace{0.03\textwidth}
  \includegraphics[width=0.46\textwidth]{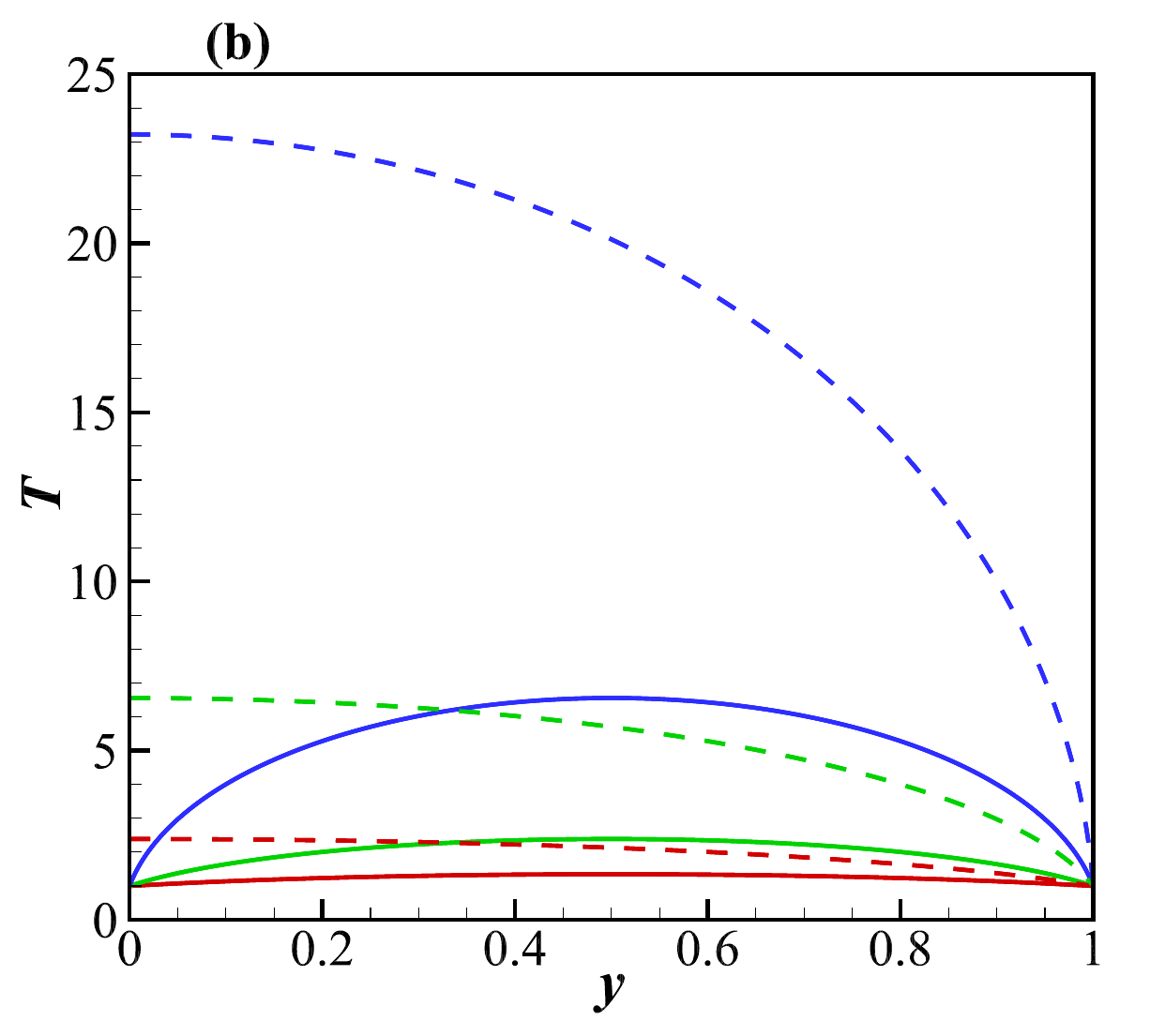}
  \caption{(a) Velocity and (b) temperature profiles at $\mathit{Ma}=2.5$, 5 and 10 and $\mathit{Kn}=10^{-4}$ (solid lines: isothermal lower wall; dashed lines: adiabatic lower wall).}
  \label{fig:6}
\end{figure}

Figure~\ref{fig:6} shows the velocity and temperature profiles computed by the GHE equations with slip boundary conditions at $\mathit{Ma}=2.5$, 5 and 10 and $\mathit{Kn}=10^{-4}$, with either an isothermal or adiabatic lower wall.
For these cases, the flow is in the continuum regime, and the GHE-slip solutions are completely identical to the NS-noslip and NS-slip solutions.
As can be seen, for the isothermal lower-wall cases, the velocity profiles are antisymmetric along the channel centerline and the temperature profile is symmetric, with a peak temperature of approximately $T_{m}=1+0.5\left( \gamma-1 \right)\mathit{Pr}\left( \mathit{Ma}/2 \right)^{2}$ at the channel center $y=0.5$.
Meanwhile, a generalized inflection point exists at $y=0.5$, i.e., $d\left( \rho du/dy \right)/dy=0$.
Under the adiabatic lower-wall conditions, the antisymmetry of the velocity profile is broken, and the temperature profile takes its extremum at the wall. The adiabatic wall temperature is approximately $T_{m}=1+0.5\left( \gamma-1 \right)\mathit{Pr}\,\mathit{Ma}^{2}$, and no generalized inflection point exists in the flow.

\begin{figure}[htbp]
  \centering
  \includegraphics[width=0.32\textwidth]{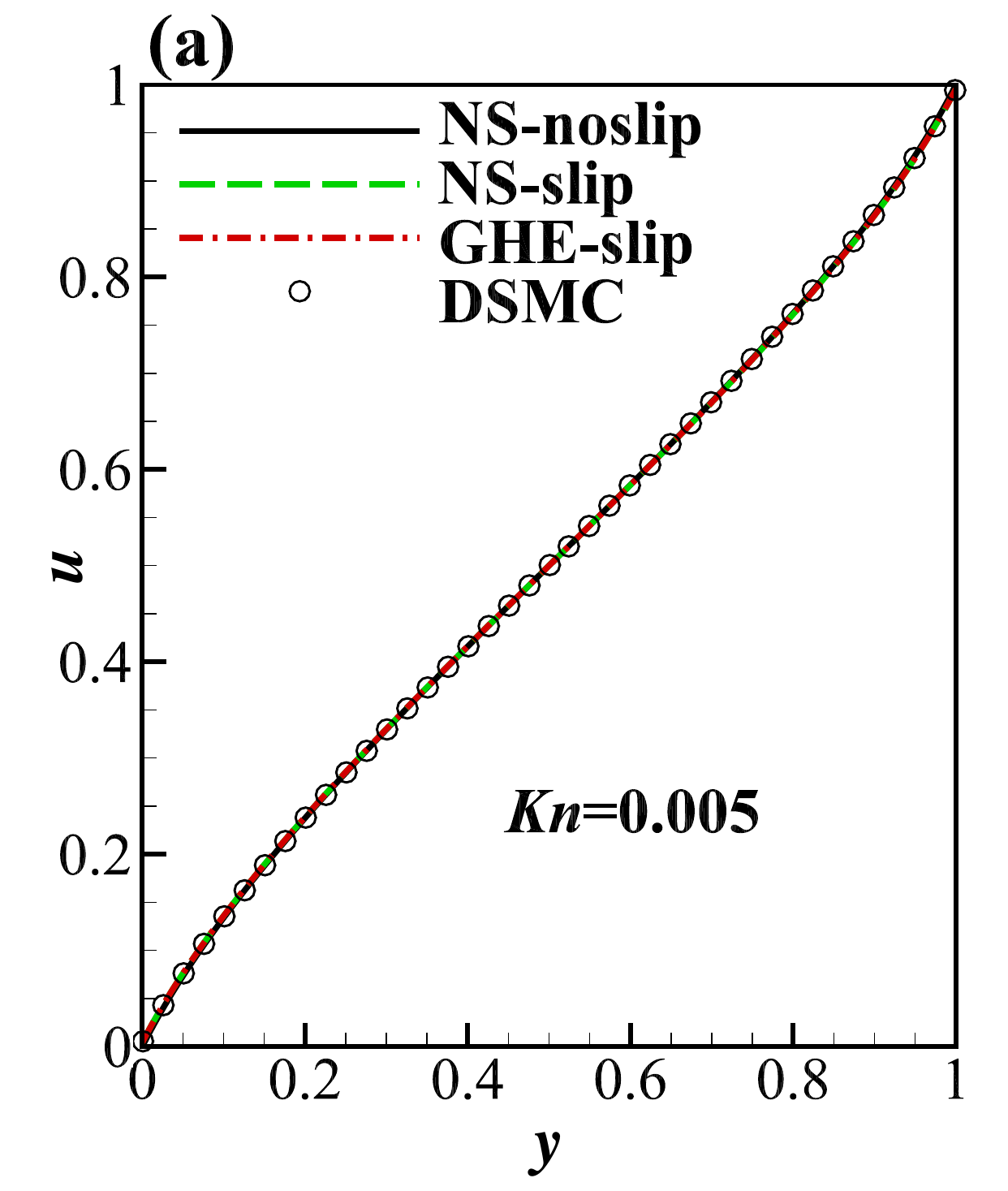}\hfill
  \includegraphics[width=0.32\textwidth]{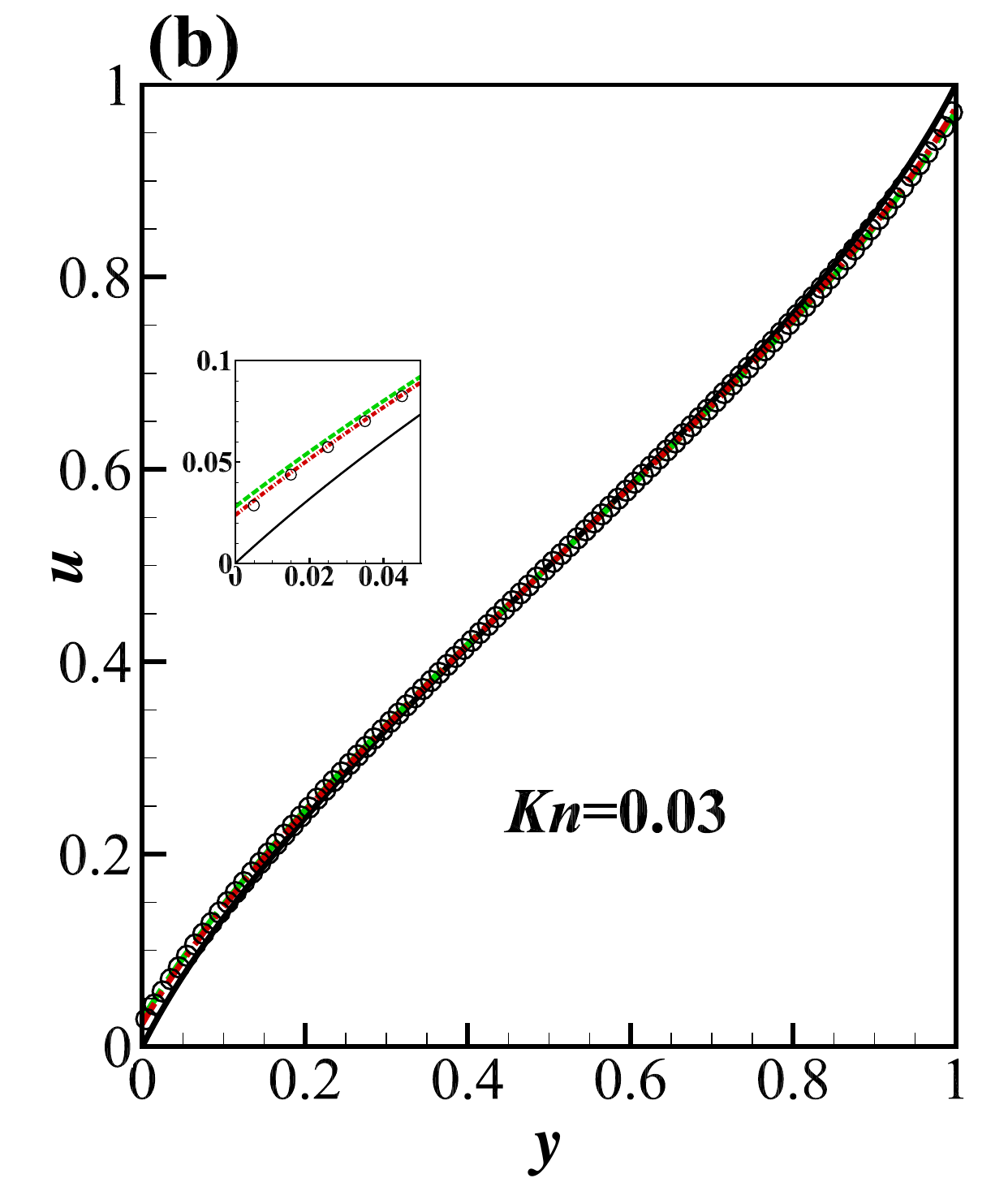}\hfill
  \includegraphics[width=0.32\textwidth]{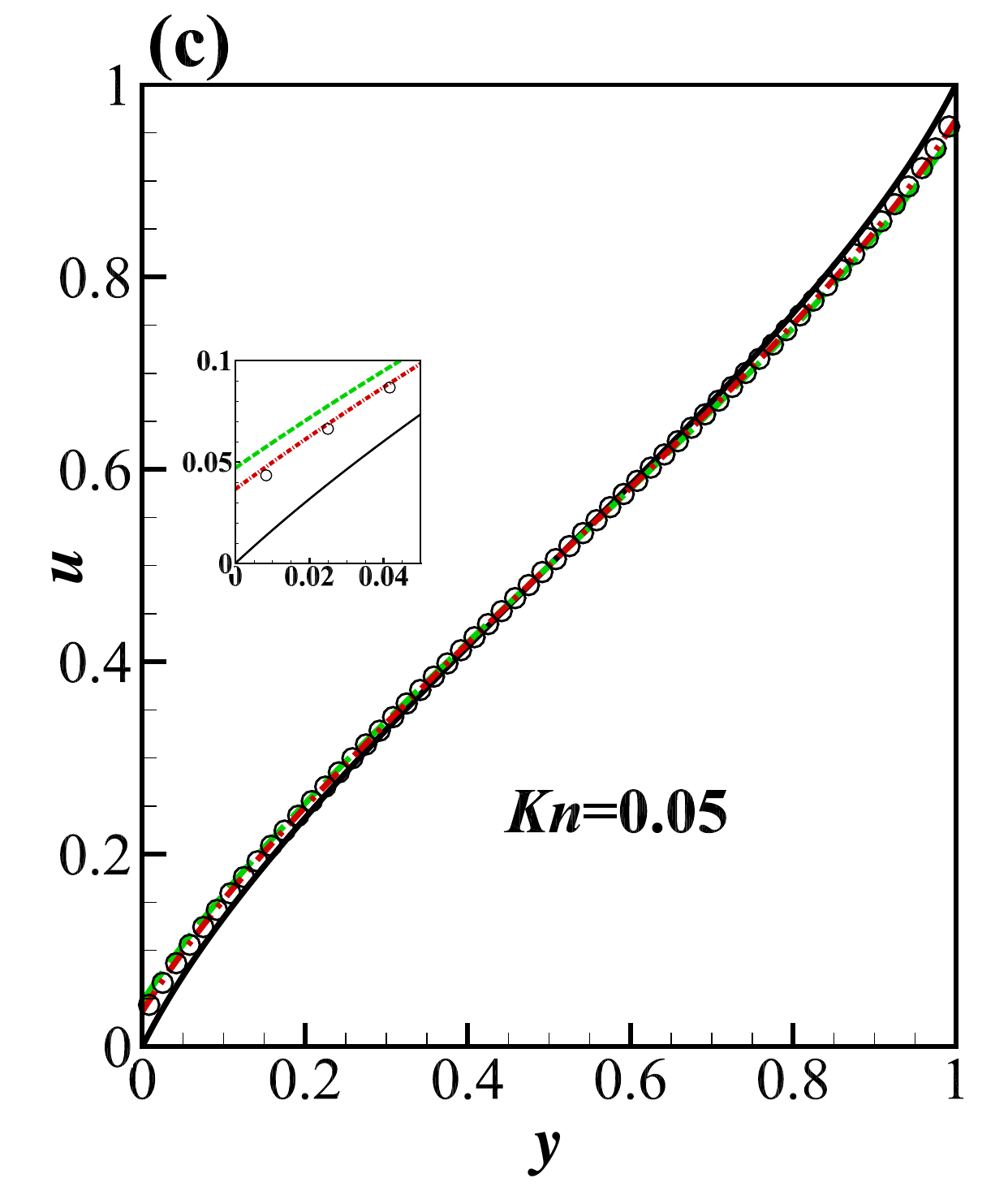}\\[6pt]
  \includegraphics[width=0.32\textwidth]{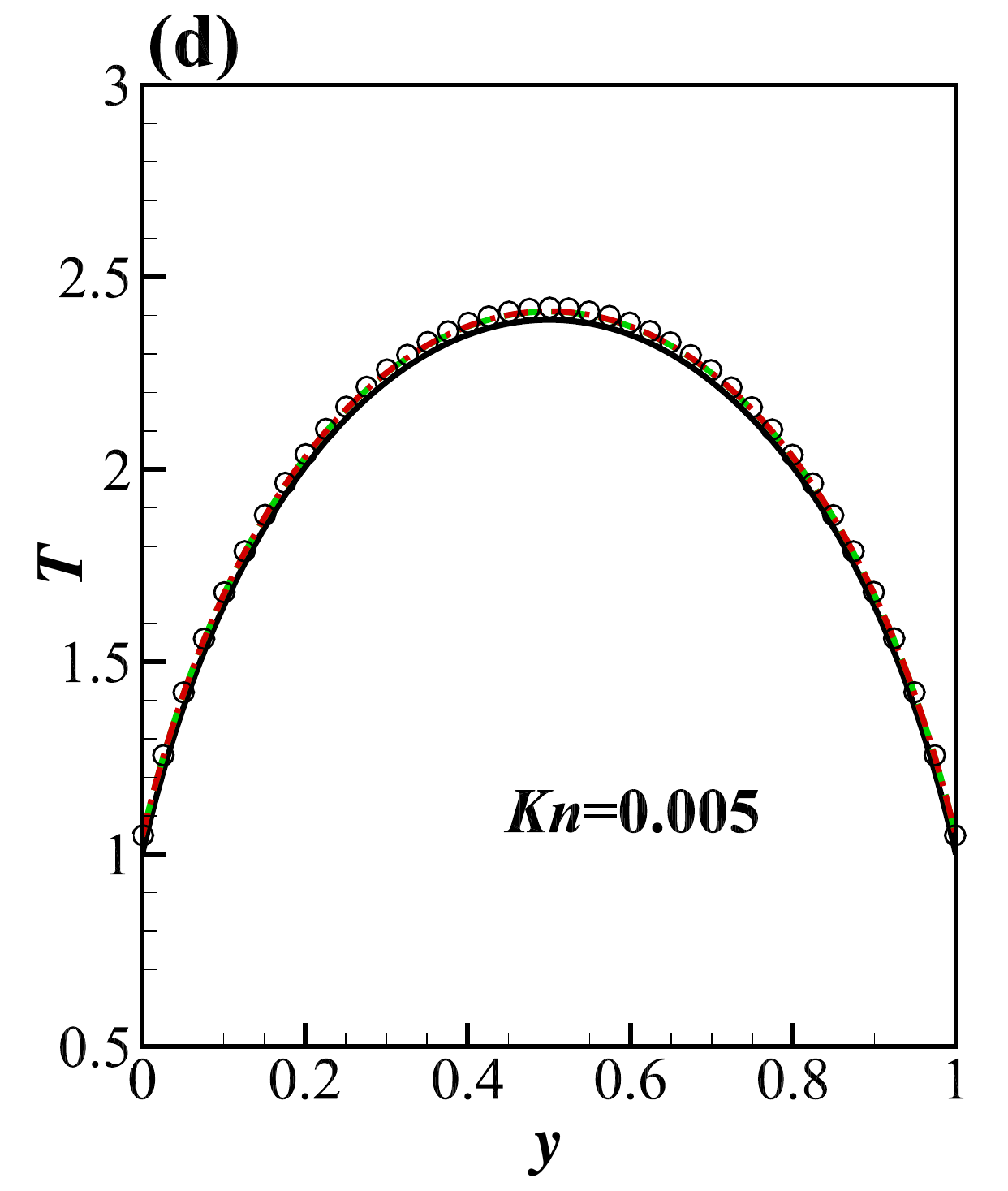}\hfill
  \includegraphics[width=0.32\textwidth]{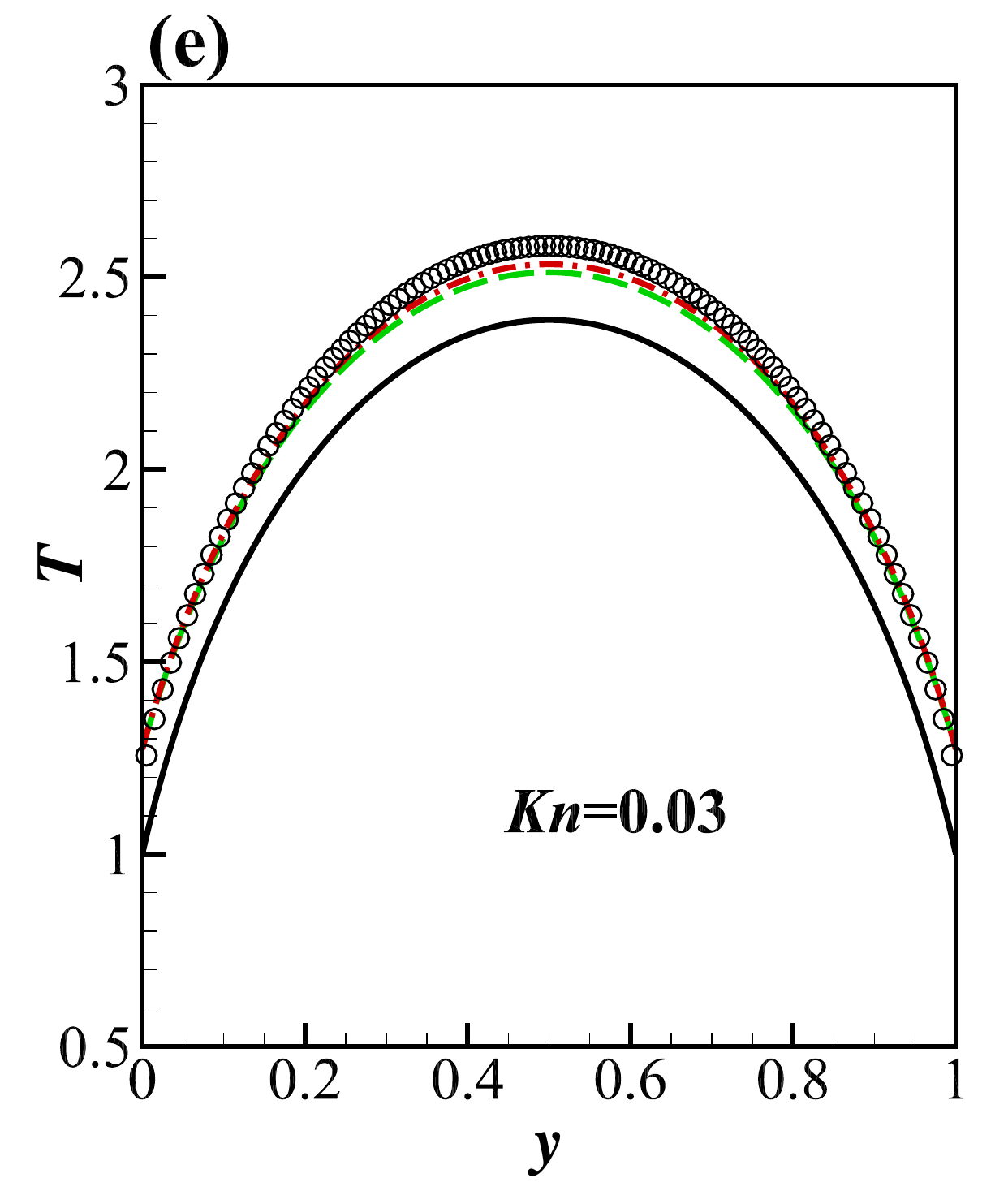}\hfill
  \includegraphics[width=0.32\textwidth]{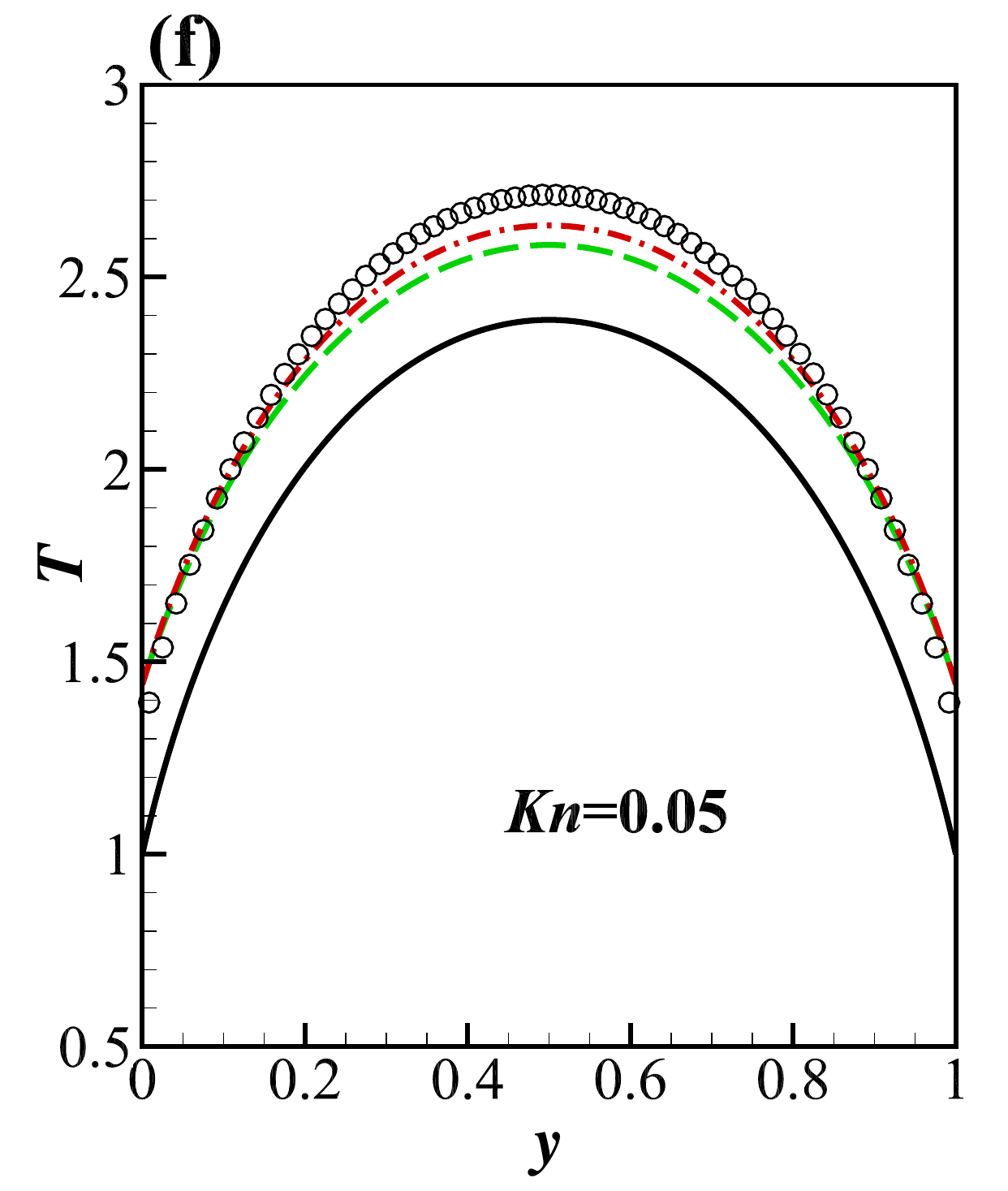}
  \caption{(a--c) Velocity and (d--f) temperature profiles at $\mathit{Ma}=5$ and $\mathit{Kn}=0.005$, 0.03 and 0.05.}
  \label{fig:7}
\end{figure}

Figure~\ref{fig:7} compares the velocity and temperature profiles at $\mathit{Ma}=5$ and different Knudsen numbers computed by the NS-noslip, NS-slip and GHE-slip models, where the no-slip or slip boundary conditions are imposed at both walls. The DSMC results are also given in the figure to assess the accuracy of different models.
At $\mathit{Kn}=0.005$, the rarefaction degree of the flow is weak, and the velocity and temperature profiles computed by the NS-noslip, NS-slip and GHE-slip models agree well with each other, and also with the DSMC results.
For this $\mathit{Kn}=0.005$ case, the surface slip velocity is approximately 0.0044 (about 0.44\% of the upper-wall velocity), and the deviations of the peak temperature between the NS/GHE models and DSMC are less than 0.5\%.
As the Knudsen number increases, the rarefaction effects gradually become prominent, and the velocity and temperature profiles computed by the NS-noslip model deviate increasingly from the DSMC results.
The slip boundary conditions can largely improve the no-slip results, and the GHE equations with slip boundary conditions show the best agreement with DSMC.
At $\mathit{Kn}=0.05$, the surface slip velocity and jump temperature reach 0.037 and 0.44, respectively. The peak temperature at the channel center predicted by the NS-noslip model deviates from the DSMC result by 12\%, whereas the deviation of the GHE-slip model is only 3\%.
The slip velocity in plane Couette flow can be estimated as $u_{s}\sim (2-\sigma)/\sigma\cdot\mathit{Kn}$, indicating that the slip effects become prominent as the Knudsen number increases. The nonlinear transport induced by shear nonequilibrium effects is also very important, which will be shown in the following.

\begin{figure}[htbp]
  \centering
  \includegraphics[width=0.46\textwidth]{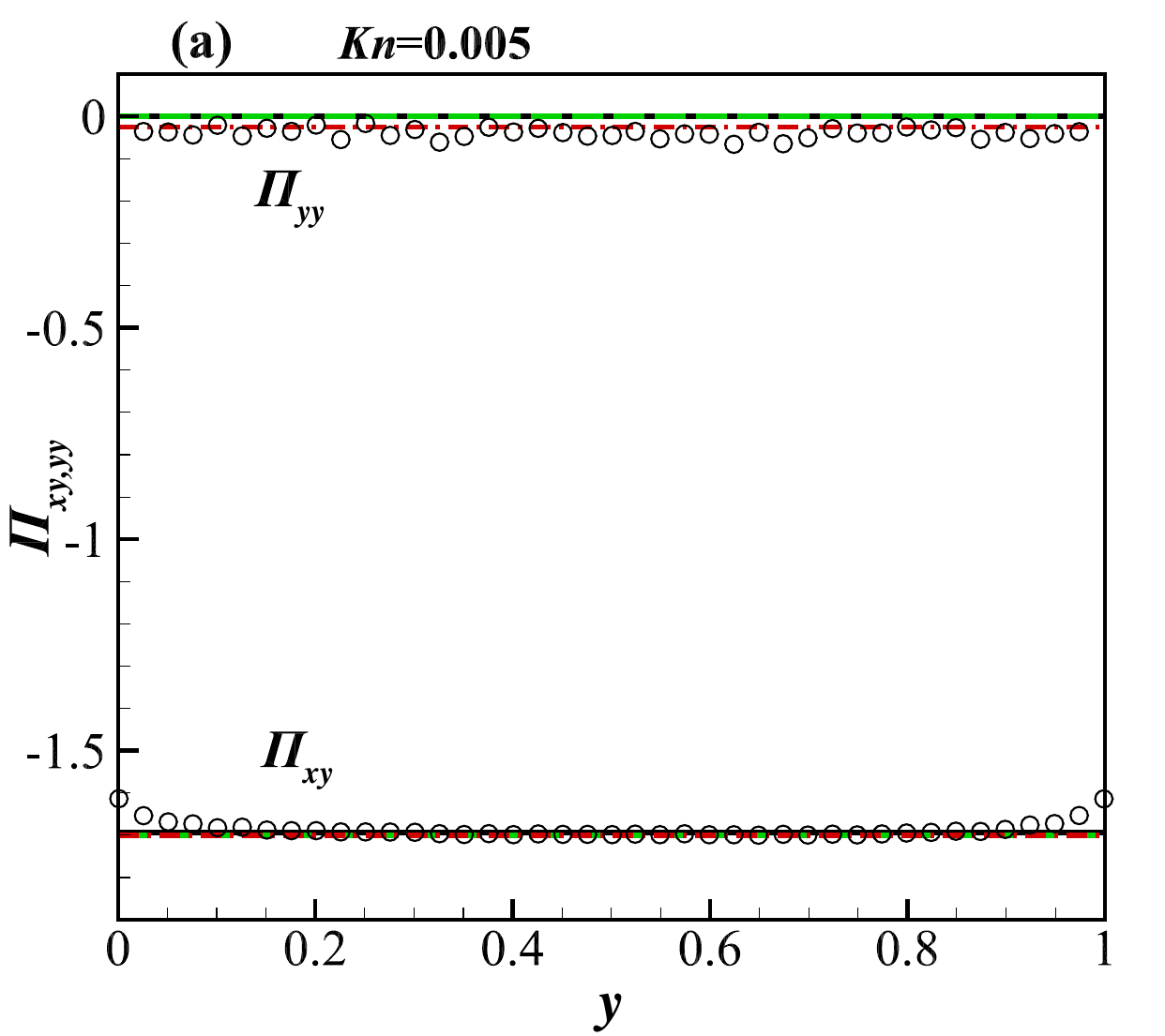}\hspace{0.03\textwidth}
  \includegraphics[width=0.46\textwidth]{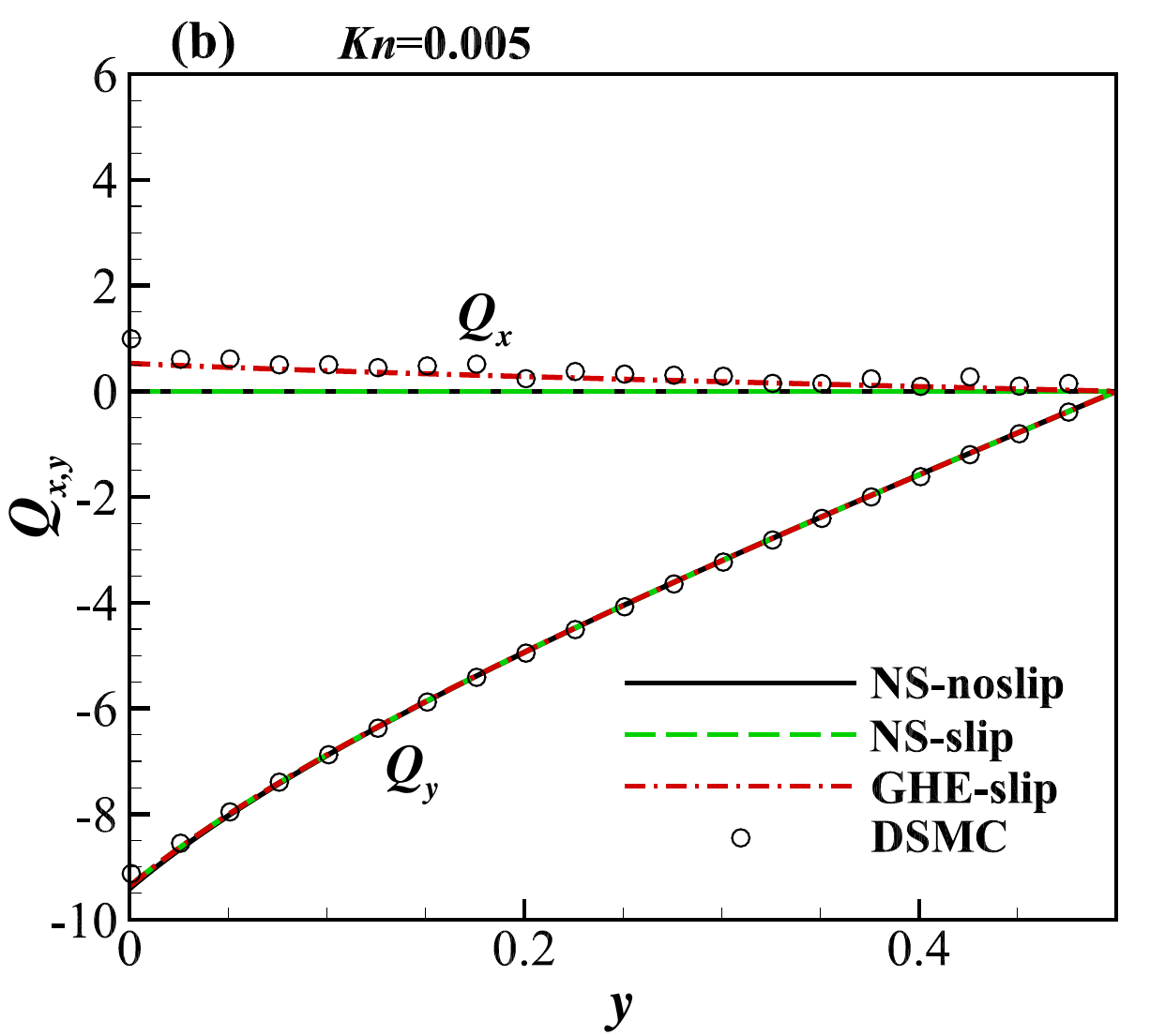}\\[6pt]
  \includegraphics[width=0.46\textwidth]{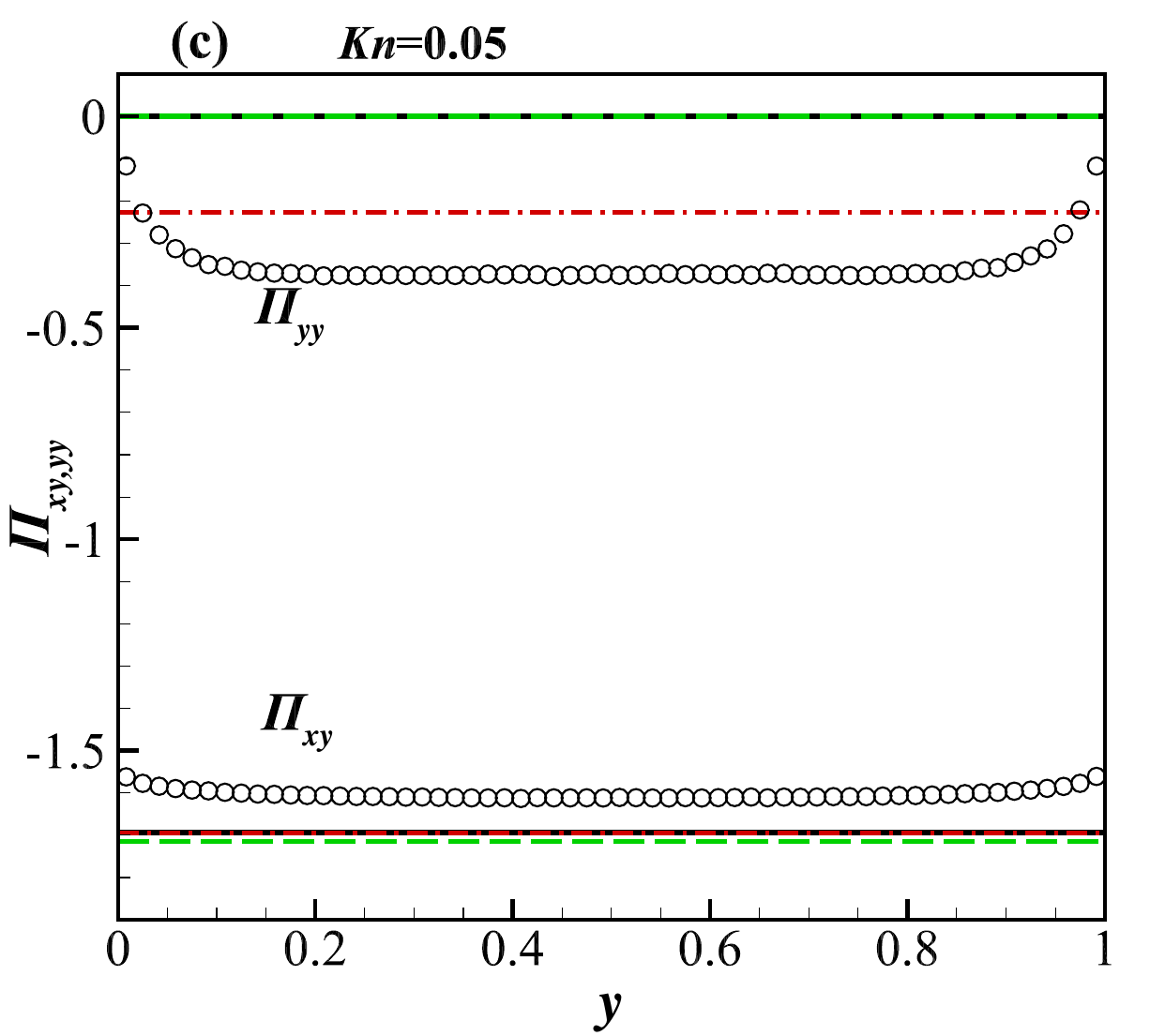}\hspace{0.03\textwidth}
  \includegraphics[width=0.46\textwidth]{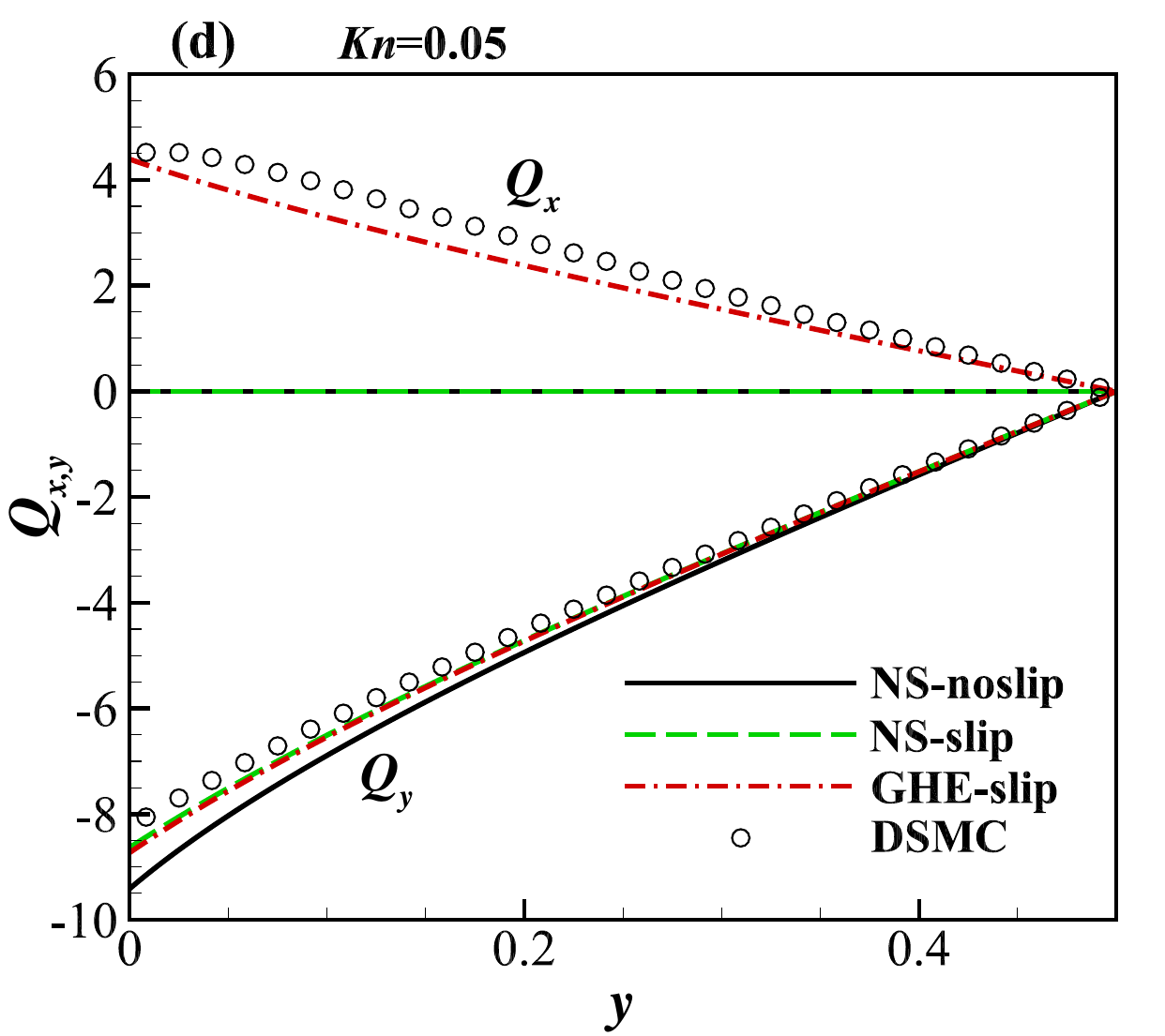}
  \caption{(a, c) Stress and (b, d) heat-flux profiles at $\mathit{Ma}=5$: (a, b) $\mathit{Kn}=0.005$; (c, d) $\mathit{Kn}=0.05$.}
  \label{fig:8}
\end{figure}

Figure~\ref{fig:8} shows the profiles of the shear stress $\Pi_{xy}$, normal stress $\Pi_{yy}$, normal heat flux $Q_{y}$ and tangential heat flux $Q_{x}$ computed by the NS-noslip, NS-slip, GHE-slip and DSMC at $\mathit{Ma}=5$ and $\mathit{Kn}=0.005$ and 0.05.
At $\mathit{Kn}=0.005$ (Figs.~\ref{fig:8}(a) and \ref{fig:8}(b)), the rarefaction effects are weak, and the shear stress and normal heat flux computed by NS, GHE and DSMC agree well.
For the normal stress and tangential heat flux, the NS model predicts zero values whether the slip conditions are used or not. However, the DSMC simulations show that the normal stress and tangential heat flux have small nonzero values, which are induced by weak rarefaction effects. The GHE model can capture such nonlinear transport features and agrees with the DSMC results.
When the Knudsen number increases to 0.05 (see Figs.~\ref{fig:8}(c) and \ref{fig:8}(d)), the deviations between NS-noslip and DSMC become more pronounced. The slip boundary conditions can improve the no-slip solution, however, the normal stress and tangential heat flux predicted by the NS model are always zero. For the DSMC results, the normal stress and tangential heat flux reach the same order as the shear stress and normal heat flux, and the ratios are about 23\% and 50\% respectively.
The GHE model can well capture such nonlinear transport behavior, and show good agreements with DSMC for all stress and heat flux components.

\subsection{Linear stability analysis}\label{sec:5.2}

In this section, linear stability analyses are performed for the above NS and GHE base flows in Sec. \ref{sec:5.1} by the consistent stability equations, to investigate the rarefaction effects on the plane Couette-flow stability, including eigenspectra, mode synchronization and the least stable mode (i.e., Mode II).
The main goal is to reveal qualitative and quantitative effects of surface slip and shear nonequilibrium effects on the stability. 

\begin{figure}[htbp]
  \centering
  \includegraphics[width=0.46\textwidth]{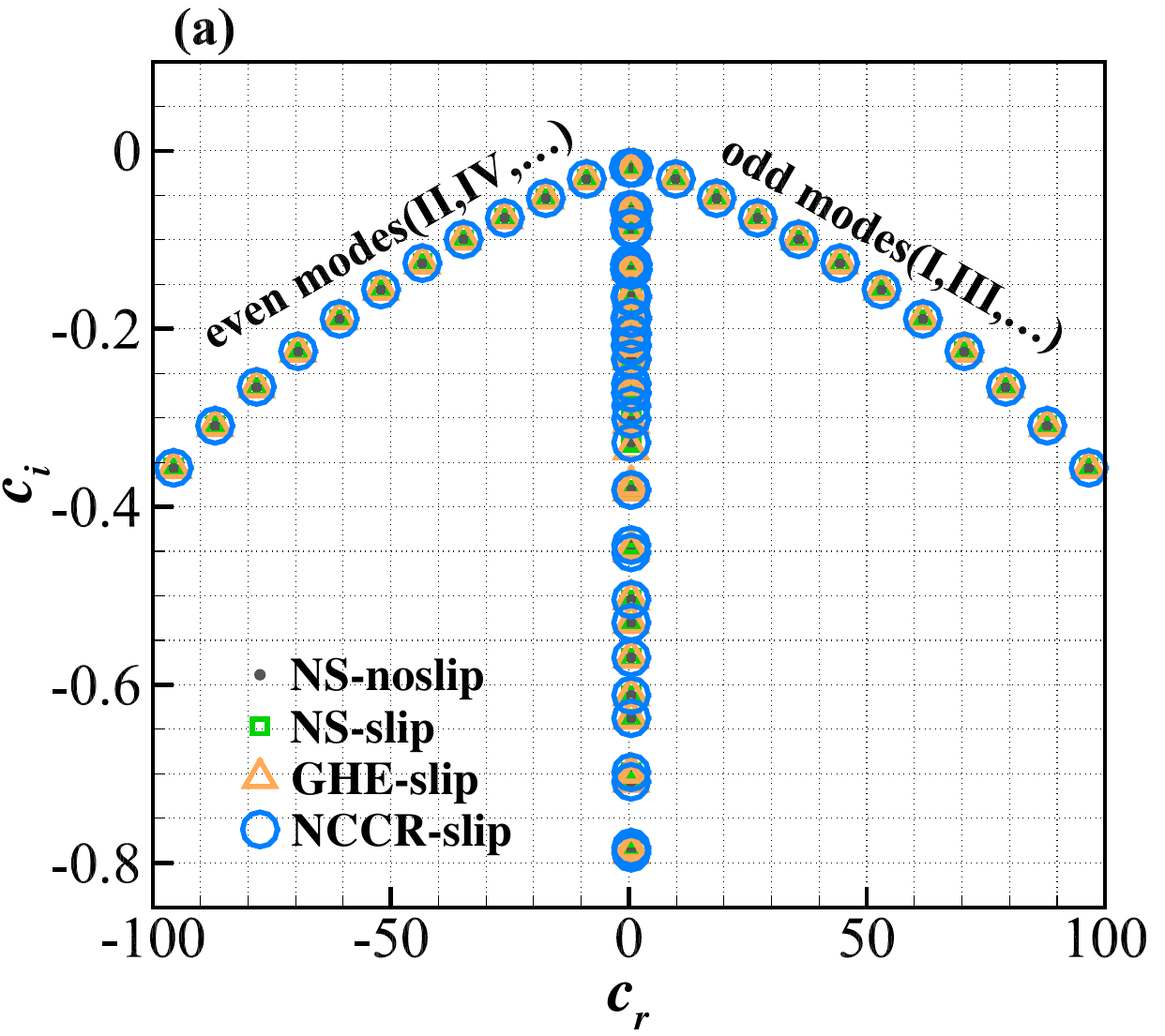}\hspace{0.03\textwidth}
  \includegraphics[width=0.46\textwidth]{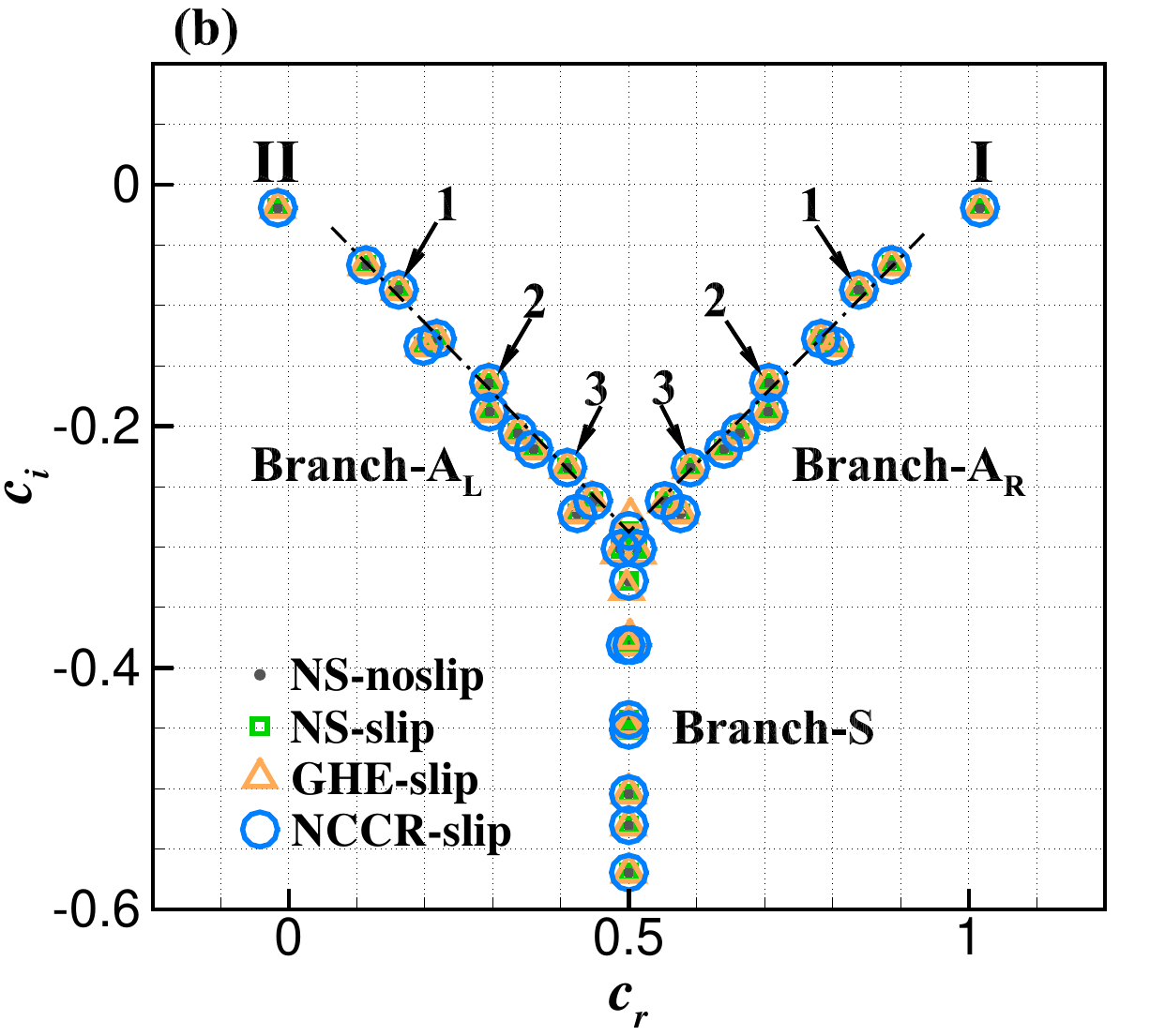}
  \caption{Eigenvalue spectrum at $\mathit{Ma}=5$, $\mathit{Kn}=8.09\times 10^{-5}$ and $\alpha=0.1$: (a) full spectrum; (b) enlarged view of $-0.2<c_{r}<1.2$.}
  \label{fig:9}
\end{figure}

First, Fig.~\ref{fig:9} shows the eigenspectra at $\mathit{Ma}=5$, $\mathit{Re}=1\times10^5$ ($\mathit{Kn}=8.09\times 10^{-5}$) and $\alpha=0.1$ computed by the NS-noslip, NS-slip, NCCR-slip and GHE-slip models, and Fig.~\ref{fig:9}(b) gives an enlarged view of $-0.2<c_{r}<1.2$.
For this continuum flow, the eigenspectra predicted by the NS, NCCR and GHE models are identical.
As shown in the figure, the compressible plane Couette flow admits both acoustic modes and viscous modes.
The acoustic modes are discrete modes located near $c_{i}=0$ and separated relatively far from each other.
They are divided into two families according to their phase speeds in the long-wave limit ($\alpha\to 0$), namely the odd modes I, III, \ldots{} ($c_{r}>1$) and the even modes II, IV, \ldots{} ($c_{r}<0$)~\cite{ref4}.
These modes are formed by acoustic waves reflected between the wall and the relative sonic line.
The viscous modes are those with phase speeds in the range $0<c_{r}<1$, forming a `Y'-shaped structure. The viscous modes are divided into the ``A'' branch and the ``S'' branch, and the discrete viscous modes on the inner side of the A branch are labeled as $A_1$, $A_2$, $A_3$, \ldots{} according to the value of $c_{i}$ from large to small~\cite{ref6}.
All viscous modes are usually stable, but they can be synchronized with the acoustic modes and lead to viscous instability~\cite{ref6}. The rarefaction effects on the eigenspectra and mode synchronization will be analyzed in the following.

\begin{figure}[htbp]
  \centering
  \includegraphics[width=0.46\textwidth]{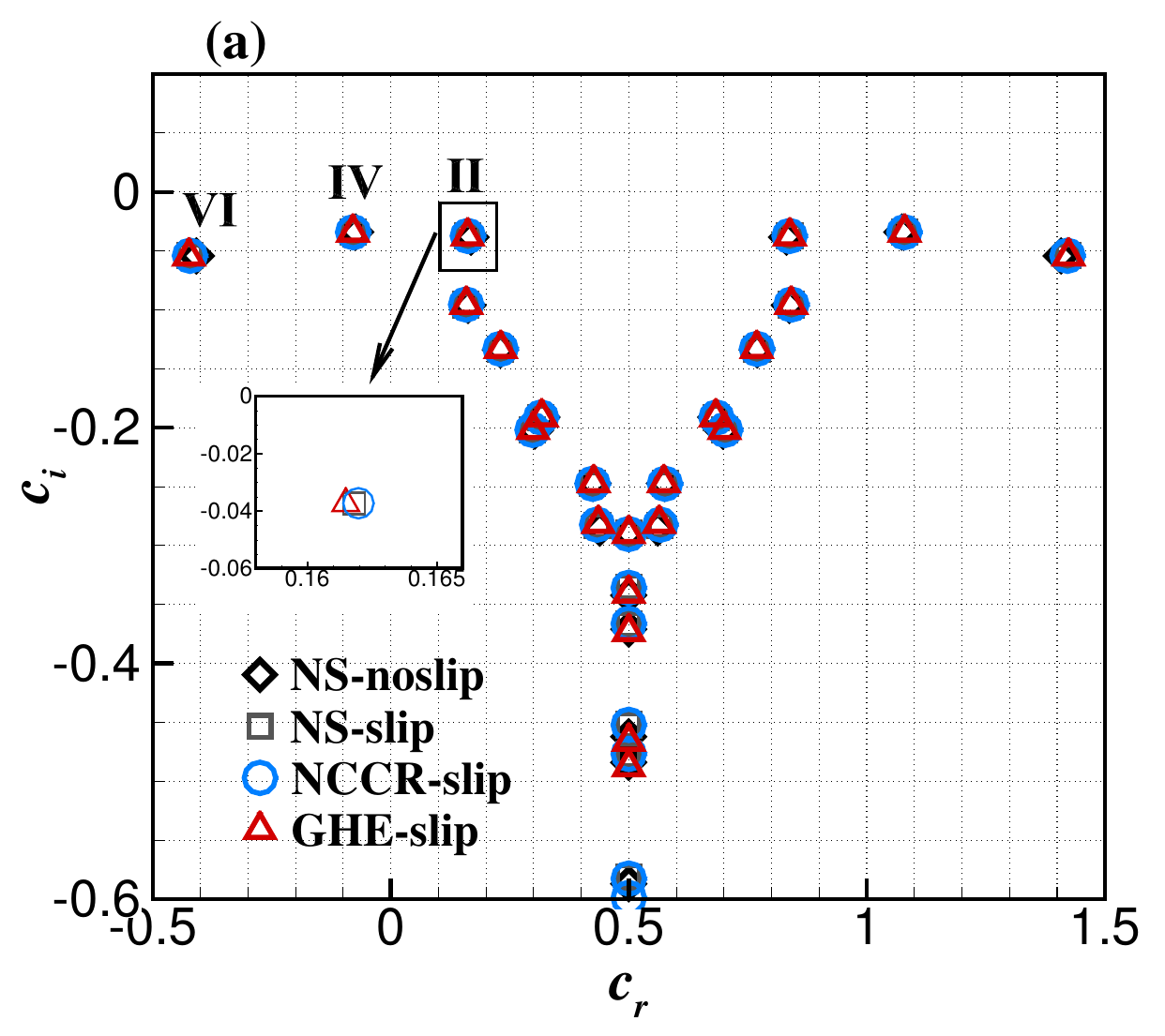}\hspace{0.03\textwidth}
  \includegraphics[width=0.46\textwidth]{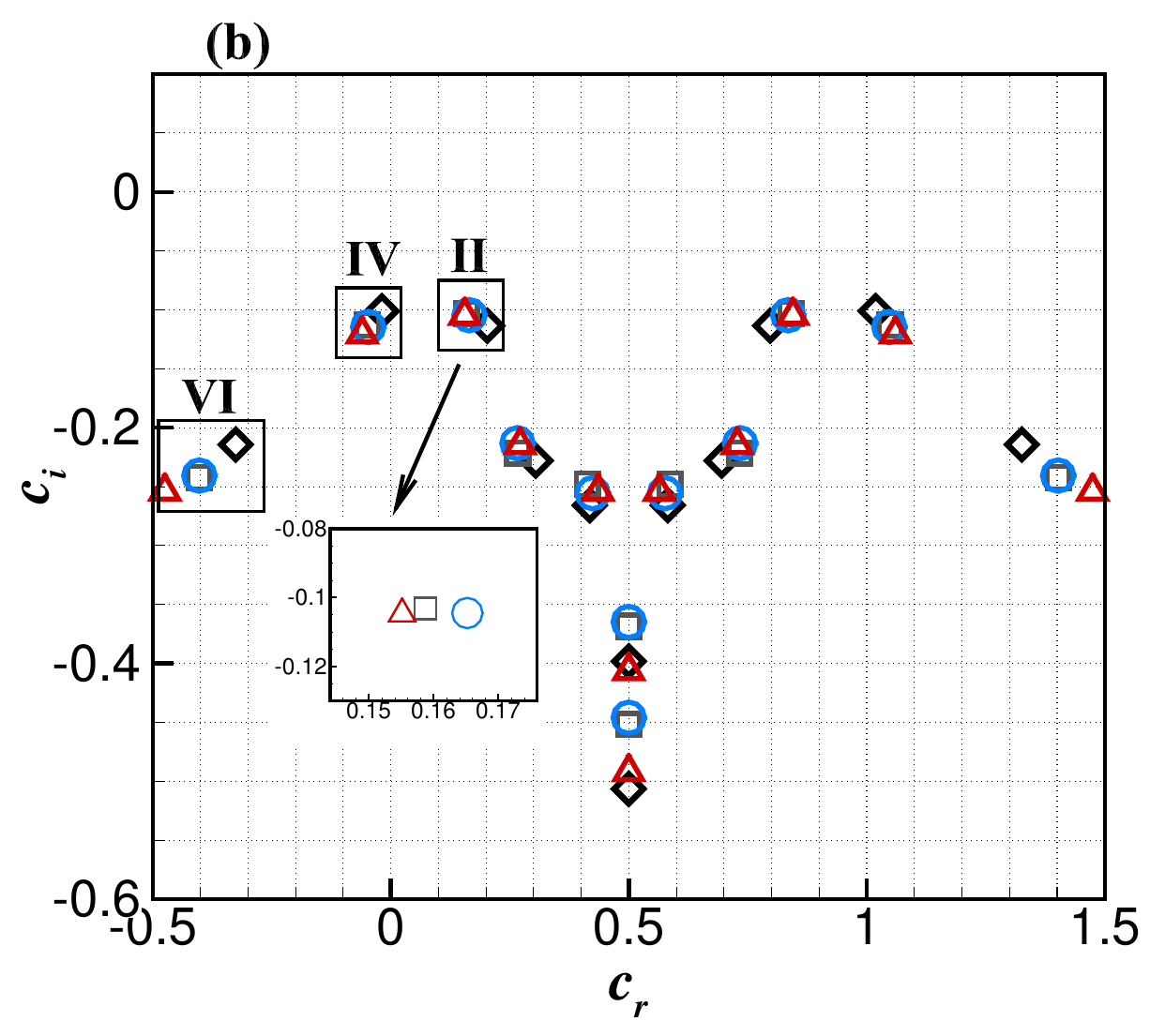}\\[6pt]
  \includegraphics[width=0.46\textwidth]{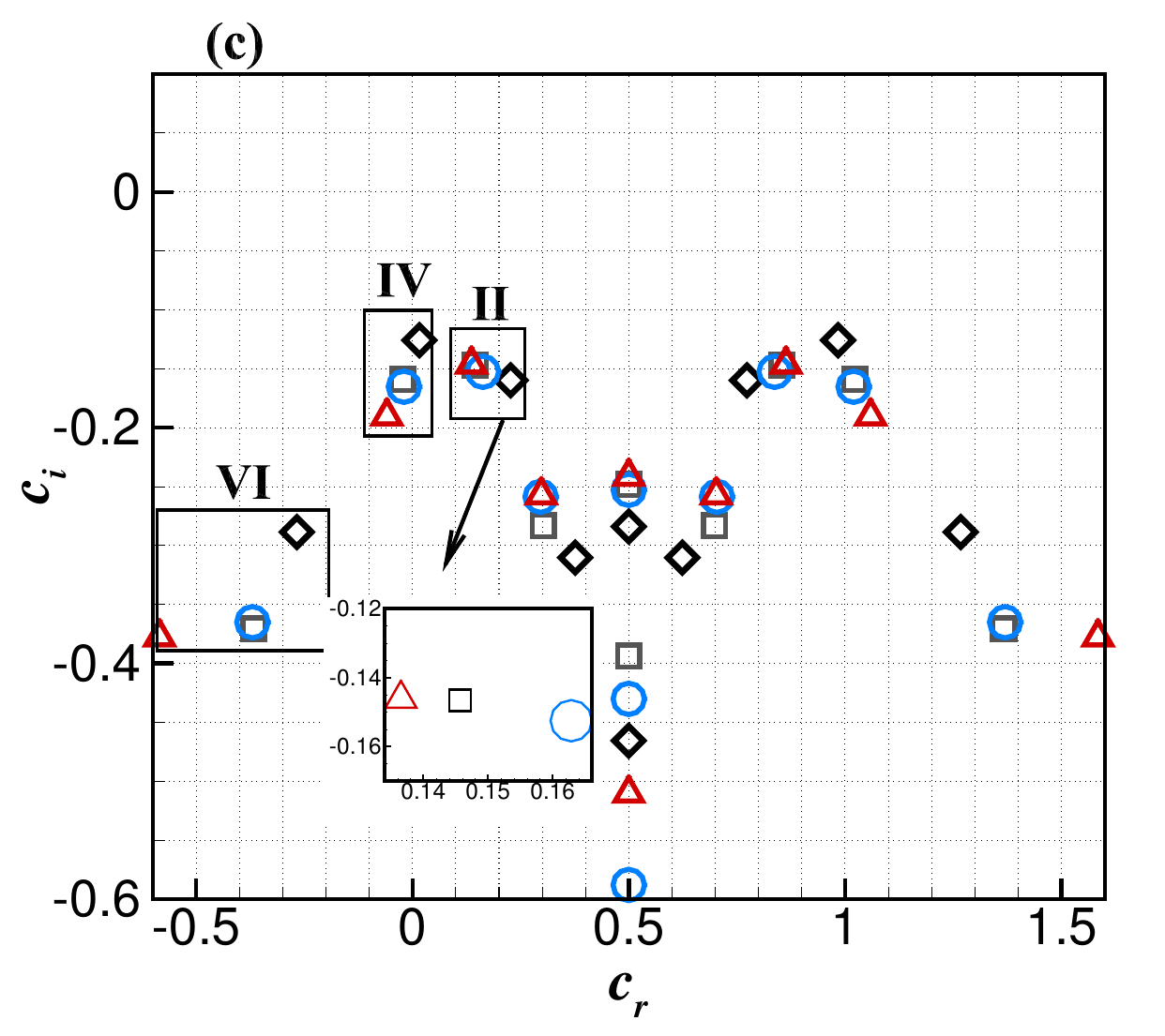}\hspace{0.03\textwidth}
  \includegraphics[width=0.46\textwidth]{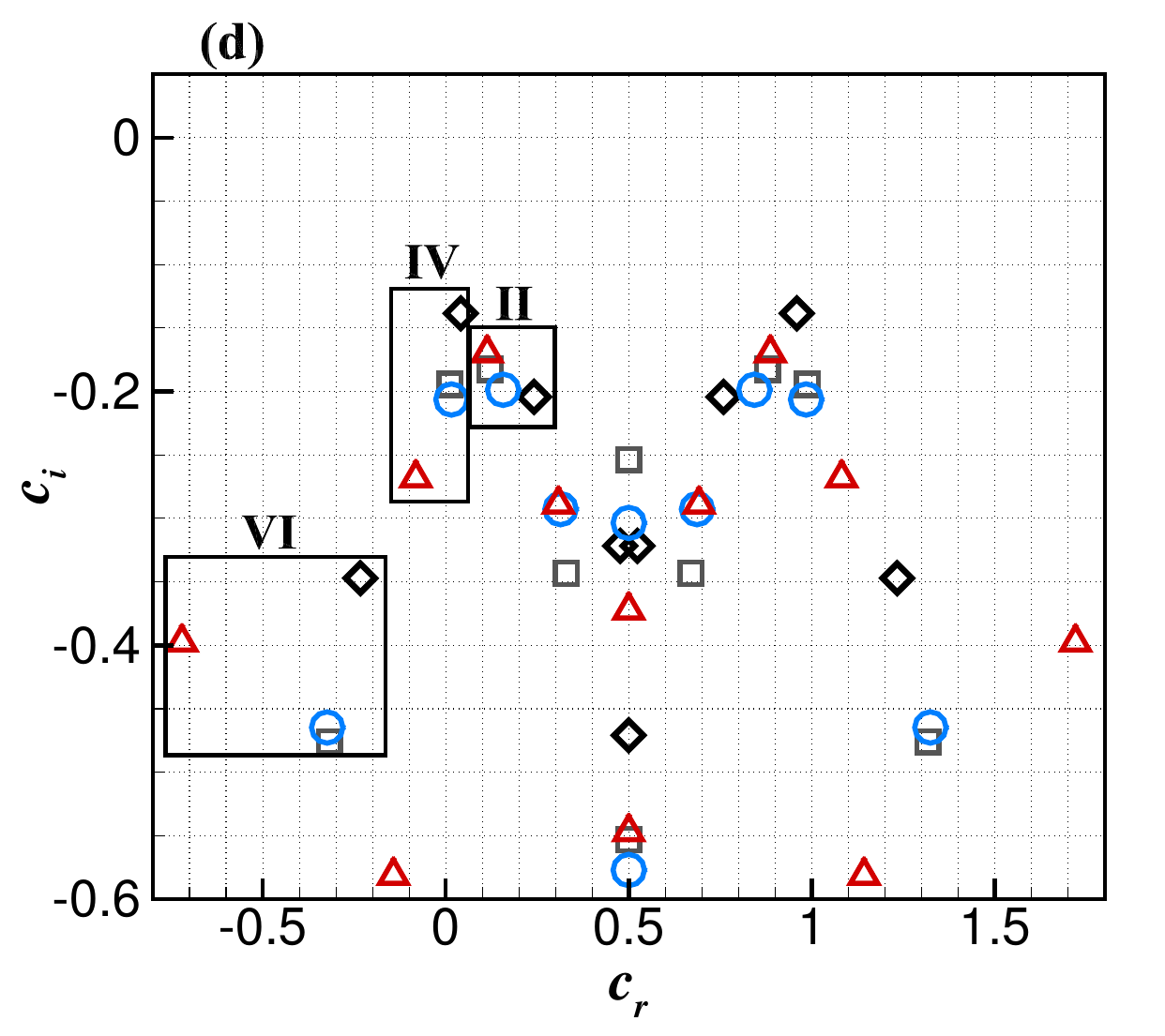}
  \caption{Eigenspectra at $\mathit{Ma}=5$ and $\alpha=2$ for different Knudsen numbers: (a) $\mathit{Kn}=0.005$; (b) $\mathit{Kn}=0.03$; (c) $\mathit{Kn}=0.05$; (d) $\mathit{Kn}=0.07$.}
  \label{fig:10}
\end{figure}

Figure~\ref{fig:10} shows the eigenspectra at $\mathit{Ma}=5$, $\mathit{Kn}=0.005$--0.07 and $\alpha=2$ by NS and GHE methods.
At $\mathit{Kn}=0.005$, the spectra by the NS-noslip, NS-slip, NCCR-slip and GHE-slip agree well with each other, since the rarefaction effects are weak and have little influence on both the base-flow profiles (see Fig.~\ref{fig:7}) and the stability.
As the Knudsen number increases, on the one hand, the number of discrete modes decreases owing to the low-Reynolds-number effect; on the other hand, the rarefaction effects gradually become prominent, and the differences among different models increase.
At $\mathit{Kn}=0.03$, the GHE-slip model predicts a larger phase speed $c_r$ and a smaller growth rate $c_i$ than the NS-noslip model, while the NS-slip and NCCR-slip results differ only slightly from the GHE-slip results. This is because the rarefaction effects in this case are dominated by surface slip effects, and the influence of the nonlinear constitutive relations is not pronounced. To accurately predict the disturbance spectrum, the surface slip effects must be taken into account. 
As $\mathit{Kn}$ further increases to 0.05 and 0.07, the deviations among NS-noslip, NS-slip models and GHE-slip model become increasingly large. Surface slip and nonlinear transport generally increase the phase speeds and decrease the growth rates of the acoustic modes, and such effects are more pronounced for higher modes. This is because higher modes have higher frequencies and shorter wavelengths (i.e., small temporal and spatial characteristic scales), and are therefore subject to more pronounced rarefaction effects.
On the whole, the GHE stability results recover to the NS stability results in the continuum regime, and as the flow becomes rarefied, both the surface slip and nonlinear transport have important effects on the stability of Couette flow. 

\begin{figure}[htbp]
  \centering
  \includegraphics[width=0.46\textwidth]{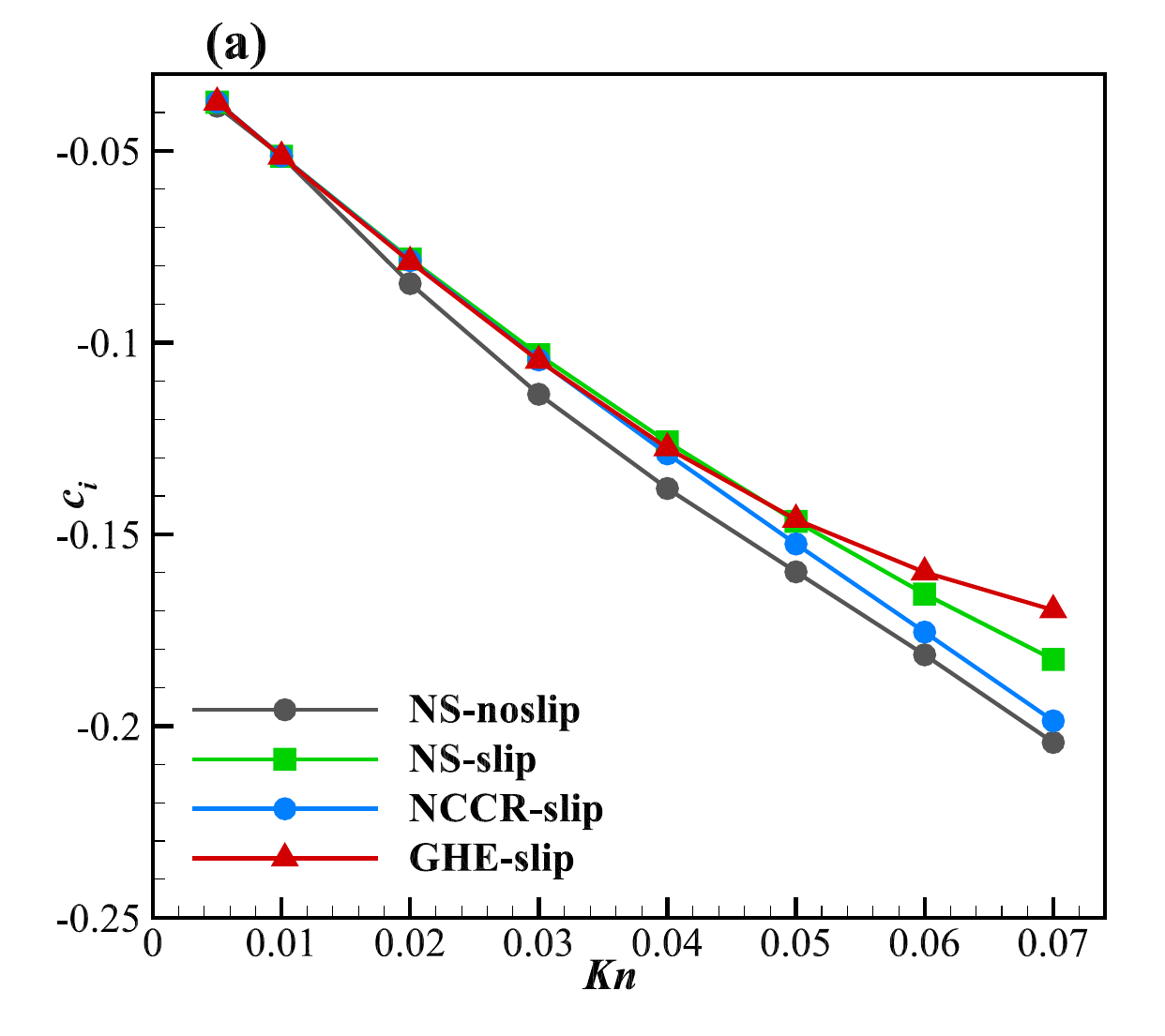}\hspace{0.03\textwidth}
  \includegraphics[width=0.46\textwidth]{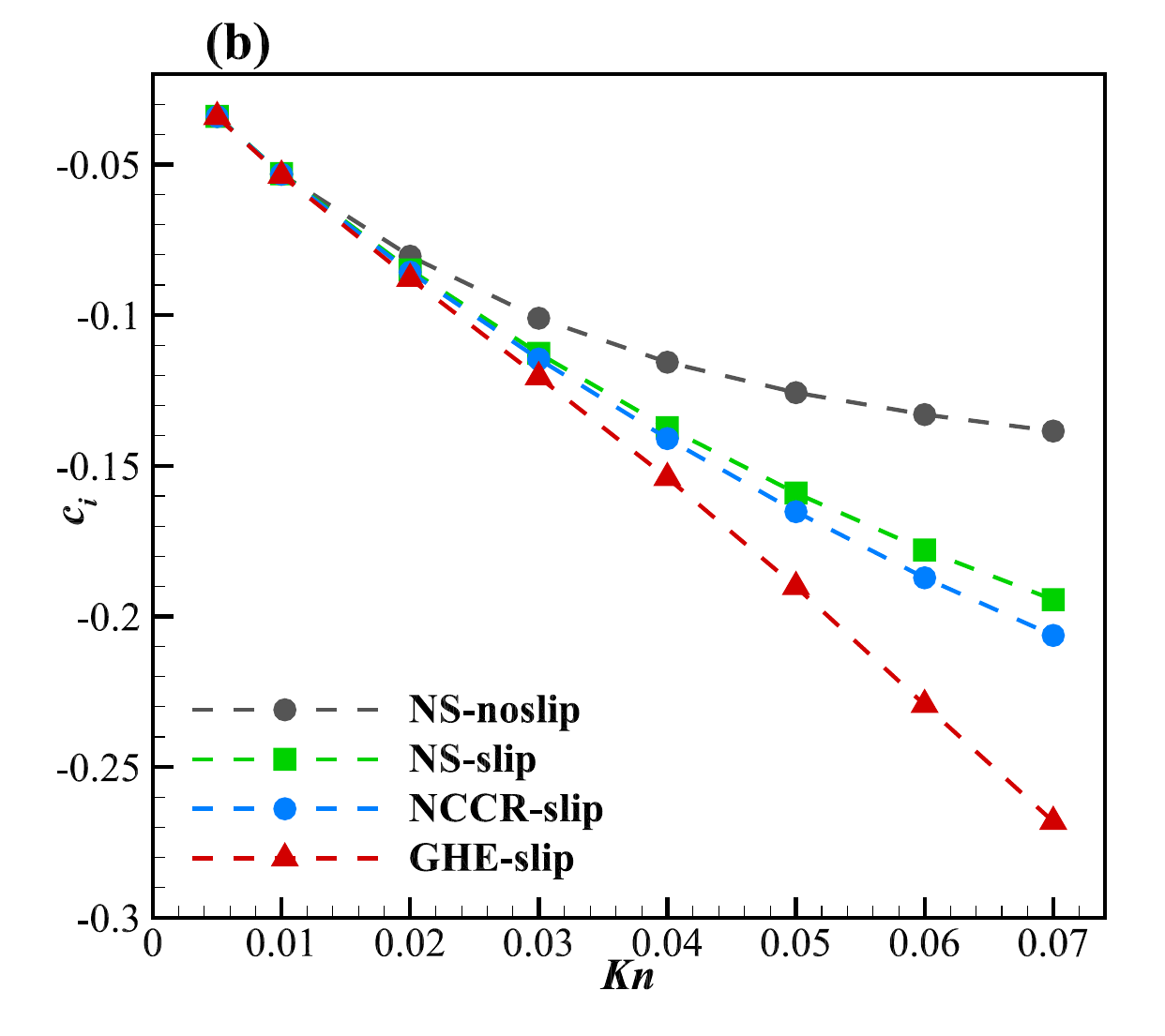}
  \caption{Variations of the imaginary part of the phase speed $c_{i}$ with $\mathit{Kn}$ predicted by different models at $\mathit{Ma}=5$ and $\alpha=2$: (a) Mode II; (b) Mode IV.}
  \label{fig:11}
\end{figure}

Figure~\ref{fig:11} shows the variations of $c_{i}$ of Mode II and Mode IV with $\mathit{Kn}$ at $\mathit{Ma}=5$ and $\alpha=2$, where $\mathit{Kn}$ ranges from 0.005 to 0.07.
For $\mathit{Kn}\le 0.01$, the results of the NS-noslip, NS-slip, NCCR-slip and GHE-slip models agree well with each other.
As $\mathit{Kn}$ increases, the differences among different models gradually increase, and the rarefaction effects have different impacts on the growth rates of different modes.
For Mode II, surface slip and nonlinear transport increase the growth rate (or decrease the decay rate). The NS-slip stability analysis shows good agreements with GHE results up to about $\mathit{Kn}=0.05$.
For Mode IV, the rarefaction effects decrease the growth rate (or increase the decay rate), and when $\mathit{Kn}$ exceeds 0.02, the NS-slip stability analysis shows marked differences from the GHE-slip model, indicating that Mode IV is more sensitive to rarefaction effects. In fact, the higher modes are more sensitive to rarefaction effects. In addition, it is noticed that compared with the GHE-slip results, the NCCR-slip results are closer to the NS-slip stability results, meaning that the material derivative terms of stress and heat flux in GHE can largely influence disturbance evolution under rarefied conditions.

\begin{figure}[htbp]
  \centering
  \includegraphics[width=0.46\textwidth]{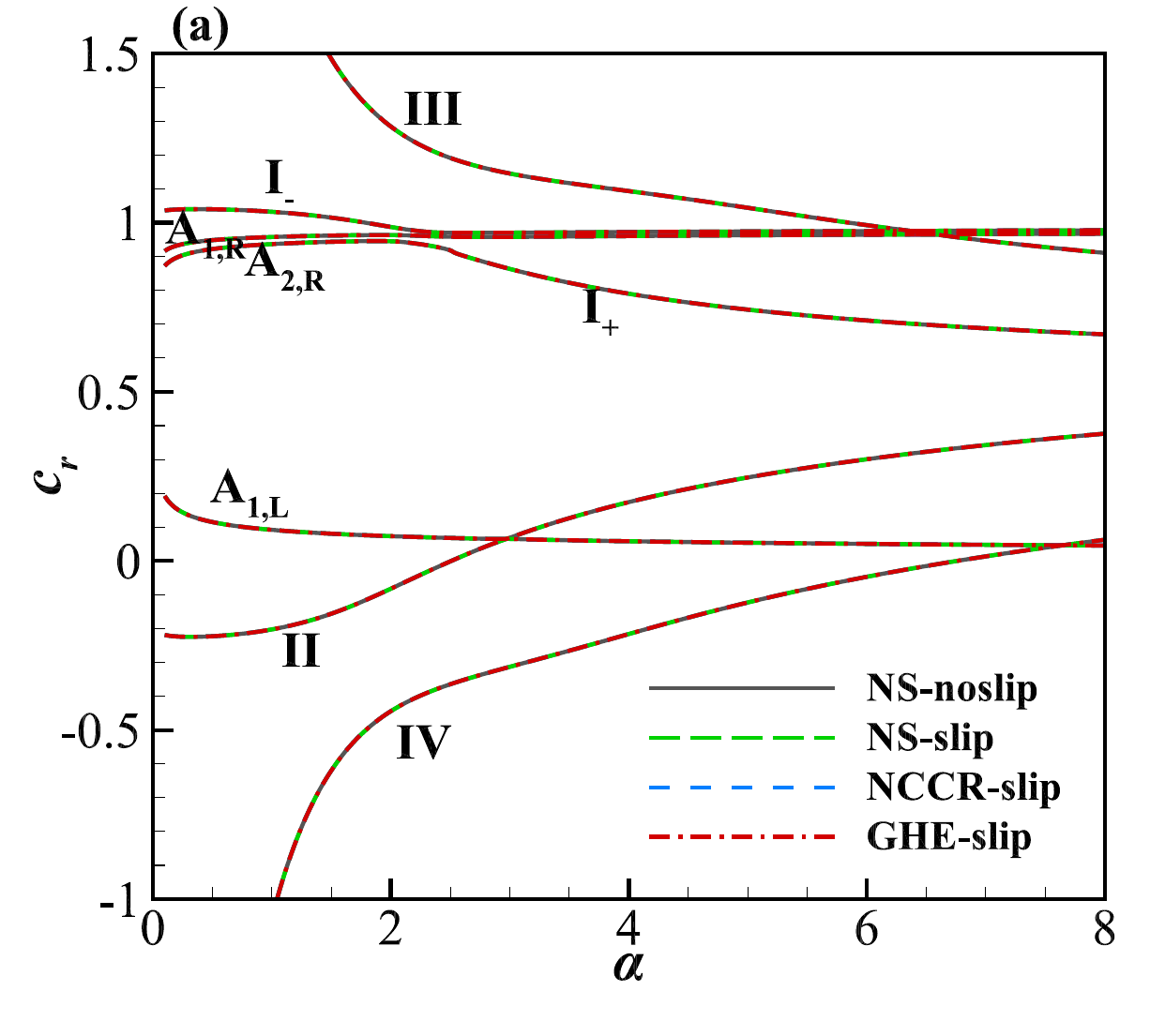}\hspace{0.03\textwidth}
  \includegraphics[width=0.46\textwidth]{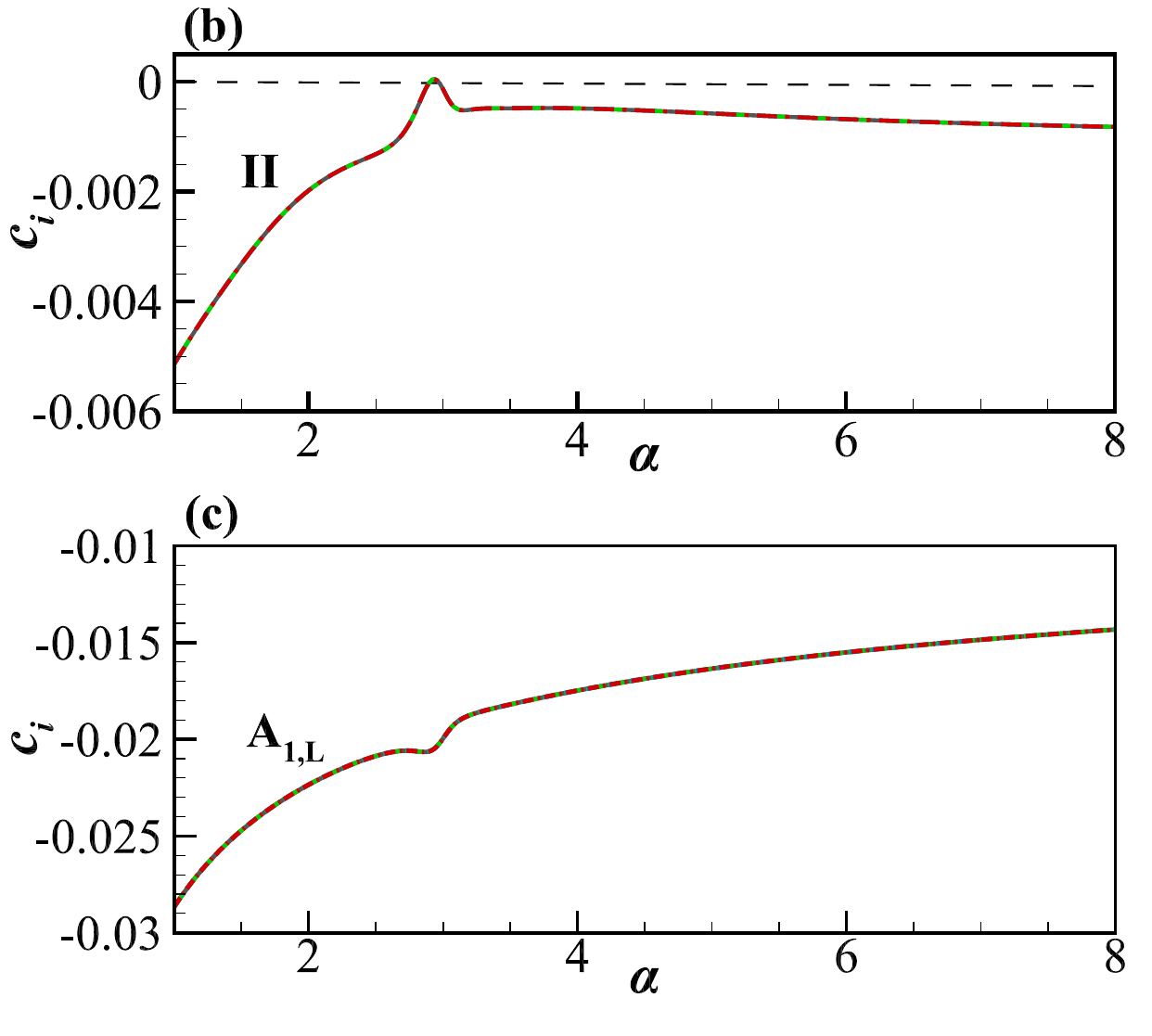}
  \caption{Variations of the real and imaginary parts of the phase speed with $\alpha$ at $\mathit{Ma}=5$ and $\mathit{Re}=3\times10^5$ ($\mathit{Kn}=2.7\times 10^{-5}$): (a) $c_{r}$ of different modes; (b) $c_{i}$ of Mode II; (c) $c_{i}$ of Mode $A_{1,L}$.}
  \label{fig:12}
\end{figure}

\begin{figure}[b]
  \centering
  \includegraphics[width=0.46\textwidth]{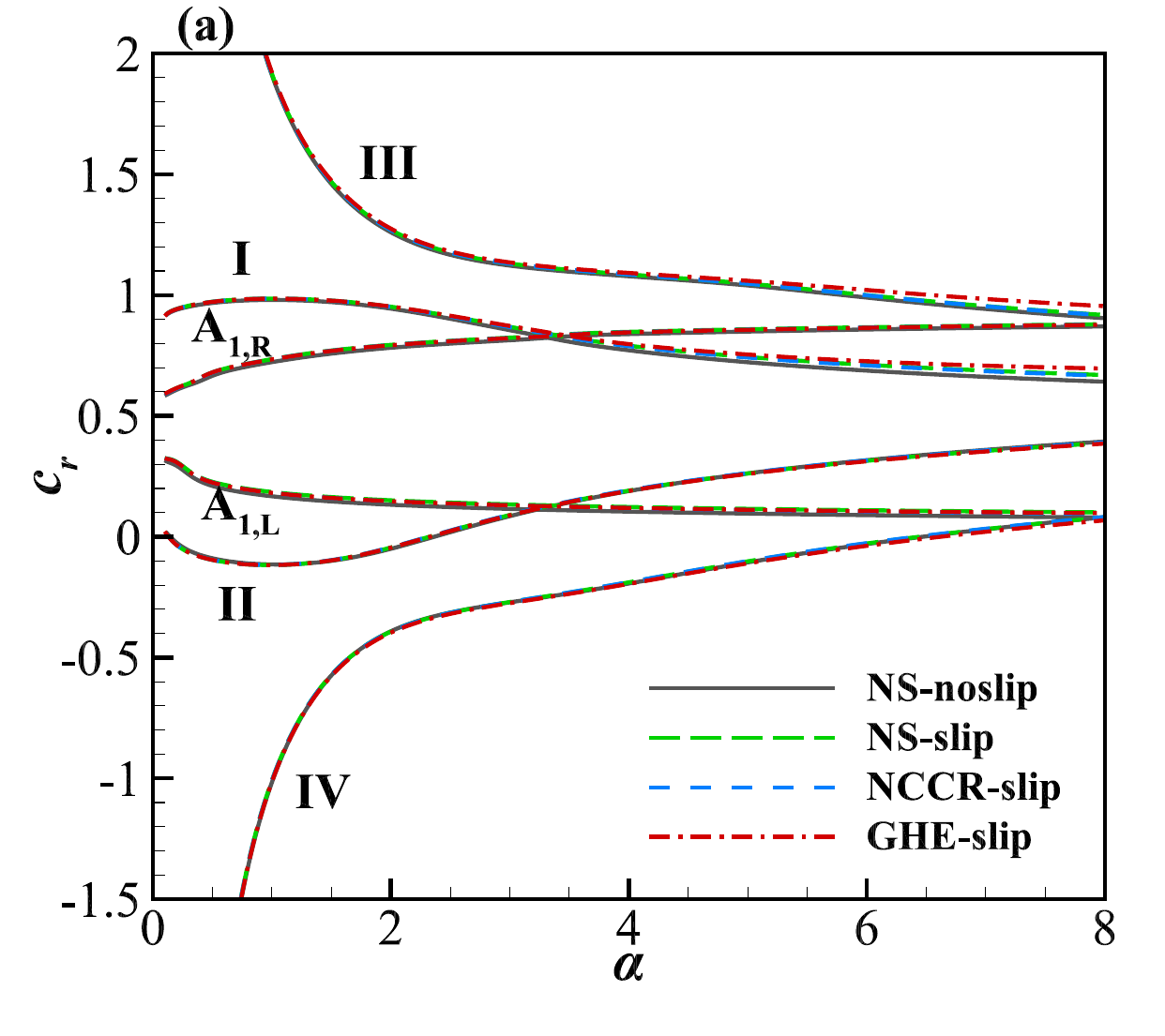}\hspace{0.03\textwidth}
  \includegraphics[width=0.46\textwidth]{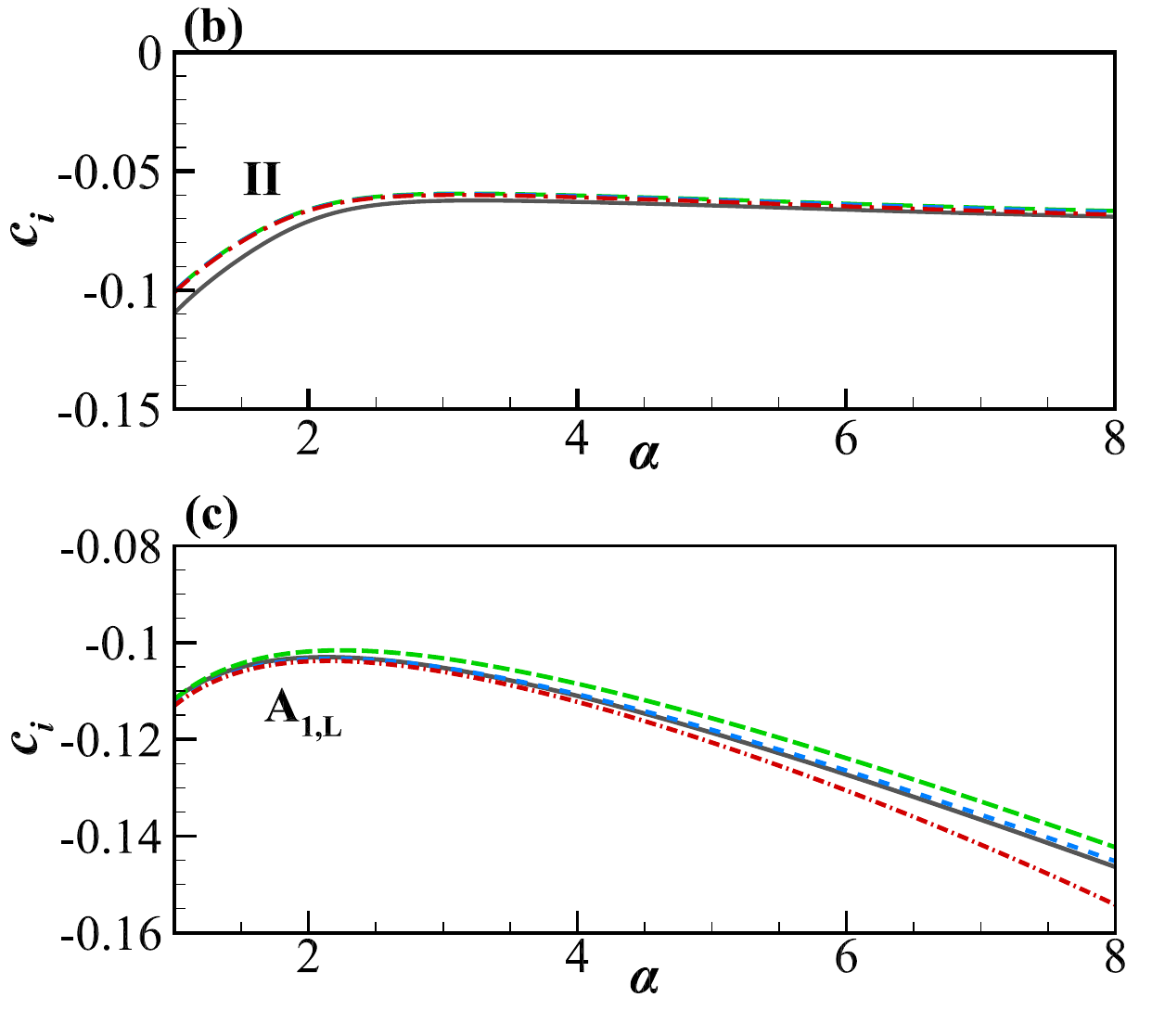}
  \caption{Variations of the real and imaginary parts of the phase speed with $\alpha$ at $\mathit{Ma}=5$ and $\mathit{Kn}=0.01$: (a) $c_{r}$ of different modes; (b) $c_{i}$ of Mode II; (c) $c_{i}$ of Mode $A_{1,L}$.}
  \label{fig:13}
\end{figure}

\begin{figure}[htbp]
  \centering
  \includegraphics[width=0.46\textwidth]{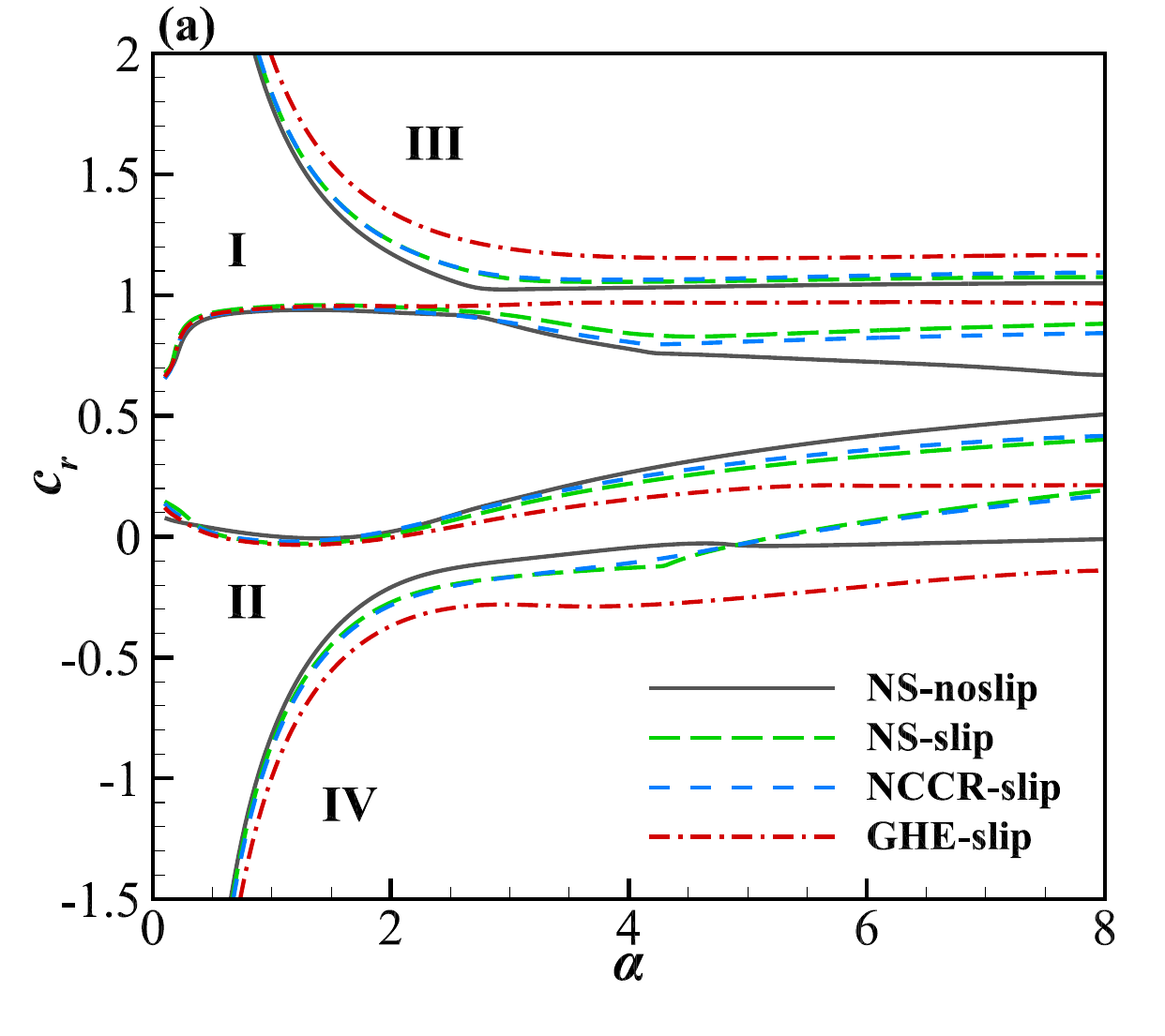}\hspace{0.03\textwidth}
  \includegraphics[width=0.46\textwidth]{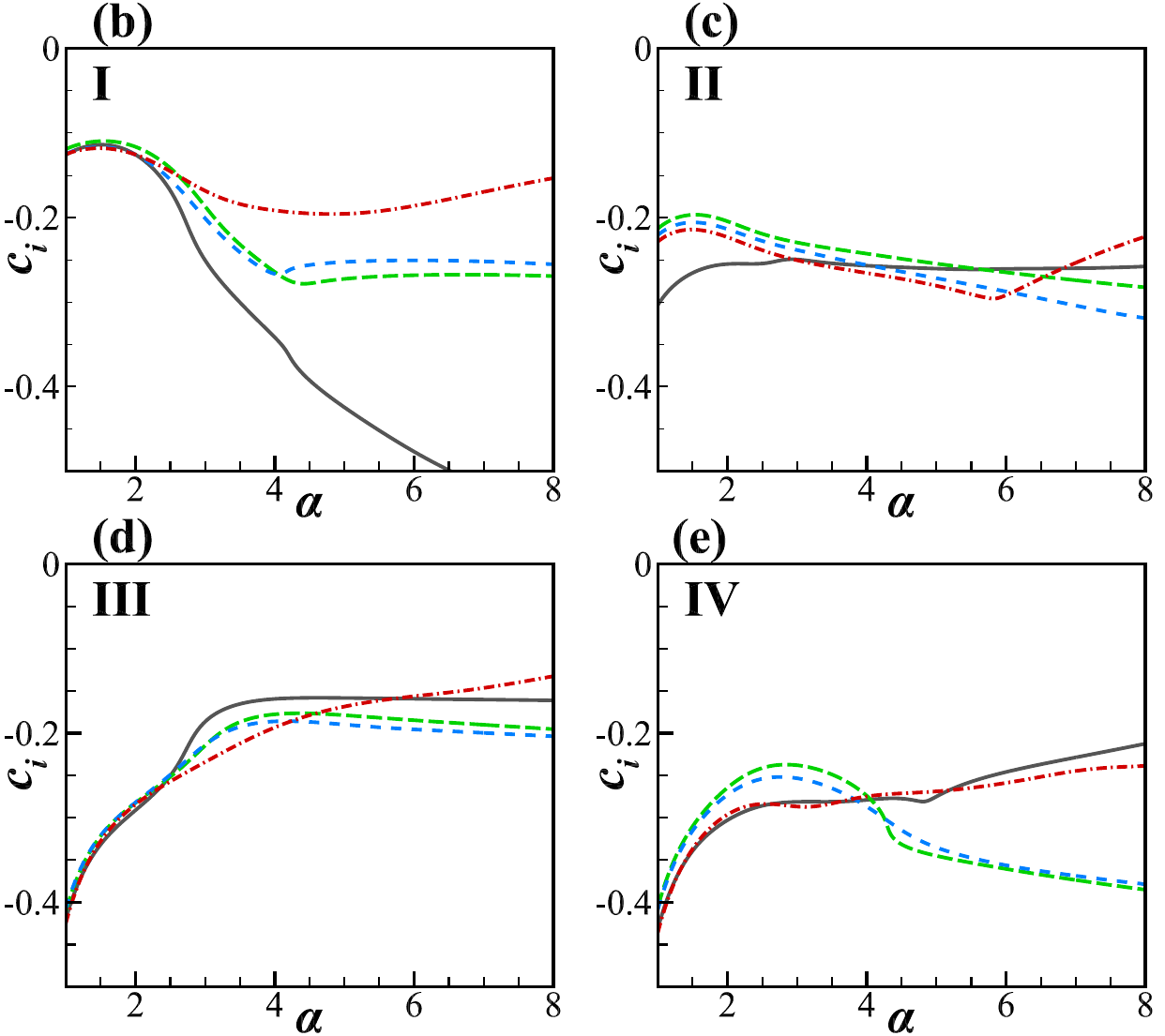}
  \caption{Variations of the real and imaginary parts of the phase speed with $\alpha$ at $\mathit{Ma}=5$ and $\mathit{Kn}=0.05$: (a) $c_{r}$ of different modes; (b--e) $c_{i}$ of Modes I--IV.}
  \label{fig:14}
\end{figure}

Existing study showed that the acoustic modes and viscous modes in supersonic plane Couette flow can be synchronized with each other, leading to instability modes~\cite{ref6}.
Figures~\ref{fig:12}--\ref{fig:14} show the variations of $c_{r}$ and $c_{i}$ of the even modes (II and IV), the odd modes (I and III) and the viscous modes $A_1$ and $A_2$ with the wavenumber at $\mathit{Ma}=5$ and different Knudsen numbers (from continuum to rarefied conditions). For these cases, an adiabatic condition is imposed at the lower wall.
First see the Fig.~\ref{fig:12}, under the continuum condition ($\mathit{Kn}=2.7\times 10^{-5}$), the GHE and NCCR models fully recover to the NS results.
As the wavenumber tends to zero, the phase speeds of the even modes (II and IV) are less than 0, while those of the odd modes (I and III) are greater than 1. The phase speeds of viscous modes ($A_1$ and $A_2$) lie between 0 and 1.
With the wavenumber increasing, the phase speeds of the even modes increase while those of the odd modes decrease. When their phase speeds approach 0 or 1, they are synchronized with the viscous branches $A_{L}$ or $A_{R}$, leading to unstable acoustic Mode II.
As shown in Figs.~\ref{fig:12} (b) and (c), $c_{i}$ of Mode II and the viscous mode $A_{1,L}$ exhibit a peak and a trough, respectively, near $\alpha=3$, and the local peak of $c_{i}$ for Mode II is positive, indicating that it is an unstable mode. The Mode I always remains stable, so its $c_{i}$ curve is not shown here.
Note that the synchronization here is different from the synchronization between the fast and slow modes in high-speed boundary layers.
Ou~\cite{ref6} pointed out that an even (or odd) mode can be synchronized with multiple viscous modes, leading to complex branching topology and viscous instability of acoustic modes.
For this Mach 5 case, the even Mode II is only synchronized with the viscous mode $A_{1,L}$, forming a `crossing over' topology, whereas the odd Mode I is synchronized with both $A_{1,R}$ and $A_{2,R}$, forming an `elastic reflecting' topology.
This is because an adiabatic lower wall breaks the symmetry of the base flow, so that the synchronizations of the even and odd modes differ much. 
In fact, the branching topology becomes simpler as Mach number decreases while more complex at high Mach numbers. 

\begin{figure}[htbp]
  \centering
  \includegraphics[width=0.46\textwidth]{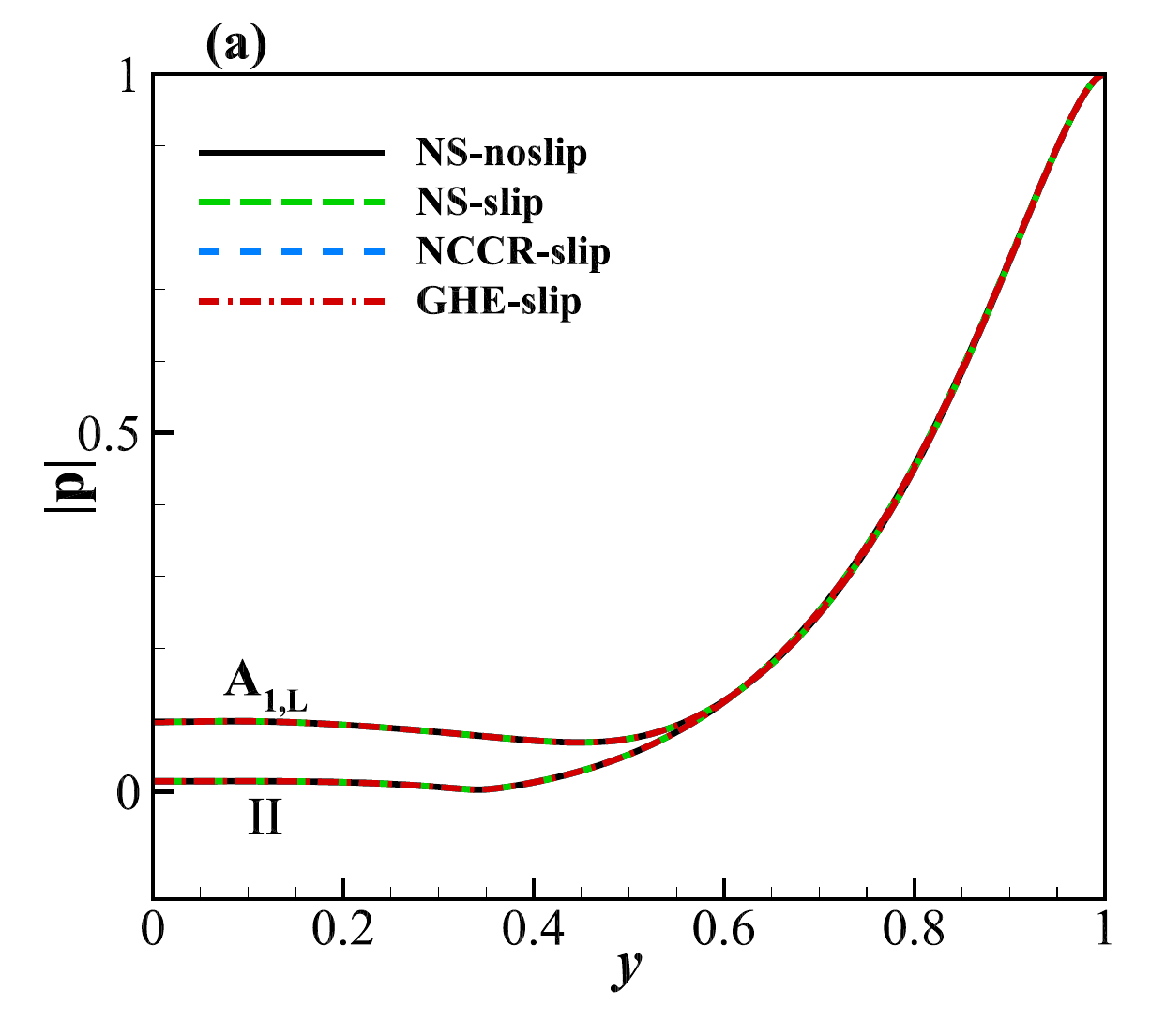}\hspace{0.03\textwidth}
  \includegraphics[width=0.46\textwidth]{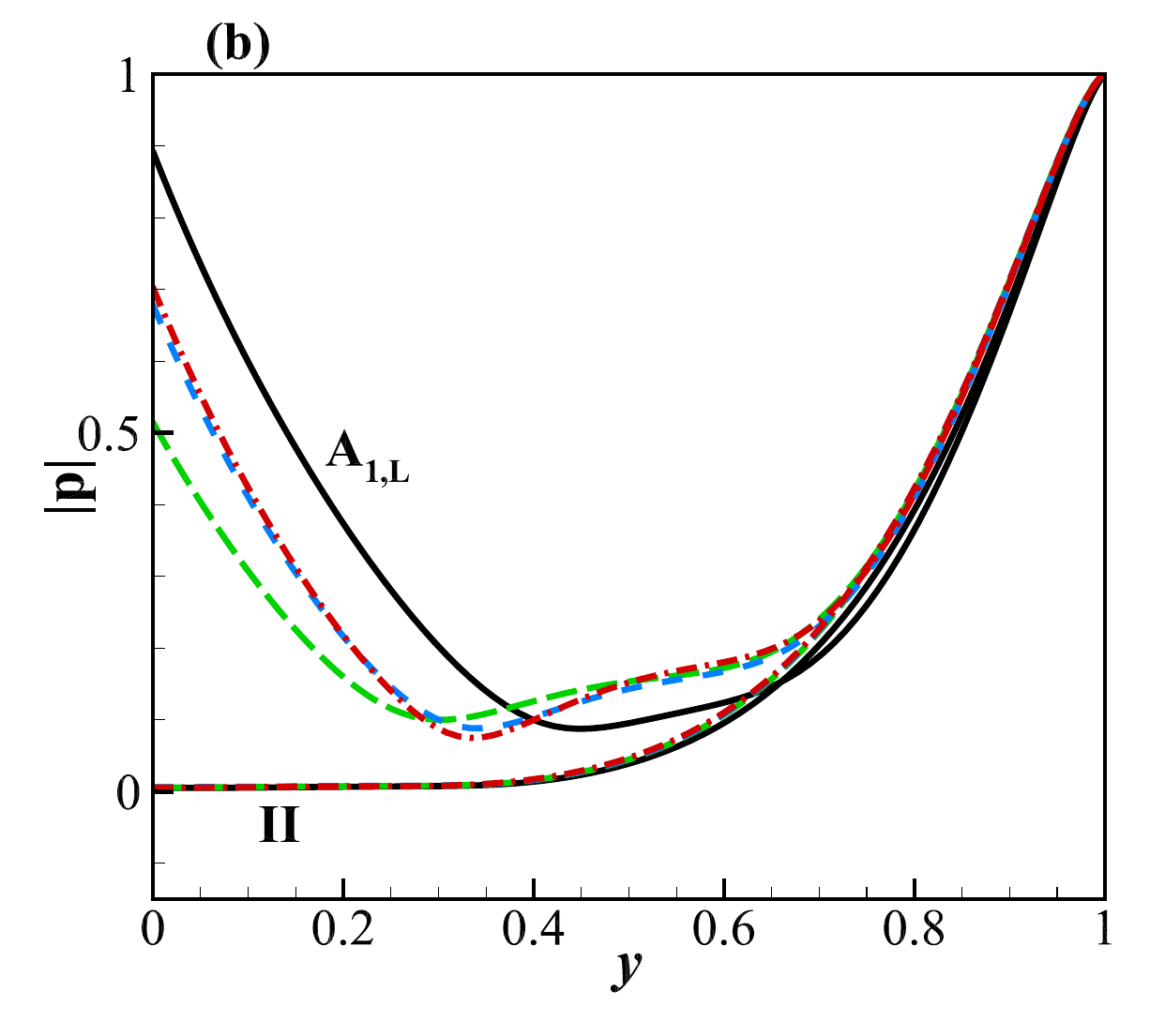}
  \caption{Eigenfunctions of pressure disturbances for Mode II and Mode $A_{1,L}$: (a) $\mathit{Ma}=5$, $\mathit{Kn}=2.7\times 10^{-5}$ and $\alpha=3$; (b) $\mathit{Ma}=5$, $\mathit{Kn}=0.01$ and $\alpha=3.3$.}
  \label{fig:15}
\end{figure}

Figures~\ref{fig:13} and \ref{fig:14} show the variations of $c_{r}$ and $c_{i}$ of different acoustic and viscous modes with the wavenumber at $\mathit{Ma}=5$, $\mathit{Kn}=0.01$ and 0.05, to investigate the rarefaction effects on mode synchronization.
At $\mathit{Kn}=0.01$, the phase speeds $c_{r}$ predicted by the NS-noslip, NS-slip, NCCR-slip and GHE-slip models differ only slightly with each other, and some differences appear in $c_{i}$. Compared with the NS-noslip results, the GHE-slip model gives a larger $c_{i}$ for Mode II and a smaller $c_{i}$ for Mode $A_{1,L}$.
Compared with the continuum case at $\mathit{Kn}=2.7\times 10^{-5}$, the phase-speed curves of the acoustic Modes I and II and the viscous mode $A_1$ are markedly different for this $\mathit{Kn}=0.01$ case.
As shown in Fig.~\ref{fig:13}(a), although the phase speeds of Mode II and the viscous branch $A_{L}$ still intersect with each other, with the intersection point at $\alpha\approx 3.3$, slightly higher than that in the $\mathit{Kn}=2.7\times 10^{-5}$ case, the $c_{i}$ distributions of Mode II and the viscous mode $A_{1,L}$ show no sharp peak or trough, and both modes are damped over the entire wavenumber ranges, indicating that there exists no synchronization between these two modes. This can be seen more clearly from the eigenfunctions in Fig.~\ref{fig:15}.
For Mode I, there is no synchronization between mode I and multiple viscous modes in branch $A_R$, and the phase-speed distribution becomes simple, which is much different from Fig.~\ref{fig:12}.
When $\mathit{Kn}$ increases to 0.05, the differences in $c_{r}$ and $c_{i}$ predicted by the NS, NCCR and GHE models become more pronounced.
Since the acoustic modes are not synchronized with the viscous modes, only the stability results of the acoustic Modes I, II, III and IV are shown.
The rarefaction effects increase the disturbance phase speed at a given wavenumber, while they have different effects on the growth rate for different modes. For Modes I, II, III and IV, the rarefaction effects increase the growth rates for $\alpha < 3$, while at high wavenumbers (e.g., $\alpha > 4$), the rarefaction effects increase the growth rate of Mode I but decrease those of Modes II, III and IV. 
In addition, some small kinks are observed in the $c_{r}$ and $c_{i}$ curves in Fig.~\ref{fig:14}, all located at which the phase speeds of the odd and even modes (I, II, III and IV) approach or equal those of the viscous modes. In fact, there exists weak interaction between these modes at these locations, but it is too weak to excite an unstable mode.

Figure~\ref{fig:15} shows the eigenfunctions of Mode II and Mode $A_{1,L}$ computed by the NS, NCCR and GHE stability equations for the $\mathit{Kn}=2.7\times 10^{-5}$ and $\mathit{Kn}=0.01$ cases, where the wavenumbers are chosen at the locations where the phase speeds of the two modes are equal, i.e., $\alpha=3$ and $\alpha=3.3$, respectively.
For the continuum case ($\mathit{Kn}=2.7\times 10^{-5}$), the disturbance eigenfunctions computed by the NCCR and GHE models recover to the NS results, and the eigenfunctions of Mode II and Mode $A_{1,L}$ are nearly identical, with some differences near the lower wall, which indicates that the two modes are synchronized.
As $\mathit{Kn}$ increases to 0.01, the eigenfunctions of Mode II predicted by the different stability equations are almost identical, whereas those of the viscous mode $A_{1,L}$ differ noticeably, indicating that the rarefaction effects have little influence on the eigenfunctions of the acoustic mode but a relatively pronounced influence on those of the viscous mode.
At $\alpha=3.3$, the eigenfunctions of Mode II and Mode $A_{1,L}$ differ considerably, showing that these two modes are hardly synchronized.

\begin{figure}[t]
  \centering
  \includegraphics[width=0.46\textwidth]{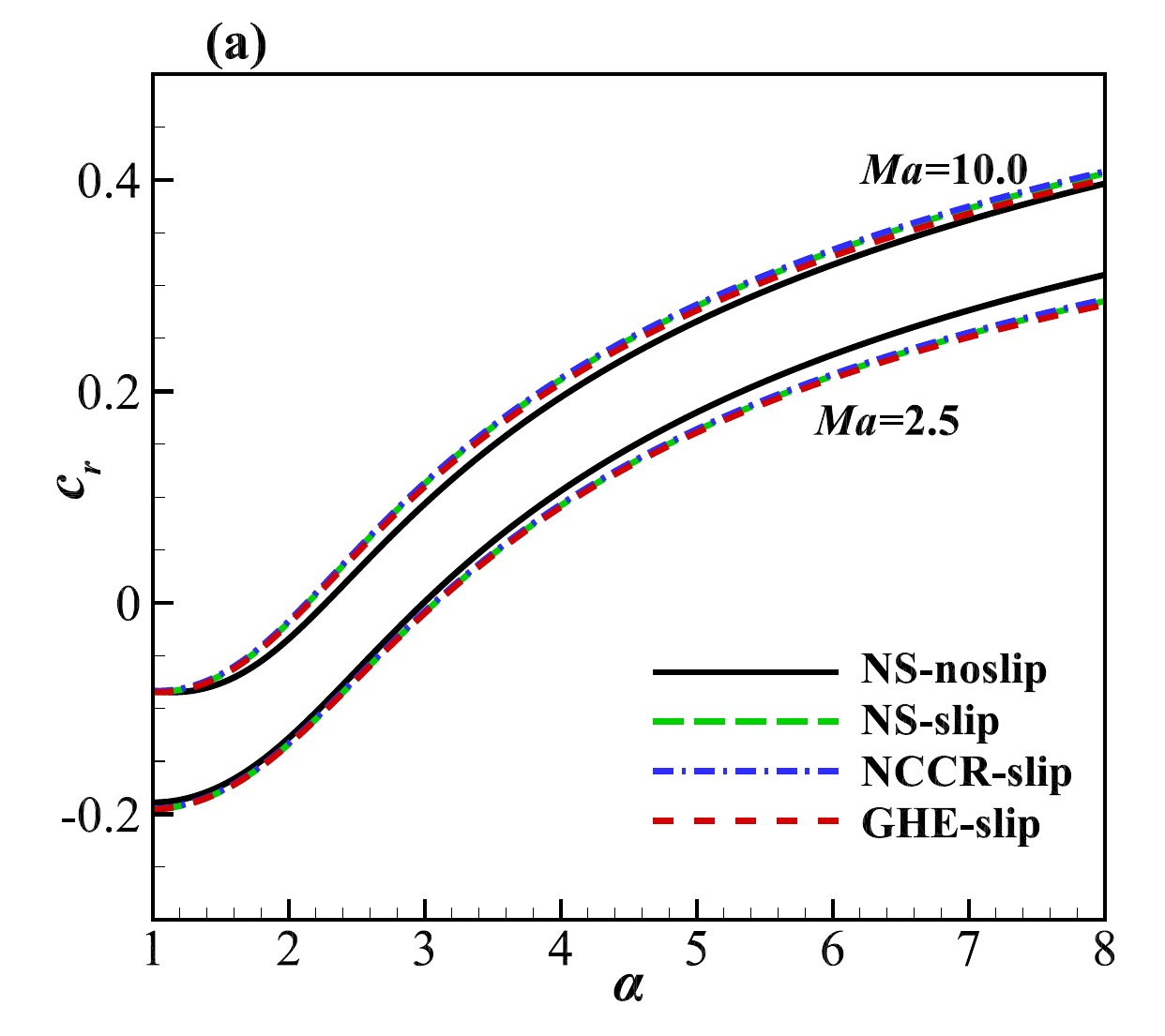}\hspace{0.03\textwidth}
  \includegraphics[width=0.46\textwidth]{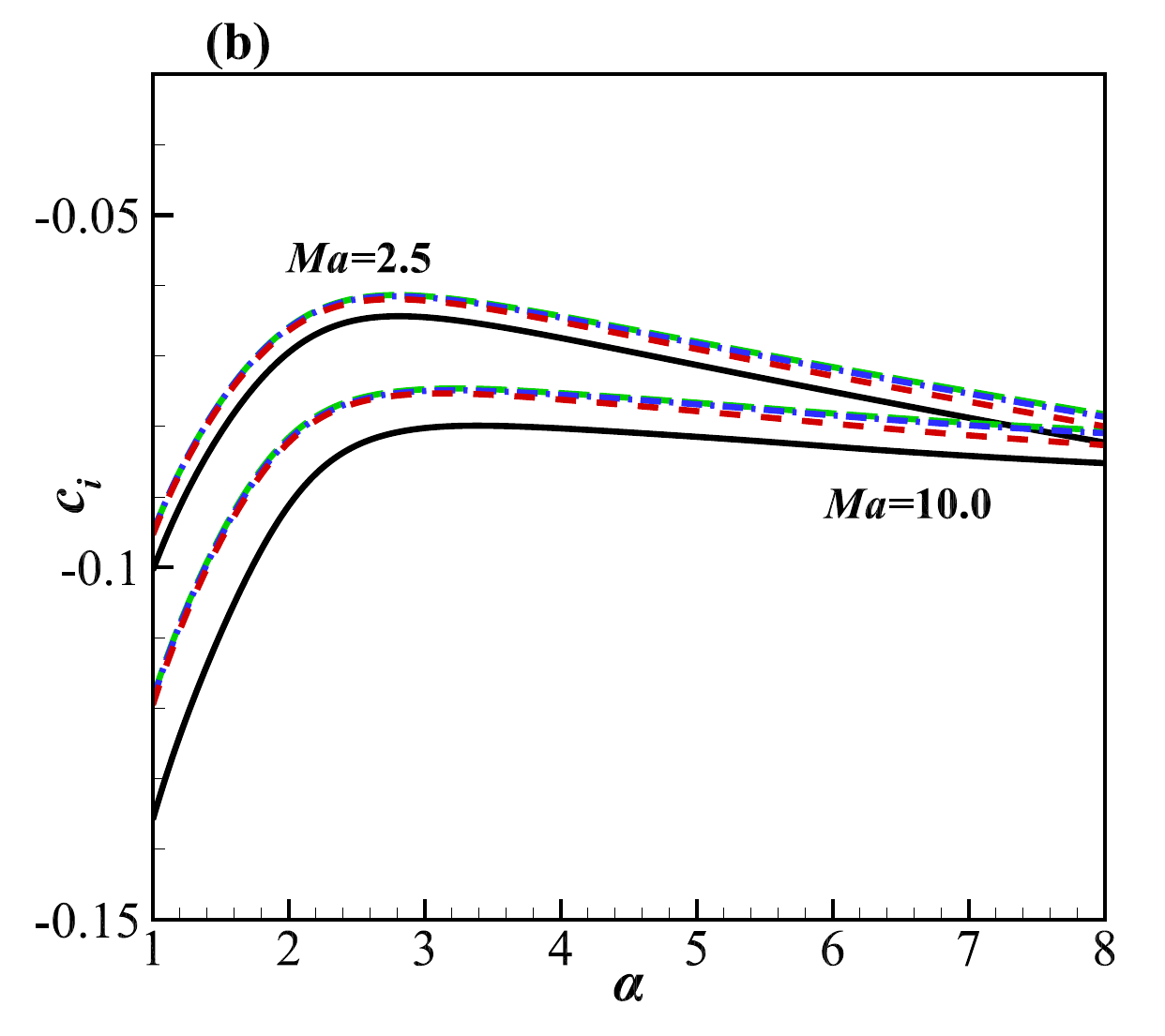}
  \caption{Variations of (a) the real part $c_{r}$ and (b) the imaginary part $c_{i}$ of the phase speed of Mode II with $\alpha$ at $\mathit{Kn}=0.01$ for $\mathit{Ma}=2.5$ and 10.}
  \label{fig:16}
\end{figure}

Finally, Fig.~\ref{fig:16} shows the variations of $c_{r}$ and $c_{i}$ of Mode II with $\alpha$ at $\mathit{Kn}=0.01$ for $\mathit{Ma}=2.5$ and 10, to examine the effects of Mach number.
At both Mach numbers, the slip models predict larger $c_{i}$ than the NS-noslip model, and such difference becomes larger at higher Mach numbers. Compared with the NS-slip model, the NCCR-slip result is close to NS-slip result, and the GHE-slip model gives smaller $c_{i}$. The differences between the GHE-slip and NCCR-slip results become larger with the wavenumber increasing.
Since the GHE and NCCR models share the same base flow, their differences arise from the time-derivative and convective terms of the stress and heat-flux in stability equations, which are retained in the GHE but neglected in the NCCR model. The effects of these terms become important at high wavenumbers.

\section{Conclusions}\label{sec:6}

In this paper, the linear stability of high-speed plane Couette flow in near-continuum regime is addressed. A linear stability theory based on the generalized hydrodynamic equations with slip boundary conditions is developed within the 13-moment framework. Meanwhile, the linear stability equations of the NCCR and NS equations are also given within the moment framework, and the corresponding numerical methods are formulated. Based on the developed methods, the base flows and linear stability of high-speed plane Couette flows at $\mathit{Ma}=2.5$--10 and $\mathit{Kn}=2.7\times 10^{-5}$--0.07 are analyzed, with main focus on the nonlinear transport, eigenspectrum, mode synchronization and least stable mode. The main findings are as follows:

(1) In near-continuum plane Couette flows, the surface slip effects can be characterized by $K_{s}\sim\left( 2-\sigma \right)/\sigma\cdot\mathit{Kn}$, and the rarefied nonequilibrium effects are characterized by $\kappa\sim \mathit{Ma}\cdot\mathit{Kn}$.
As the Mach and Knudsen numbers increase, the base-flow profiles predicted by the NS equations with no-slip conditions deviate increasingly from the DSMC results.
The slip boundary models can improve the no-slip solutions, but cannot accurately capture the nonlinear transport, such as the nonzero normal stress and tangential heat flux. The GHE with consistent slip boundary conditions can well describe the nonlinear stress and heat flux, as well as the velocity and temperature profiles, and show good agreements with DSMC.

(2) The GHE-based LST recovers to the conventional NS-based LST in the continuum limit, and captures surface slip and shear nonequilibrium effects in rarefied flows. For all test cases, no new unstable modes are found in the present GHE and NCCR stability analyses compared with the NS stability results. However, the NS-LST and GHE-LST do exhibit large differences as Mach and Knudsen numbers increase. The conventional NS-noslip theory deviates noticeably from the GHE-slip model at $\mathit{Kn}=O(0.01)$. When $\mathit{Kn}$ reaches 0.05, the effects of nonlinear transport on the stability become important and can no longer be neglected. There exist larger deviations between GHE and NCCR under rarefied conditions, which is owing to the time-derivative and convective terms of the stress and heat-flux in the GHE stability equations.

(3) The acoustic modes are synchronized with the viscous branch in the continuum regime, leading to viscous instabilities of acoustic modes and multimode branching topology.
As the gas becomes rarefied (e.g., $\mathit{Kn}\sim O(0.01)$), such synchronization becomes weak, and all modes are stable.
The surface slip effects and nonlinear transport increase the disturbance phase speeds, while have different effects on the growth rates for different modes. For instance, rarefaction effects increase the growth rate of Mode II and decrease that of Mode IV. Moreover, the rarefaction effects are more pronounced for higher modes and high-wavenumber disturbances, for which the continuum assumption may break down at lower Knudsen numbers.

We should state that the present study provides a novel linear stability method for near-continuum shear flows based on the generalized hydrodynamic equations, and quantifies the effects of surface slip and nonlinear transport on the Couette-flow stability. However, no unstable mode exists in the present cases at $\mathit{Kn}\sim O(0.01)$. The extension of present work to hypersonic near-continuum boundary layer, which admits unstable second mode, is our ongoing work.

\appendix*
\allowdisplaybreaks[1]
\setlength{\jot}{5pt}%
\section{Coefficient matrices in GHE, NCCR and NS-based stability equations}
In this paper, the linear stability equations of GHE, NCCR and NS are given within the 13-moment framework, written in a unified form as (also see Eq.~\eqref{eq:54})
\begin{equation}
    {\bf{A}}\frac{{d\hat \phi }}{{dy}} + {\bf{B}}\hat \phi  = \omega \,{\bf{C}}\hat \phi
\end{equation}
Two-dimensional stability equations are formulated, and nine disturbance variables are chosen as unknowns, i.e., $\hat \phi  = \left[ {\hat u,\;\hat v,\;\hat T,\;\hat p,\;{{\hat \Pi }_{xx}},\;{{\hat \Pi }_{yy}},\;{{\hat \Pi }_{xy}},\;{{\hat Q}_x},\;{{\hat Q}_y}} \right]$. 
Noted that the first four rows of $\mathbf{A}$, $\mathbf{B}$ and $\mathbf{C}$ come from the conservation equations, and they are same for GHE, NCCR and NS-based stability equations. The fifth to ninth rows of $\mathbf{A}$, $\mathbf{B}$ and $\mathbf{C}$ come from the constitutive relations, which are different for different models. In the following, the first to ninth rows of $\mathbf{A}$, $\mathbf{B}$ and $\mathbf{C}$ in GHE-LST are given, and for NCCR-LST and NS-LST, only the fifth to ninth rows are given. The nonzero elements in $\mathbf{A}$, $\mathbf{B}$ and $\mathbf{C}$ are as follows.

\noindent\begin{minipage}{\linewidth}
    \centering
    (1) GHE-LST
\end{minipage}
\begin{align*}
A_{1,7}&=\frac1{\mathit{Re}},\\
A_{2,4}&=1,\qquad A_{2,6}=\frac1{\mathit{Re}},\\
A_{3,1}&=\frac{(\gamma-1)\mathit{Ma}^2}{\mathit{Re}}\Pi_{xy0},\qquad A_{3,2}=\frac{(\gamma-1)\mathit{Ma}^2}{\mathit{Re}}\Pi_{yy0},\qquad A_{3,9}=\frac1{\mathit{Pr}\,\mathit{Re}},\\
A_{4,2}&=1,\\
A_{5,1}&=\frac43\Pi_{xy0},\qquad A_{5,2}=\Pi_{xx0}-\frac23(\mathit{Re}\,p_0+\Pi_{yy0}),\\
A_{6,1}&=-\frac23\Pi_{xy0},\qquad A_{6,2}=\frac43\mathit{Re}\,p_0+\frac73\Pi_{yy0},\\
A_{7,1}&=\mathit{Re}\,p_0+\Pi_{yy0},\qquad A_{7,2}=2\Pi_{xy0},\\
A_{8,1}&=Q_{y0},\qquad A_{8,2}=Q_{x0},\qquad A_{8,3}=\mathit{Pr}\,\Pi_{xy0},\\
A_{9,2}&=2Q_{y0},\qquad A_{9,3}=\mathit{Pr}(\mathit{Re}\,p_0+\Pi_{yy0}).\\[8pt]
B_{1,1}&=i\alpha\rho_0u_0,\qquad B_{1,2}=\rho_0u_0',\qquad B_{1,4}=i\alpha,\qquad B_{1,5}=\frac{i\alpha}{\mathit{Re}},\\
B_{2,2}&=i\alpha\rho_0u_0,\qquad B_{2,7}=\frac{i\alpha}{\mathit{Re}}.\\[8pt]
B_{3,1}&=\frac{(\gamma-1)\mathit{Ma}^2}{\mathit{Re}}i\alpha\Pi_{xx0},\qquad B_{3,2}=\rho_0T_0'-(\gamma-1)\mathit{Ma}^2p_0' +\frac{(\gamma-1)\mathit{Ma}^2}{\mathit{Re}}i\alpha\Pi_{xy0},\\
B_{3,3}&=i\alpha\rho_0u_0,\qquad B_{3,4}=-(\gamma-1)\mathit{Ma}^2i\alpha u_0,\qquad B_{3,7}=\frac{(\gamma-1)\mathit{Ma}^2}{\mathit{Re}}u_0',\qquad B_{3,8}=\frac{i\alpha}{\mathit{Pr}\,\mathit{Re}}.\\[8pt]
B_{4,1}&=i\alpha,\qquad B_{4,2}=\frac{\rho_0'}{\rho_0},\qquad B_{4,3}=-\frac{i\alpha u_0}{T_0},\qquad B_{4,4}=\frac{\gamma \mathit{Ma}^2i\alpha u_0}{\rho_0T_0}.\\[8pt]
B_{5,1}&=i\alpha\left(\frac43\mathit{Re}\,p_0+\frac73\Pi_{xx0}\right),\qquad B_{5,2}=\Pi_{xx0}'-\frac23i\alpha\Pi_{xy0},\\
B_{5,3}&=-\frac{\mathit{Re}\,p_0q(\kappa_0)\Pi_{xx0}}{\mu_0^2}\left.\frac{d\mu}{dT}\right|_0 -\Lambda_2\Pi_{xx0}\frac{\varepsilon(Q_{x0}^2+Q_{y0}^2)}{T_0^2},\qquad B_{5,4}=\left(\frac{\mathit{Re}\,q(\kappa_0)}{\mu_0}-\Lambda_1\right)\Pi_{xx0},\\
B_{5,5}&=i\alpha u_0+\frac{\mathit{Re}\,p_0q(\kappa_0)}{\mu_0} +\Lambda_2\Pi_{xx0}(2\Pi_{xx0}+\Pi_{yy0}),\qquad B_{5,6}=\Lambda_2\Pi_{xx0}(2\Pi_{yy0}+\Pi_{xx0}),\\
B_{5,7}&=\frac43u_0'+2\Lambda_2\Pi_{xx0}\Pi_{xy0},\qquad B_{5,8}=\frac{2\Lambda_2\varepsilon\Pi_{xx0}Q_{x0}}{T_0},\qquad B_{5,9}=\frac{2\Lambda_2\varepsilon\Pi_{xx0}Q_{y0}}{T_0}.\\[8pt]
B_{6,1}&=i\alpha\left[\Pi_{yy0}-\frac23(\mathit{Re}\,p_0+\Pi_{xx0})\right],\qquad B_{6,2}=\Pi_{yy0}'+\frac43i\alpha\Pi_{xy0},\\
B_{6,3}&=-\frac{\mathit{Re}\,p_0q(\kappa_0)\Pi_{yy0}}{\mu_0^2}\left.\frac{d\mu}{dT}\right|_0 -\Lambda_2\Pi_{yy0}\frac{\varepsilon(Q_{x0}^2+Q_{y0}^2)}{T_0^2},\qquad B_{6,4}=\left(\frac{\mathit{Re}\,q(\kappa_0)}{\mu_0}-\Lambda_1\right)\Pi_{yy0},\\
B_{6,5}&=\Lambda_2\Pi_{yy0}(2\Pi_{xx0}+\Pi_{yy0}),\qquad B_{6,6}=i\alpha u_0+\frac{\mathit{Re}\,p_0q(\kappa_0)}{\mu_0} +\Lambda_2\Pi_{yy0}(2\Pi_{yy0}+\Pi_{xx0}),\\
B_{6,7}&=-\frac23u_0'+2\Lambda_2\Pi_{yy0}\Pi_{xy0},\qquad B_{6,8}=\frac{2\Lambda_2\varepsilon\Pi_{yy0}Q_{x0}}{T_0},\qquad B_{6,9}=\frac{2\Lambda_2\varepsilon\Pi_{yy0}Q_{y0}}{T_0}.\\[8pt]
B_{7,1}&=2i\alpha\Pi_{xy0},\qquad B_{7,2}=\Pi_{xy0}'+i\alpha(\mathit{Re}\,p_0+\Pi_{xx0}),\\
B_{7,3}&=-\frac{\mathit{Re}\,p_0q(\kappa_0)\Pi_{xy0}}{\mu_0^2}\left.\frac{d\mu}{dT}\right|_0 -\Lambda_2\Pi_{xy0}\frac{\varepsilon(Q_{x0}^2+Q_{y0}^2)}{T_0^2},\\
B_{7,4}&=\mathit{Re}\,u_0'+\left(\frac{\mathit{Re}\,q(\kappa_0)}{\mu_0}-\Lambda_1\right)\Pi_{xy0},\qquad B_{7,5}=\Lambda_2\Pi_{xy0}(2\Pi_{xx0}+\Pi_{yy0}),\\
B_{7,6}&=u_0' +\Lambda_2\Pi_{xy0}(2\Pi_{yy0}+\Pi_{xx0}),\qquad B_{7,7}=i\alpha u_0+\frac{\mathit{Re}\,p_0q(\kappa_0)}{\mu_0}+2\Lambda_2\Pi_{xy0}^2,\\
B_{7,8}&=\frac{2\Lambda_2\varepsilon\Pi_{xy0}Q_{x0}}{T_0},\qquad B_{7,9}=\frac{2\Lambda_2\varepsilon\Pi_{xy0}Q_{y0}}{T_0}.\\[8pt]
B_{8,1}&=2i\alpha Q_{x0} +(\gamma-1)\mathit{Pr}\,\mathit{Ma}^2i\alpha u_0\Pi_{xx0},\\
B_{8,2}&=Q_{x0}' +(\gamma-1)\mathit{Pr}\,\mathit{Ma}^2\left(i\alpha u_0\Pi_{xy0}+u_0'\Pi_{xx0}\right),\\
B_{8,3}&=\mathit{Pr}\,i\alpha(\mathit{Re}\,p_0+\Pi_{xx0}) -\frac{\mathit{Pr}\,\mathit{Re}\,p_0q(\kappa_0)Q_{x0}}{\mu_0^2}\left.\frac{d\mu}{dT}\right|_0 -\mathit{Pr}\,\Lambda_2Q_{x0}\frac{\varepsilon(Q_{x0}^2+Q_{y0}^2)}{T_0^2},\\
B_{8,4}&=\mathit{Pr}\left(\frac{\mathit{Re}\,q(\kappa_0)}{\mu_0}-\Lambda_1\right)Q_{x0},\qquad B_{8,5}=\mathit{Pr}\,\Lambda_2Q_{x0}(2\Pi_{xx0}+\Pi_{yy0}),\\
B_{8,6}&=\mathit{Pr}\,\Lambda_2Q_{x0}(2\Pi_{yy0}+\Pi_{xx0}),\qquad B_{8,7}=\mathit{Pr}\,T_0'+2\mathit{Pr}\,\Lambda_2Q_{x0}\Pi_{xy0},\\
B_{8,8}&=i\alpha u_0+\frac{\mathit{Re}\,\mathit{Pr}\,p_0q(\kappa_0)}{\mu_0}
+\frac{2\mathit{Pr}\,\Lambda_2\varepsilon Q_{x0}^2}{T_0},\qquad B_{8,9}=u_0'+\frac{2\mathit{Pr}\,\Lambda_2\varepsilon Q_{x0}Q_{y0}}{T_0}.\\[8pt]
B_{9,1}&=i\alpha Q_{y0} +(\gamma-1)\mathit{Pr}\,\mathit{Ma}^2i\alpha u_0\Pi_{xy0},\\
B_{9,2}&=Q_{y0}'+i\alpha Q_{x0} +(\gamma-1)\mathit{Pr}\,\mathit{Ma}^2\left(i\alpha u_0\Pi_{yy0}+u_0'\Pi_{xy0}\right),\\
B_{9,3}&=\mathit{Pr}\,i\alpha\Pi_{xy0} -\frac{\mathit{Pr}\,\mathit{Re}\,p_0q(\kappa_0)Q_{y0}}{\mu_0^2}\left.\frac{d\mu}{dT}\right|_0 -\mathit{Pr}\,\Lambda_2Q_{y0}\frac{\varepsilon(Q_{x0}^2+Q_{y0}^2)}{T_0^2},\\
B_{9,4}&=\mathit{Pr}\,\mathit{Re}\,T_0'+\mathit{Pr}\left(\frac{\mathit{Re}\,q(\kappa_0)}{\mu_0}-\Lambda_1\right)Q_{y0},\qquad B_{9,5}=\mathit{Pr}\,\Lambda_2Q_{y0}(2\Pi_{xx0}+\Pi_{yy0}),\\
B_{9,6}&=\mathit{Pr}\,T_0'+\mathit{Pr}\,\Lambda_2Q_{y0}(2\Pi_{yy0}+\Pi_{xx0}),\qquad B_{9,7}=2\mathit{Pr}\,\Lambda_2Q_{y0}\Pi_{xy0},\\
B_{9,8}&=\frac{2\mathit{Pr}\,\Lambda_2\varepsilon Q_{x0}Q_{y0}}{T_0},\qquad B_{9,9}=i\alpha u_0+\frac{\mathit{Re}\,\mathit{Pr}\,p_0q(\kappa_0)}{\mu_0}
+\frac{2\mathit{Pr}\,\Lambda_2\varepsilon Q_{y0}^2}{T_0}.\\[8pt]
C_{1,1}&=C_{2,2}=C_{3,3}=i\rho_0,\qquad C_{3,4}=-i(\gamma-1)\mathit{Ma}^2,\\
C_{4,3}&=-\frac{i}{T_0},\qquad C_{4,4}=\frac{i\gamma \mathit{Ma}^2}{\rho_0T_0},\\
C_{5,5}&=C_{6,6}=C_{7,7}=i,\\
C_{8,8}&=C_{9,9}=i,\qquad C_{8,1}=i(\gamma-1)\mathit{Pr}\,\mathit{Ma}^2\Pi_{xx0},\qquad C_{8,2}=i(\gamma-1)\mathit{Pr}\,\mathit{Ma}^2\Pi_{xy0},\\
C_{9,1}&=i(\gamma-1)\mathit{Pr}\,\mathit{Ma}^2\Pi_{xy0},\qquad C_{9,2}=i(\gamma-1)\mathit{Pr}\,\mathit{Ma}^2\Pi_{yy0}.
\end{align*}

\noindent\begin{minipage}{\linewidth}
    \centering
    (2) NCCR-LST
\end{minipage}
\begin{align*}
A_{5,1}&=\frac43\Pi_{xy0},\qquad A_{5,2}=-\frac23(\mathit{Re}\,p_0+\Pi_{yy0}),\\
A_{6,1}&=-\frac23\Pi_{xy0},\qquad A_{6,2}=\frac43(\mathit{Re}\,p_0+\Pi_{yy0}),\\
A_{7,1}&=\mathit{Re}\,p_0+\Pi_{yy0},\qquad A_{7,2}=\Pi_{xy0},\\
A_{8,1}&=Q_{y0},\qquad A_{8,3}=\mathit{Pr}\,\Pi_{xy0},\\
A_{9,2}&=Q_{y0},\qquad A_{9,3}=\mathit{Pr}(\mathit{Re}\,p_0+\Pi_{yy0}).\\[8pt]
B_{5,1}&=\frac43i\alpha(\mathit{Re}\,p_0+\Pi_{xx0}),\qquad B_{5,2}=-\frac23i\alpha\Pi_{xy0},\\
B_{5,3}&=-\frac{\mathit{Re}\,p_0q(\kappa_0)\Pi_{xx0}}{\mu_0^2}\left.\frac{d\mu}{dT}\right|_0 -\Lambda_2\Pi_{xx0}\frac{\varepsilon(Q_{x0}^2+Q_{y0}^2)}{T_0^2},\qquad B_{5,4}=\left(\frac{\mathit{Re}\,q(\kappa_0)}{\mu_0}-\Lambda_1\right)\Pi_{xx0},\\
B_{5,5}&=\frac{\mathit{Re}\,p_0q(\kappa_0)}{\mu_0} +\Lambda_2\Pi_{xx0}(2\Pi_{xx0}+\Pi_{yy0}),\qquad B_{5,6}=\Lambda_2\Pi_{xx0}(2\Pi_{yy0}+\Pi_{xx0}),\\
B_{5,7}&=\frac43u_0'+2\Lambda_2\Pi_{xx0}\Pi_{xy0},\qquad B_{5,8}=\frac{2\Lambda_2\varepsilon\Pi_{xx0}Q_{x0}}{T_0},\qquad B_{5,9}=\frac{2\Lambda_2\varepsilon\Pi_{xx0}Q_{y0}}{T_0}.\\[8pt]
B_{6,1}&=-\frac23i\alpha(\mathit{Re}\,p_0+\Pi_{xx0}),\qquad B_{6,2}=\frac43i\alpha\Pi_{xy0},\\
B_{6,3}&=-\frac{\mathit{Re}\,p_0q(\kappa_0)\Pi_{yy0}}{\mu_0^2}\left.\frac{d\mu}{dT}\right|_0 -\Lambda_2\Pi_{yy0}\frac{\varepsilon(Q_{x0}^2+Q_{y0}^2)}{T_0^2},\qquad B_{6,4}=\left(\frac{\mathit{Re}\,q(\kappa_0)}{\mu_0}-\Lambda_1\right)\Pi_{yy0},\\
B_{6,5}&=\Lambda_2\Pi_{yy0}(2\Pi_{xx0}+\Pi_{yy0}),\qquad B_{6,6}=\frac{\mathit{Re}\,p_0q(\kappa_0)}{\mu_0} +\Lambda_2\Pi_{yy0}(2\Pi_{yy0}+\Pi_{xx0}),\\
B_{6,7}&=-\frac23u_0'+2\Lambda_2\Pi_{yy0}\Pi_{xy0},\qquad B_{6,8}=\frac{2\Lambda_2\varepsilon\Pi_{yy0}Q_{x0}}{T_0},\qquad B_{6,9}=\frac{2\Lambda_2\varepsilon\Pi_{yy0}Q_{y0}}{T_0}.\\[8pt]
B_{7,1}&=i\alpha\Pi_{xy0},\qquad B_{7,2}=i\alpha(\mathit{Re}\,p_0+\Pi_{xx0}),\\
B_{7,3}&=-\frac{\mathit{Re}\,p_0q(\kappa_0)\Pi_{xy0}}{\mu_0^2}\left.\frac{d\mu}{dT}\right|_0 -\Lambda_2\Pi_{xy0}\frac{\varepsilon(Q_{x0}^2+Q_{y0}^2)}{T_0^2},\\
B_{7,4}&=\mathit{Re}\,u_0'+\left(\frac{\mathit{Re}\,q(\kappa_0)}{\mu_0}-\Lambda_1\right)\Pi_{xy0},\qquad B_{7,5}=\Lambda_2\Pi_{xy0}(2\Pi_{xx0}+\Pi_{yy0}),\\
B_{7,6}&=u_0' +\Lambda_2\Pi_{xy0}(2\Pi_{yy0}+\Pi_{xx0}),\qquad B_{7,7}=\frac{\mathit{Re}\,p_0q(\kappa_0)}{\mu_0}+2\Lambda_2\Pi_{xy0}^2,\\
B_{7,8}&=\frac{2\Lambda_2\varepsilon\Pi_{xy0}Q_{x0}}{T_0},\qquad B_{7,9}=\frac{2\Lambda_2\varepsilon\Pi_{xy0}Q_{y0}}{T_0}.\\[8pt]
B_{8,1}&=i\alpha Q_{x0},\\
B_{8,3}&=\mathit{Pr}\,i\alpha(\mathit{Re}\,p_0+\Pi_{xx0}) -\frac{\mathit{Pr}\,\mathit{Re}\,p_0q(\kappa_0)Q_{x0}}{\mu_0^2}\left.\frac{d\mu}{dT}\right|_0 -\mathit{Pr}\,\Lambda_2Q_{x0}\frac{\varepsilon(Q_{x0}^2+Q_{y0}^2)}{T_0^2},\\
B_{8,4}&=\mathit{Pr}\left(\frac{\mathit{Re}\,q(\kappa_0)}{\mu_0}-\Lambda_1\right)Q_{x0},\qquad B_{8,5}=\mathit{Pr}\,\Lambda_2Q_{x0}(2\Pi_{xx0}+\Pi_{yy0}),\\
B_{8,6}&=\mathit{Pr}\,\Lambda_2Q_{x0}(2\Pi_{yy0}+\Pi_{xx0}),\qquad B_{8,7}=\mathit{Pr}\,T_0'+2\mathit{Pr}\,\Lambda_2Q_{x0}\Pi_{xy0},\\
B_{8,8}&=\frac{\mathit{Re}\,\mathit{Pr}\,p_0q(\kappa_0)}{\mu_0}
+\frac{2\mathit{Pr}\,\Lambda_2\varepsilon Q_{x0}^2}{T_0},\qquad B_{8,9}=u_0'+\frac{2\mathit{Pr}\,\Lambda_2\varepsilon Q_{x0}Q_{y0}}{T_0}.\\[8pt]
B_{9,2}&=i\alpha Q_{x0},\qquad B_{9,3}=\mathit{Pr}\,i\alpha\Pi_{xy0} -\frac{\mathit{Pr}\,\mathit{Re}\,p_0q(\kappa_0)Q_{y0}}{\mu_0^2}\left.\frac{d\mu}{dT}\right|_0 -\mathit{Pr}\,\Lambda_2Q_{y0}\frac{\varepsilon(Q_{x0}^2+Q_{y0}^2)}{T_0^2},\\
B_{9,4}&=\mathit{Pr}\,\mathit{Re}\,T_0'+\mathit{Pr}\left(\frac{\mathit{Re}\,q(\kappa_0)}{\mu_0}-\Lambda_1\right)Q_{y0},\qquad B_{9,5}=\mathit{Pr}\,\Lambda_2Q_{y0}(2\Pi_{xx0}+\Pi_{yy0}),\\
B_{9,6}&=\mathit{Pr}\,T_0'+\mathit{Pr}\,\Lambda_2Q_{y0}(2\Pi_{yy0}+\Pi_{xx0}),\qquad B_{9,7}=2\mathit{Pr}\,\Lambda_2Q_{y0}\Pi_{xy0},\\
B_{9,8}&=\frac{2\mathit{Pr}\,\Lambda_2\varepsilon Q_{x0}Q_{y0}}{T_0},\qquad B_{9,9}=\frac{\mathit{Re}\,\mathit{Pr}\,p_0q(\kappa_0)}{\mu_0}
+\frac{2\mathit{Pr}\,\Lambda_2\varepsilon Q_{y0}^2}{T_0}.\\[8pt]
C_{i,j}&=0,\qquad\text{where}\qquad i=5,\ldots,9,\;\;j=1,\ldots,9.
\end{align*}

\noindent\begin{minipage}{\linewidth}
    \centering
    (3) NS-LST
\end{minipage}
\begin{align*}
A_{5,2}&=-\frac23\mu_0,\qquad A_{6,2}=\frac43\mu_0,\qquad A_{7,1}=\mu_0,\qquad A_{9,3}=k_0.\\[8pt]
B_{5,1}&=\frac43i\alpha\mu_0,\qquad B_{5,5}=1,\\
B_{6,1}&=-\frac23i\alpha\mu_0,\qquad B_{6,6}=1,\\
B_{7,2}&=i\alpha\mu_0,\qquad B_{7,3}=u_0'\left.\frac{d\mu}{dT}\right|_0,\qquad B_{7,7}=1,\\
B_{8,3}&=i\alpha k_0,\qquad B_{8,8}=1,\\
B_{9,3}&=T_0'\left.\frac{dk}{dT}\right|_0,\qquad B_{9,9}=1.\\[8pt]
C_{i,j}&=0,\qquad\text{where}\qquad i=5,\ldots,9,\;\;j=1,\ldots,9.
\end{align*}

In the above equations, the subscript `0' denotes base-flow quantities, and the prime `$\prime$' denotes the first-order derivative with respect to $y$, e.g., $u_0^\prime=du_0/dy$. The auxiliary variables are
\begin{align*}
\Lambda_1&=\frac{\mathit{Re}\,\kappa_0}{\mu_0}
\left.\frac{dq}{d\kappa}\right|_{\kappa_0},\\
\Lambda_2&=\frac{\mathit{Re}\,p_0}{\mu_0\kappa_0}
\left.\frac{dq}{d\kappa}\right|_{\kappa_0}
\left(\frac{c}{\mathit{Re}\,p_0}\right)^2.
\end{align*}


\begin{thebibliography}{41}%
\makeatletter
\providecommand \@ifxundefined [1]{%
 \@ifx{#1\undefined}
}%
\providecommand \@ifnum [1]{%
 \ifnum #1\expandafter \@firstoftwo
 \else \expandafter \@secondoftwo
 \fi
}%
\providecommand \@ifx [1]{%
 \ifx #1\expandafter \@firstoftwo
 \else \expandafter \@secondoftwo
 \fi
}%
\providecommand \natexlab [1]{#1}%
\providecommand \enquote  [1]{``#1''}%
\providecommand \bibnamefont  [1]{#1}%
\providecommand \bibfnamefont [1]{#1}%
\providecommand \citenamefont [1]{#1}%
\providecommand \href@noop [0]{\@secondoftwo}%
\providecommand \href [0]{\begingroup \@sanitize@url \@href}%
\providecommand \@href[1]{\@@startlink{#1}\@@href}%
\providecommand \@@href[1]{\endgroup#1\@@endlink}%
\providecommand \@sanitize@url [0]{\catcode `\\12\catcode `\$12\catcode
  `\&12\catcode `\#12\catcode `\^12\catcode `\_12\catcode `\%12\relax}%
\providecommand \@@startlink[1]{}%
\providecommand \@@endlink[0]{}%
\providecommand \url  [0]{\begingroup\@sanitize@url \@url }%
\providecommand \@url [1]{\endgroup\@href {#1}{\urlprefix }}%
\providecommand \urlprefix  [0]{URL }%
\providecommand \Eprint [0]{\href }%
\providecommand \doibase [0]{https://doi.org/}%
\providecommand \selectlanguage [0]{\@gobble}%
\providecommand \bibinfo  [0]{\@secondoftwo}%
\providecommand \bibfield  [0]{\@secondoftwo}%
\providecommand \translation [1]{[#1]}%
\providecommand \BibitemOpen [0]{}%
\providecommand \bibitemStop [0]{}%
\providecommand \bibitemNoStop [0]{.\EOS\space}%
\providecommand \EOS [0]{\spacefactor3000\relax}%
\providecommand \BibitemShut  [1]{\csname bibitem#1\endcsname}%
\let\auto@bib@innerbib\@empty
\bibitem [{\citenamefont {Schmid}\ and\ \citenamefont
  {Henningson}(2001)}]{ref1}%
  \BibitemOpen
  \bibfield  {author} {\bibinfo {author} {\bibfnamefont {P.~J.}\ \bibnamefont
  {Schmid}}\ and\ \bibinfo {author} {\bibfnamefont {D.~S.}\ \bibnamefont
  {Henningson}},\ }\href@noop {} {\emph {\bibinfo {title} {Stability and
  Transition in Shear Flows}}}\ (\bibinfo  {publisher} {Springer},\ \bibinfo
  {address} {New York},\ \bibinfo {year} {2001})\BibitemShut {NoStop}%
\bibitem [{\citenamefont {Fedorov}(2011)}]{ref2}%
  \BibitemOpen
  \bibfield  {author} {\bibinfo {author} {\bibfnamefont {A.}~\bibnamefont
  {Fedorov}},\ }\bibfield  {title} {\bibinfo {title} {{Transition and stability
  of high-speed boundary layers}},\ }\href@noop {} {\bibfield  {journal}
  {\bibinfo  {journal} {Annual Review of Fluid Mechanics}\ }\textbf {\bibinfo
  {volume} {43}},\ \bibinfo {pages} {79} (\bibinfo {year} {2011})}\BibitemShut
  {NoStop}%
\bibitem [{\citenamefont {Chen}\ and\ \citenamefont {Zhou}(2021)}]{ref3}%
  \BibitemOpen
  \bibfield  {author} {\bibinfo {author} {\bibfnamefont {J.}~\bibnamefont
  {Chen}}\ and\ \bibinfo {author} {\bibfnamefont {H.}~\bibnamefont {Zhou}},\
  }\bibfield  {title} {\bibinfo {title} {{Rarefied gas effect in hypersonic
  shear flows}},\ }\href@noop {} {\bibfield  {journal} {\bibinfo  {journal}
  {Acta Mechanica Sinica}\ }\textbf {\bibinfo {volume} {37}},\ \bibinfo {pages}
  {2} (\bibinfo {year} {2021})}\BibitemShut {NoStop}%
\bibitem [{\citenamefont {Duck}\ \emph {et~al.}(1994)\citenamefont {Duck},
  \citenamefont {Erlebacher},\ and\ \citenamefont {Hussaini}}]{ref4}%
  \BibitemOpen
  \bibfield  {author} {\bibinfo {author} {\bibfnamefont {P.~W.}\ \bibnamefont
  {Duck}}, \bibinfo {author} {\bibfnamefont {G.}~\bibnamefont {Erlebacher}},\
  and\ \bibinfo {author} {\bibfnamefont {M.~Y.}\ \bibnamefont {Hussaini}},\
  }\bibfield  {title} {\bibinfo {title} {{On the linear stability of
  compressible plane Couette flow}},\ }\href@noop {} {\bibfield  {journal}
  {\bibinfo  {journal} {Journal of Fluid Mechanics}\ }\textbf {\bibinfo
  {volume} {258}},\ \bibinfo {pages} {131} (\bibinfo {year}
  {1994})}\BibitemShut {NoStop}%
\bibitem [{\citenamefont {Hu}\ and\ \citenamefont {Zhong}(1998)}]{ref5}%
  \BibitemOpen
  \bibfield  {author} {\bibinfo {author} {\bibfnamefont {S.}~\bibnamefont
  {Hu}}\ and\ \bibinfo {author} {\bibfnamefont {X.}~\bibnamefont {Zhong}},\
  }\bibfield  {title} {\bibinfo {title} {{Linear stability of viscous
  supersonic plane Couette flow}},\ }\href@noop {} {\bibfield  {journal}
  {\bibinfo  {journal} {Physics of Fluids}\ }\textbf {\bibinfo {volume} {10}},\
  \bibinfo {pages} {709} (\bibinfo {year} {1998})}\BibitemShut {NoStop}%
\bibitem [{\citenamefont {Ou}(2024)}]{ref6}%
  \BibitemOpen
  \bibfield  {author} {\bibinfo {author} {\bibfnamefont {J.}~\bibnamefont
  {Ou}},\ }\bibfield  {title} {\bibinfo {title} {{Revisit compressible plane
  Couette and Poiseuille instability: Spectrum features of disturbances and
  dominant instability}},\ }\href@noop {} {\bibfield  {journal} {\bibinfo
  {journal} {Physics of Fluids}\ }\textbf {\bibinfo {volume} {36}},\ \bibinfo
  {pages} {114119} (\bibinfo {year} {2024})}\BibitemShut {NoStop}%
\bibitem [{\citenamefont {Mack}(1984)}]{ref7}%
  \BibitemOpen
  \bibfield  {author} {\bibinfo {author} {\bibfnamefont {L.~M.}\ \bibnamefont
  {Mack}},\ }\href@noop {} {\emph {\bibinfo {title} {{Boundary-layer linear
  stability theory}}}},\ \bibinfo {type} {Report No.}\ \bibinfo {number} {709}\
  (\bibinfo  {institution} {AGARD},\ \bibinfo {year} {1984})\BibitemShut
  {NoStop}%
\bibitem [{\citenamefont {Reed}\ \emph {et~al.}(1996)\citenamefont {Reed},
  \citenamefont {Saric},\ and\ \citenamefont {Arnal}}]{ref8}%
  \BibitemOpen
  \bibfield  {author} {\bibinfo {author} {\bibfnamefont {H.~L.}\ \bibnamefont
  {Reed}}, \bibinfo {author} {\bibfnamefont {W.~S.}\ \bibnamefont {Saric}},\
  and\ \bibinfo {author} {\bibfnamefont {D.}~\bibnamefont {Arnal}},\ }\bibfield
   {title} {\bibinfo {title} {{Linear stability theory applied to boundary
  layers}},\ }\href@noop {} {\bibfield  {journal} {\bibinfo  {journal} {Annual
  Review of Fluid Mechanics}\ }\textbf {\bibinfo {volume} {28}},\ \bibinfo
  {pages} {389} (\bibinfo {year} {1996})}\BibitemShut {NoStop}%
\bibitem [{\citenamefont {Sandham}\ and\ \citenamefont
  {Reynolds}(1990)}]{ref9}%
  \BibitemOpen
  \bibfield  {author} {\bibinfo {author} {\bibfnamefont {N.~D.}\ \bibnamefont
  {Sandham}}\ and\ \bibinfo {author} {\bibfnamefont {W.~C.}\ \bibnamefont
  {Reynolds}},\ }\bibfield  {title} {\bibinfo {title} {{Compressible mixing
  layer: Linear theory and direct simulation}},\ }\href@noop {} {\bibfield
  {journal} {\bibinfo  {journal} {AIAA Journal}\ }\textbf {\bibinfo {volume}
  {28}},\ \bibinfo {pages} {618} (\bibinfo {year} {1990})}\BibitemShut
  {NoStop}%
\bibitem [{\citenamefont {Tam}\ and\ \citenamefont {Hu}(1989)}]{ref10}%
  \BibitemOpen
  \bibfield  {author} {\bibinfo {author} {\bibfnamefont {C.~K.~W.}\
  \bibnamefont {Tam}}\ and\ \bibinfo {author} {\bibfnamefont {F.~Q.}\
  \bibnamefont {Hu}},\ }\bibfield  {title} {\bibinfo {title} {{The instability
  and acoustic wave modes of supersonic mixing layers inside a rectangular
  channel}},\ }\href@noop {} {\bibfield  {journal} {\bibinfo  {journal}
  {Journal of Fluid Mechanics}\ }\textbf {\bibinfo {volume} {203}},\ \bibinfo
  {pages} {51} (\bibinfo {year} {1989})}\BibitemShut {NoStop}%
\bibitem [{\citenamefont {Rees}\ and\ \citenamefont {Juniper}(2010)}]{ref11}%
  \BibitemOpen
  \bibfield  {author} {\bibinfo {author} {\bibfnamefont {S.~J.}\ \bibnamefont
  {Rees}}\ and\ \bibinfo {author} {\bibfnamefont {M.~P.}\ \bibnamefont
  {Juniper}},\ }\bibfield  {title} {\bibinfo {title} {{The effect of
  confinement on the stability of viscous planar jets and wakes}},\ }\href@noop
  {} {\bibfield  {journal} {\bibinfo  {journal} {Journal of Fluid Mechanics}\
  }\textbf {\bibinfo {volume} {656}},\ \bibinfo {pages} {309} (\bibinfo {year}
  {2010})}\BibitemShut {NoStop}%
\bibitem [{\citenamefont {Wang}\ \emph {et~al.}(2025)\citenamefont {Wang},
  \citenamefont {Ou},\ and\ \citenamefont {Chen}}]{refWang2025}%
  \BibitemOpen
  \bibfield  {author} {\bibinfo {author} {\bibfnamefont {C.}~\bibnamefont
  {Wang}}, \bibinfo {author} {\bibfnamefont {J.}~\bibnamefont {Ou}},\ and\
  \bibinfo {author} {\bibfnamefont {J.}~\bibnamefont {Chen}},\ }\bibfield
  {title} {\bibinfo {title} {{Rarefaction effects on hypersonic boundary-layer
  stability over a blunt cone at varying degrees of wall cooling and nose
  bluntness in near-continuum regime}},\ }\href@noop {} {\bibfield  {journal}
  {\bibinfo  {journal} {Physical Review Fluids}\ }\textbf {\bibinfo {volume}
  {10}},\ \bibinfo {pages} {093901} (\bibinfo {year} {2025})}\BibitemShut
  {NoStop}%
\bibitem [{\citenamefont {Morawetz}(1952)}]{ref12}%
  \BibitemOpen
  \bibfield  {author} {\bibinfo {author} {\bibfnamefont {C.~S.}\ \bibnamefont
  {Morawetz}},\ }\bibfield  {title} {\bibinfo {title} {{The eigenvalues of some
  stability problems involving viscosity}},\ }\href@noop {} {\bibfield
  {journal} {\bibinfo  {journal} {Indiana University Mathematics Journal}\
  }\textbf {\bibinfo {volume} {1}},\ \bibinfo {pages} {579} (\bibinfo {year}
  {1952})}\BibitemShut {NoStop}%
\bibitem [{\citenamefont {Romanov}(1973)}]{ref13}%
  \BibitemOpen
  \bibfield  {author} {\bibinfo {author} {\bibfnamefont {V.~A.}\ \bibnamefont
  {Romanov}},\ }\bibfield  {title} {\bibinfo {title} {{Stability of
  plane-parallel Couette flow}},\ }\href@noop {} {\bibfield  {journal}
  {\bibinfo  {journal} {Functional Analysis and Its Applications}\ }\textbf
  {\bibinfo {volume} {7}},\ \bibinfo {pages} {137} (\bibinfo {year}
  {1973})}\BibitemShut {NoStop}%
\bibitem [{\citenamefont {Malik}\ \emph {et~al.}(2008)\citenamefont {Malik},
  \citenamefont {Dey},\ and\ \citenamefont {Alam}}]{ref14}%
  \BibitemOpen
  \bibfield  {author} {\bibinfo {author} {\bibfnamefont {M.}~\bibnamefont
  {Malik}}, \bibinfo {author} {\bibfnamefont {J.}~\bibnamefont {Dey}},\ and\
  \bibinfo {author} {\bibfnamefont {M.}~\bibnamefont {Alam}},\ }\bibfield
  {title} {\bibinfo {title} {{Linear stability, transient energy growth, and
  the role of viscosity stratification in compressible plane Couette flow}},\
  }\href@noop {} {\bibfield  {journal} {\bibinfo  {journal} {Physical Review
  E}\ }\textbf {\bibinfo {volume} {77}},\ \bibinfo {pages} {036322} (\bibinfo
  {year} {2008})}\BibitemShut {NoStop}%
\bibitem [{\citenamefont {Ramachandran}\ \emph {et~al.}(2016)\citenamefont
  {Ramachandran}, \citenamefont {Saikia}, \citenamefont {Sinha},\ and\
  \citenamefont {Govindarajan}}]{ref15}%
  \BibitemOpen
  \bibfield  {author} {\bibinfo {author} {\bibfnamefont {A.}~\bibnamefont
  {Ramachandran}}, \bibinfo {author} {\bibfnamefont {B.}~\bibnamefont
  {Saikia}}, \bibinfo {author} {\bibfnamefont {K.}~\bibnamefont {Sinha}},\ and\
  \bibinfo {author} {\bibfnamefont {R.}~\bibnamefont {Govindarajan}},\
  }\bibfield  {title} {\bibinfo {title} {{Effect of Prandtl number on the
  linear stability of compressible Couette flow}},\ }\href@noop {} {\bibfield
  {journal} {\bibinfo  {journal} {International Journal of Heat and Fluid
  Flow}\ }\textbf {\bibinfo {volume} {61}},\ \bibinfo {pages} {553} (\bibinfo
  {year} {2016})}\BibitemShut {NoStop}%
\bibitem [{\citenamefont {Deka}\ \emph
  {et~al.}(2023{\natexlab{a}})\citenamefont {Deka}, \citenamefont {Tomar},\
  and\ \citenamefont {Kumaran}}]{ref16}%
  \BibitemOpen
  \bibfield  {author} {\bibinfo {author} {\bibfnamefont {M.}~\bibnamefont
  {Deka}}, \bibinfo {author} {\bibfnamefont {G.}~\bibnamefont {Tomar}},\ and\
  \bibinfo {author} {\bibfnamefont {V.}~\bibnamefont {Kumaran}},\ }\bibfield
  {title} {\bibinfo {title} {{Linear stability of a compressible flow in a
  channel}},\ }\href@noop {} {\bibfield  {journal} {\bibinfo  {journal}
  {Quarterly Journal of Mechanics and Applied Mathematics}\ }\textbf {\bibinfo
  {volume} {76}},\ \bibinfo {pages} {405} (\bibinfo {year}
  {2023}{\natexlab{a}})}\BibitemShut {NoStop}%
\bibitem [{\citenamefont {Deka}\ \emph
  {et~al.}(2023{\natexlab{b}})\citenamefont {Deka}, \citenamefont {Tomar},\
  and\ \citenamefont {Kumaran}}]{ref17}%
  \BibitemOpen
  \bibfield  {author} {\bibinfo {author} {\bibfnamefont {M.}~\bibnamefont
  {Deka}}, \bibinfo {author} {\bibfnamefont {G.}~\bibnamefont {Tomar}},\ and\
  \bibinfo {author} {\bibfnamefont {V.}~\bibnamefont {Kumaran}},\ }\bibfield
  {title} {\bibinfo {title} {{Linear stability analysis of compressible pipe
  flow}},\ }\href@noop {} {\bibfield  {journal} {\bibinfo  {journal}
  {Theoretical and Computational Fluid Dynamics}\ }\textbf {\bibinfo {volume}
  {37}},\ \bibinfo {pages} {589} (\bibinfo {year}
  {2023}{\natexlab{b}})}\BibitemShut {NoStop}%
\bibitem [{\citenamefont {Yoshida}\ and\ \citenamefont {Aoki}(2006)}]{ref18}%
  \BibitemOpen
  \bibfield  {author} {\bibinfo {author} {\bibfnamefont {H.}~\bibnamefont
  {Yoshida}}\ and\ \bibinfo {author} {\bibfnamefont {K.}~\bibnamefont {Aoki}},\
  }\bibfield  {title} {\bibinfo {title} {{Linear stability of the cylindrical
  Couette flow of a rarefied gas}},\ }\href@noop {} {\bibfield  {journal}
  {\bibinfo  {journal} {Physical Review E}\ }\textbf {\bibinfo {volume} {73}},\
  \bibinfo {pages} {021201} (\bibinfo {year} {2006})}\BibitemShut {NoStop}%
\bibitem [{\citenamefont {Zou}\ \emph {et~al.}(2022)\citenamefont {Zou},
  \citenamefont {Zhong}, \citenamefont {Bi}, \citenamefont {Yuan},\ and\
  \citenamefont {Tang}}]{ref19}%
  \BibitemOpen
  \bibfield  {author} {\bibinfo {author} {\bibfnamefont {S.}~\bibnamefont
  {Zou}}, \bibinfo {author} {\bibfnamefont {C.}~\bibnamefont {Zhong}}, \bibinfo
  {author} {\bibfnamefont {L.}~\bibnamefont {Bi}}, \bibinfo {author}
  {\bibfnamefont {X.}~\bibnamefont {Yuan}},\ and\ \bibinfo {author}
  {\bibfnamefont {Z.}~\bibnamefont {Tang}},\ }\bibfield  {title} {\bibinfo
  {title} {{A new linear stability analysis approach for microchannel flow
  based on the Boltzmann Bhatnagar--Gross--Krook equation}},\ }\href@noop {}
  {\bibfield  {journal} {\bibinfo  {journal} {Physics of Fluids}\ }\textbf
  {\bibinfo {volume} {34}},\ \bibinfo {pages} {124114} (\bibinfo {year}
  {2022})}\BibitemShut {NoStop}%
\bibitem [{\citenamefont {Zou}\ \emph {et~al.}(2023)\citenamefont {Zou},
  \citenamefont {Bi}, \citenamefont {Zhong}, \citenamefont {Yuan},\ and\
  \citenamefont {Tang}}]{ref20}%
  \BibitemOpen
  \bibfield  {author} {\bibinfo {author} {\bibfnamefont {S.}~\bibnamefont
  {Zou}}, \bibinfo {author} {\bibfnamefont {L.}~\bibnamefont {Bi}}, \bibinfo
  {author} {\bibfnamefont {C.}~\bibnamefont {Zhong}}, \bibinfo {author}
  {\bibfnamefont {X.}~\bibnamefont {Yuan}},\ and\ \bibinfo {author}
  {\bibfnamefont {Z.}~\bibnamefont {Tang}},\ }\bibfield  {title} {\bibinfo
  {title} {{A novel linear stability analysis method for plane Couette flow
  considering rarefaction effects}},\ }\href@noop {} {\bibfield  {journal}
  {\bibinfo  {journal} {Journal of Fluid Mechanics}\ }\textbf {\bibinfo
  {volume} {963}},\ \bibinfo {pages} {A33} (\bibinfo {year}
  {2023})}\BibitemShut {NoStop}%
\bibitem [{\citenamefont {Zou}\ \emph {et~al.}(2024)\citenamefont {Zou},
  \citenamefont {Bi}, \citenamefont {Zhong}, \citenamefont {Yuan},\ and\
  \citenamefont {Tang}}]{ref21}%
  \BibitemOpen
  \bibfield  {author} {\bibinfo {author} {\bibfnamefont {S.}~\bibnamefont
  {Zou}}, \bibinfo {author} {\bibfnamefont {L.}~\bibnamefont {Bi}}, \bibinfo
  {author} {\bibfnamefont {C.}~\bibnamefont {Zhong}}, \bibinfo {author}
  {\bibfnamefont {X.}~\bibnamefont {Yuan}},\ and\ \bibinfo {author}
  {\bibfnamefont {Z.}~\bibnamefont {Tang}},\ }\bibfield  {title} {\bibinfo
  {title} {{A normal-mode approach for high-speed rarefied plane Couette
  flow}},\ }\href@noop {} {\bibfield  {journal} {\bibinfo  {journal} {Physics
  of Fluids}\ }\textbf {\bibinfo {volume} {36}},\ \bibinfo {pages} {014121}
  (\bibinfo {year} {2024})}\BibitemShut {NoStop}%
\bibitem [{\citenamefont {Karpuzcu}\ \emph {et~al.}(2026)\citenamefont
  {Karpuzcu}, \citenamefont {Levin},\ and\ \citenamefont
  {Theofilis}}]{refKarpuzcu}%
  \BibitemOpen
  \bibfield  {author} {\bibinfo {author} {\bibfnamefont {I.~T.}\ \bibnamefont
  {Karpuzcu}}, \bibinfo {author} {\bibfnamefont {D.~A.}\ \bibnamefont
  {Levin}},\ and\ \bibinfo {author} {\bibfnamefont {V.}~\bibnamefont
  {Theofilis}},\ }\bibfield  {title} {\bibinfo {title} {{Kinetic linear
  stability theory for high-speed compressible flows: A high performance
  computing framework}},\ }\href@noop {} {\bibfield  {journal} {\bibinfo
  {journal} {arXiv:2607.27440}\ } (\bibinfo {year} {2026})}\BibitemShut
  {NoStop}%
\bibitem [{\citenamefont {Wu}(2026)}]{ref22}%
  \BibitemOpen
  \bibfield  {author} {\bibinfo {author} {\bibfnamefont {L.}~\bibnamefont
  {Wu}},\ }\bibfield  {title} {\bibinfo {title} {{Efficient solutions of
  eigenvalue problems in rarefied gas flows}},\ }\href@noop {} {\bibfield
  {journal} {\bibinfo  {journal} {Journal of Computational Physics}\ }\textbf
  {\bibinfo {volume} {549}},\ \bibinfo {pages} {114607} (\bibinfo {year}
  {2026})}\BibitemShut {NoStop}%
\bibitem [{\citenamefont {Lauga}\ and\ \citenamefont {Cossu}(2005)}]{ref23}%
  \BibitemOpen
  \bibfield  {author} {\bibinfo {author} {\bibfnamefont {E.}~\bibnamefont
  {Lauga}}\ and\ \bibinfo {author} {\bibfnamefont {C.}~\bibnamefont {Cossu}},\
  }\bibfield  {title} {\bibinfo {title} {{A note on the stability of slip
  channel flows}},\ }\href@noop {} {\bibfield  {journal} {\bibinfo  {journal}
  {Physics of Fluids}\ }\textbf {\bibinfo {volume} {17}},\ \bibinfo {pages}
  {088106} (\bibinfo {year} {2005})}\BibitemShut {NoStop}%
\bibitem [{\citenamefont {Chai}\ and\ \citenamefont {Song}(2019)}]{ref24}%
  \BibitemOpen
  \bibfield  {author} {\bibinfo {author} {\bibfnamefont {C.}~\bibnamefont
  {Chai}}\ and\ \bibinfo {author} {\bibfnamefont {B.}~\bibnamefont {Song}},\
  }\bibfield  {title} {\bibinfo {title} {{Stability of slip channel flow
  revisited}},\ }\href@noop {} {\bibfield  {journal} {\bibinfo  {journal}
  {Physics of Fluids}\ }\textbf {\bibinfo {volume} {31}},\ \bibinfo {pages}
  {084105} (\bibinfo {year} {2019})}\BibitemShut {NoStop}%
\bibitem [{\citenamefont {Xiong}\ and\ \citenamefont {Tao}(2020)}]{ref25}%
  \BibitemOpen
  \bibfield  {author} {\bibinfo {author} {\bibfnamefont {X.}~\bibnamefont
  {Xiong}}\ and\ \bibinfo {author} {\bibfnamefont {J.}~\bibnamefont {Tao}},\
  }\bibfield  {title} {\bibinfo {title} {{Linear stability and energy stability
  of plane Poiseuille flow with isotropic and anisotropic slip boundary
  conditions}},\ }\href@noop {} {\bibfield  {journal} {\bibinfo  {journal}
  {Physics of Fluids}\ }\textbf {\bibinfo {volume} {32}},\ \bibinfo {pages}
  {094104} (\bibinfo {year} {2020})}\BibitemShut {NoStop}%
\bibitem [{\citenamefont {Liu}\ and\ \citenamefont {Zhang}(2020)}]{ref26}%
  \BibitemOpen
  \bibfield  {author} {\bibinfo {author} {\bibfnamefont {B.}~\bibnamefont
  {Liu}}\ and\ \bibinfo {author} {\bibfnamefont {Y.}~\bibnamefont {Zhang}},\
  }\bibfield  {title} {\bibinfo {title} {{A numerical study on the natural
  transition locations in the flat-plate boundary layers on superhydrophobic
  surfaces}},\ }\href@noop {} {\bibfield  {journal} {\bibinfo  {journal}
  {Physics of Fluids}\ }\textbf {\bibinfo {volume} {32}},\ \bibinfo {pages}
  {124103} (\bibinfo {year} {2020})}\BibitemShut {NoStop}%
\bibitem [{\citenamefont {Ou}\ and\ \citenamefont {Chen}(2021)}]{ref27}%
  \BibitemOpen
  \bibfield  {author} {\bibinfo {author} {\bibfnamefont {J.}~\bibnamefont
  {Ou}}\ and\ \bibinfo {author} {\bibfnamefont {J.}~\bibnamefont {Chen}},\
  }\bibfield  {title} {\bibinfo {title} {{Numerical study of supersonic
  boundary-layer modal stability for a slightly rarefied gas using
  Navier--Stokes approach}},\ }\href@noop {} {\bibfield  {journal} {\bibinfo
  {journal} {Physics of Fluids}\ }\textbf {\bibinfo {volume} {33}},\ \bibinfo
  {pages} {114107} (\bibinfo {year} {2021})}\BibitemShut {NoStop}%
\bibitem [{\citenamefont {Ou}\ \emph {et~al.}(2024)\citenamefont {Ou},
  \citenamefont {Wang},\ and\ \citenamefont {Chen}}]{ref28}%
  \BibitemOpen
  \bibfield  {author} {\bibinfo {author} {\bibfnamefont {J.}~\bibnamefont
  {Ou}}, \bibinfo {author} {\bibfnamefont {C.}~\bibnamefont {Wang}},\ and\
  \bibinfo {author} {\bibfnamefont {J.}~\bibnamefont {Chen}},\ }\bibfield
  {title} {\bibinfo {title} {{Rarefaction effects on hypersonic boundary-layer
  stability}},\ }\href@noop {} {\bibfield  {journal} {\bibinfo  {journal} {Acta
  Mechanica Sinica}\ }\textbf {\bibinfo {volume} {40}},\ \bibinfo {pages}
  {123184} (\bibinfo {year} {2024})}\BibitemShut {NoStop}%
\bibitem [{\citenamefont {Burnett}(1935)}]{ref29}%
  \BibitemOpen
  \bibfield  {author} {\bibinfo {author} {\bibfnamefont {D.}~\bibnamefont
  {Burnett}},\ }\bibfield  {title} {\bibinfo {title} {{The distribution of
  velocities in a slightly non-uniform gas}},\ }\href@noop {} {\bibfield
  {journal} {\bibinfo  {journal} {Proceedings of the London Mathematical
  Society}\ }\textbf {\bibinfo {volume} {s2-39}},\ \bibinfo {pages} {385}
  (\bibinfo {year} {1935})}\BibitemShut {NoStop}%
\bibitem [{\citenamefont {Zhong}\ \emph {et~al.}(1993)\citenamefont {Zhong},
  \citenamefont {MacCormack},\ and\ \citenamefont {Chapman}}]{ref30}%
  \BibitemOpen
  \bibfield  {author} {\bibinfo {author} {\bibfnamefont {X.}~\bibnamefont
  {Zhong}}, \bibinfo {author} {\bibfnamefont {R.~W.}\ \bibnamefont
  {MacCormack}},\ and\ \bibinfo {author} {\bibfnamefont {D.~R.}\ \bibnamefont
  {Chapman}},\ }\bibfield  {title} {\bibinfo {title} {{Stabilization of the
  Burnett equations and application to hypersonic flows}},\ }\href@noop {}
  {\bibfield  {journal} {\bibinfo  {journal} {AIAA Journal}\ }\textbf {\bibinfo
  {volume} {31}},\ \bibinfo {pages} {1036} (\bibinfo {year}
  {1993})}\BibitemShut {NoStop}%
\bibitem [{\citenamefont {Grad}(1949)}]{ref31}%
  \BibitemOpen
  \bibfield  {author} {\bibinfo {author} {\bibfnamefont {H.}~\bibnamefont
  {Grad}},\ }\bibfield  {title} {\bibinfo {title} {{On the kinetic theory of
  rarefied gases}},\ }\href@noop {} {\bibfield  {journal} {\bibinfo  {journal}
  {Communications on Pure and Applied Mathematics}\ }\textbf {\bibinfo {volume}
  {2}},\ \bibinfo {pages} {331} (\bibinfo {year} {1949})}\BibitemShut {NoStop}%
\bibitem [{\citenamefont {Struchtrup}\ and\ \citenamefont
  {Torrilhon}(2003)}]{ref32}%
  \BibitemOpen
  \bibfield  {author} {\bibinfo {author} {\bibfnamefont {H.}~\bibnamefont
  {Struchtrup}}\ and\ \bibinfo {author} {\bibfnamefont {M.}~\bibnamefont
  {Torrilhon}},\ }\bibfield  {title} {\bibinfo {title} {{Regularization of
  Grad's 13 moment equations: Derivation and linear analysis}},\ }\href@noop {}
  {\bibfield  {journal} {\bibinfo  {journal} {Physics of Fluids}\ }\textbf
  {\bibinfo {volume} {15}},\ \bibinfo {pages} {2668} (\bibinfo {year}
  {2003})}\BibitemShut {NoStop}%
\bibitem [{\citenamefont {Eu}(1980)}]{ref33}%
  \BibitemOpen
  \bibfield  {author} {\bibinfo {author} {\bibfnamefont {B.~C.}\ \bibnamefont
  {Eu}},\ }\bibfield  {title} {\bibinfo {title} {{A modified moment method and
  irreversible thermodynamics}},\ }\href@noop {} {\bibfield  {journal}
  {\bibinfo  {journal} {Journal of Chemical Physics}\ }\textbf {\bibinfo
  {volume} {73}},\ \bibinfo {pages} {2958} (\bibinfo {year}
  {1980})}\BibitemShut {NoStop}%
\bibitem [{\citenamefont {Myong}(1999)}]{ref34}%
  \BibitemOpen
  \bibfield  {author} {\bibinfo {author} {\bibfnamefont {R.~S.}\ \bibnamefont
  {Myong}},\ }\bibfield  {title} {\bibinfo {title} {{Thermodynamically
  consistent hydrodynamic computational models for high-Knudsen-number gas
  flows}},\ }\href@noop {} {\bibfield  {journal} {\bibinfo  {journal} {Physics
  of Fluids}\ }\textbf {\bibinfo {volume} {11}},\ \bibinfo {pages} {2788}
  (\bibinfo {year} {1999})}\BibitemShut {NoStop}%
\bibitem [{\citenamefont {Jiang}\ \emph {et~al.}(2021)\citenamefont {Jiang},
  \citenamefont {Zhao}, \citenamefont {Chen},\ and\ \citenamefont
  {Yuan}}]{ref35}%
  \BibitemOpen
  \bibfield  {author} {\bibinfo {author} {\bibfnamefont {Z.}~\bibnamefont
  {Jiang}}, \bibinfo {author} {\bibfnamefont {W.}~\bibnamefont {Zhao}},
  \bibinfo {author} {\bibfnamefont {W.}~\bibnamefont {Chen}},\ and\ \bibinfo
  {author} {\bibfnamefont {Z.}~\bibnamefont {Yuan}},\ }\bibfield  {title}
  {\bibinfo {title} {{Eu's generalized hydrodynamics with its derived
  constitutive model: Comparison to Grad's method and linear stability
  analysis}},\ }\href@noop {} {\bibfield  {journal} {\bibinfo  {journal}
  {Physics of Fluids}\ }\textbf {\bibinfo {volume} {33}},\ \bibinfo {pages}
  {127116} (\bibinfo {year} {2021})}\BibitemShut {NoStop}%
\bibitem [{\citenamefont {Khayat}\ and\ \citenamefont {Eu}(1993)}]{ref36}%
  \BibitemOpen
  \bibfield  {author} {\bibinfo {author} {\bibfnamefont {R.~E.}\ \bibnamefont
  {Khayat}}\ and\ \bibinfo {author} {\bibfnamefont {B.~C.}\ \bibnamefont
  {Eu}},\ }\bibfield  {title} {\bibinfo {title} {{Generalized hydrodynamics and
  linear stability analysis of cylindrical Couette flow of a dilute
  Lennard-Jones fluid}},\ }\href@noop {} {\bibfield  {journal} {\bibinfo
  {journal} {Canadian Journal of Physics}\ }\textbf {\bibinfo {volume} {71}},\
  \bibinfo {pages} {518} (\bibinfo {year} {1993})}\BibitemShut {NoStop}%
\bibitem [{\citenamefont {Maxwell}(1879)}]{ref37}%
  \BibitemOpen
  \bibfield  {author} {\bibinfo {author} {\bibfnamefont {J.~C.}\ \bibnamefont
  {Maxwell}},\ }\bibfield  {title} {\bibinfo {title} {{On stresses in rarefied
  gases arising from inequalities of temperature}},\ }\href@noop {} {\bibfield
  {journal} {\bibinfo  {journal} {Philosophical Transactions of the Royal
  Society of London}\ }\textbf {\bibinfo {volume} {170}},\ \bibinfo {pages}
  {231} (\bibinfo {year} {1879})}\BibitemShut {NoStop}%
\bibitem [{\citenamefont {von Smoluchowski}(1898)}]{ref38}%
  \BibitemOpen
  \bibfield  {author} {\bibinfo {author} {\bibfnamefont {M.}~\bibnamefont {von
  Smoluchowski}},\ }\bibfield  {title} {\bibinfo {title} {{Ueber
  W{\"a}rmeleitung in verd{\"u}nnten Gasen}},\ }\href@noop {} {\bibfield
  {journal} {\bibinfo  {journal} {Annalen der Physik und Chemie}\ }\textbf
  {\bibinfo {volume} {64}},\ \bibinfo {pages} {101} (\bibinfo {year}
  {1898})}\BibitemShut {NoStop}%
\bibitem [{\citenamefont {Garz{\'o}}\ and\ \citenamefont
  {Santos}(2003)}]{ref39}%
  \BibitemOpen
  \bibfield  {author} {\bibinfo {author} {\bibfnamefont {V.}~\bibnamefont
  {Garz{\'o}}}\ and\ \bibinfo {author} {\bibfnamefont {A.}~\bibnamefont
  {Santos}},\ }\href@noop {} {\emph {\bibinfo {title} {Kinetic Theory of Gases
  in Shear Flows: Nonlinear Transport}}}\ (\bibinfo  {publisher} {Kluwer
  Academic Publishers},\ \bibinfo {address} {Dordrecht},\ \bibinfo {year}
  {2003})\BibitemShut {NoStop}%
\end{thebibliography}
\end{document}